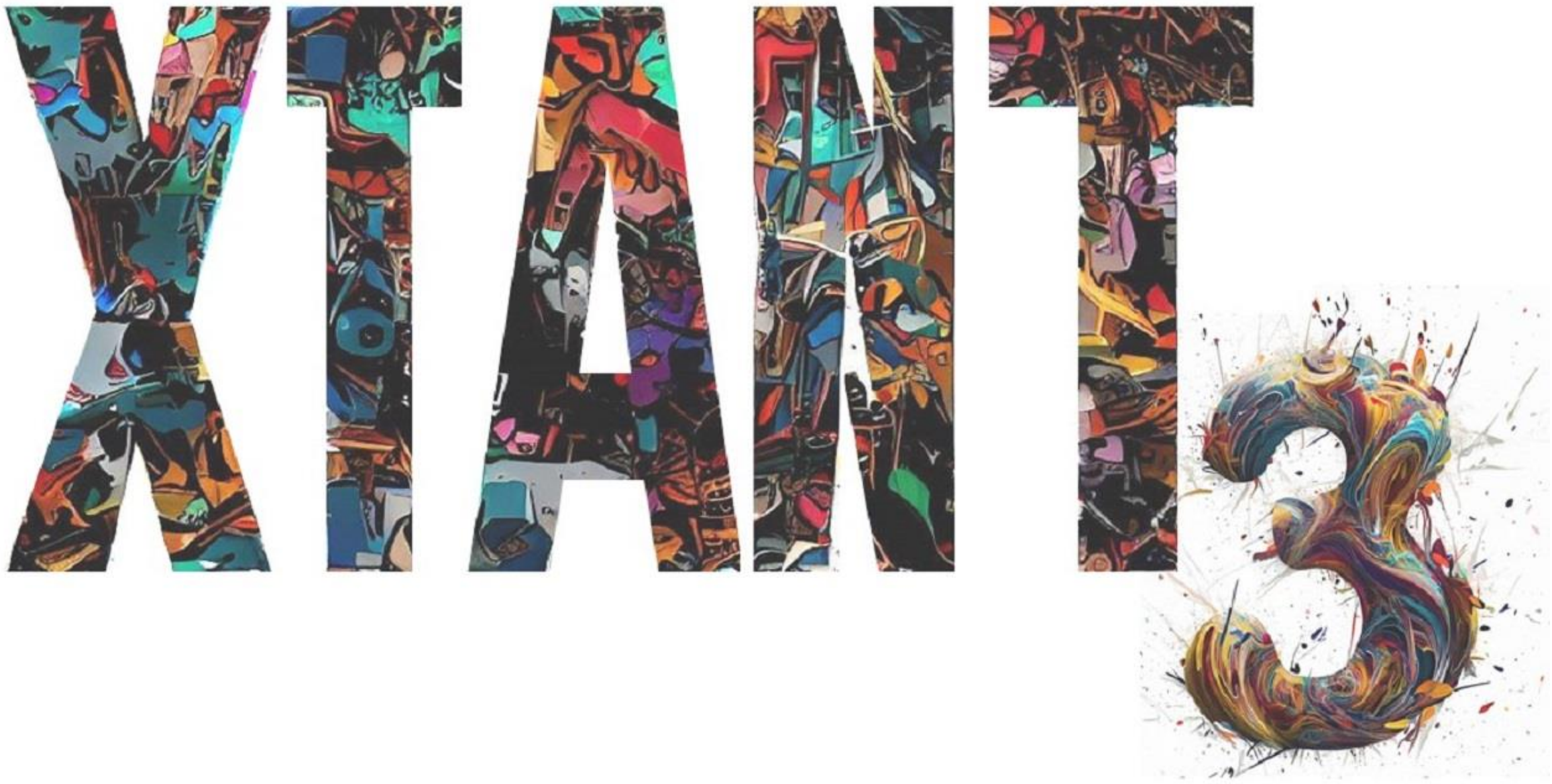

# *XTANT-3: X-ray-induced Thermal And Nonthermal Transitions in Matter*

theory, numerical details, user manual

Current version: XTANT-3 update: 28.09.2023

**Nikita Medvedev**[1]

This is the user manual for the hybrid code XTANT-3, simulating intense femtosecond X-ray irradiation of matter. The code combines a few models into one with feedbacks: transport Monte Carlo simulation, Boltzmann collision integrals, and tight binding molecular dynamics. Such a combination allows the simulation of nonequilibrium, nonadiabatic, and nonthermal effects in electronically excited matter, and the synergy and interplay of these effects. This text contains a description of the theoretical basis of the model and the practical user manual. The detailed description should allow new users, students, and non-specialists to access the ideas behind the code and make the learning curve less steep.

[1] Corresponding author: ORCID: 0000-0003-0491-1090, email: nikita.medvedev@fzu.cz

# Contents

## Disclaimer, how to cite

Although we endeavour to ensure that the code XTANT-3 and the results delivered are correct, no warranty is given as to its accuracy. We assume no responsibility for possible errors or omissions. We shall not be liable for any damage arising from the use of this code or its parts or any results produced with it, or from any action or decision taken as a result of using this code or any related material.

This code is distributed at https://github.com/N-Medvedev/XTANT-3 *as is* for non-commercial peaceful purposes only, such as research and education. It is explicitly **prohibited** to use the code, its parts, its results, or any related material for military-related and other than peaceful purposes.

By using this code or its materials, you agree with these terms and conditions.

The use of the code is at your own risk. Should you choose to use it, an appropriate citation is mandatory.

This document may be cited as:

*N. Medvedev "XTANT-3: X-ray-induced Thermal And Nonthermal Transitions in Matter: theory, numerical details, user manual"* (2023) https://doi.org/10.48550/arXiv.2307.03953

The journal citation that may be used is:

*N. Medvedev, V. Tkachenko, V. Lipp, Z. Li, B. Ziaja, "Various damage mechanisms in carbon and silicon materials under femtosecond x-ray irradiation", 4open. 1 (2018) 3.* https://doi.org/10.1051/fopen/2018003

Should you use electron-phonon coupling in the calculations, the following citation should be included in addition to the abovementioned one:

*N. Medvedev, I. Milov, "Electron-phonon coupling in metals at high electronic temperatures", Phys. Rev. B. 102 (2020) 064302.* https://doi.org/10.1103/PhysRevB.102.064302

In a publication, I recommend that at least the following parameters should be mentioned for the reproducibility of the results:

Material, its initial structure, the number of atoms in the supercell, the initial conditions (atomic and electronic temperatures), an ensemble used, a type of boundary conditions, a type of cross sections in Monte Carlo simulation, a type of tight binding parameterization, whether the electron emission was included or not and if yes, whether Coulomb potential for atoms was accounted for and what model for electron emission was used, whether an additional short-range repulsive potential was used, the time step of MD simulation, parameters of the incoming laser pulse (its photon energy, deposited dose, duration).

Most of these parameters can be found in an output file !OUTPUT_[*Material*]_Parameters.txt described below.

## Acknowledgements

This code would not exist (certainly not in the way it does now) without help of many colleagues, contributing with discussions and helpful comments. First of all, I thank B. Ziaja for valuable contribution to the general philosophy of the code and discussions on the results and various applications, which helped to developed the appropriate models in the code, in particular on the usage of the tight-binding (TB) molecular dynamics (MD) method; H. Jeschke for helpful discussions on the TBMD method; B. Rethfeld for sharing her experience on Boltzmann collision integrals, together with O. Osmani for prior collaborative works on combining MC method with other approaches; Z. Li for joint work on the nonadiabatic coupling, optical coefficients evaluation (RPA implementation in TB), and correlation analysis in MD; O. Vendrell and R. Santra for discussions on Born-Oppenheimer approximation and nonadiabatic effects; V. Tkachenko for contributions into development of the RPA module and help in debugging the code; V. Lipp for discussions on debugging the code; A.E. Volkov for discussions on the Monte Carlo (MC) part, electron-ion coupling and scattering theory, and on the multiscale approaches and applications in general; Z. Jurek for discussions and comments on programming.

The XTANT-3 logo was created with help of Bing Image Creator (powered by DALL-E).

# PART I: Theory

## I. Introduction

This document is specifically on the code XTANT-3[2] [1,2] and its features, many of which are unique to this code. Note that XTANT-3 is a separate development from XTANT+ [3], the two codes should not be confused. Although both grew out of the original XTANT code [4], years of independent development since then resulted in many differences in their philosophy, global structure, models used, and functionality (a reader interested in XTANT+ may consult its developers[3]). Problems faced in the combined code development resulted in many innovative technical solutions that had to be invented and which were not described elsewhere. The most important ones are presented here in detail with the hope that they might be useful not only to the users of XTANT-3 but also to other code developers facing similar issues. As multiscale codes are gaining popularity in recent years [5], I believe solutions specific to this class of methods may be valuable.

### I.1. Ultrafast irradiation problem

Ultrafast laser irradiation of materials plays an important role in both, fundamental and applied sciences [6–10]. Understanding of basic phenomena in the physics of solids, nonequilibrium kinetics, and highly excited states of matter, benefits from experiments accessing the natural time window of the involved processes (femtosecond timescales) such as electron kinetics, electron-ion (electron-phonon) coupling, and atomic response [10–12]. Laser irradiation experiments are vital for materials processing, nano and micro technology, and medical applications such as laser surgery [9,10,13,14]. In turn, the interpretation of experimental results requires detailed theoretical and model descriptions of the processes involved.

Those processes span multiple timescales, starting from attoseconds of the photon absorption from the incoming pulse to femtoseconds for electronic and holes kinetics, to picoseconds of the atomic/ionic response [5,15]. In photoabsorption, for photon energies above the bandgap, energy is delivered primarily to electrons of the matter [16]. The photoelectrons are excited to higher energy states, which drives the electronic ensemble out of equilibrium [16,17]. The electrons then scatter among themselves in the conduction band of material, with electrons from the valence band and deeper shell (for those shells whose ionization potential is below the electron energy), or elastically with the atoms of the target (also known as the electron-phonon scattering). Electron-electron scattering thermalizes the electronic ensemble towards equilibrium Fermi-Dirac distribution, whereas electron-atom scattering exchanges energy between the two systems, heating the lattice. Typical electronic thermalization in metals takes place at femtosecond timescales [6,18], while equilibration between the electronic and atomic temperatures may take up to tens of picoseconds [2,6].

Another important effect is, since the electrons form the interatomic potential of the material, the excitation of electrons modifies the potential energy surface [1,6,19]. Atoms, in their former equilibrium positions, now experience new forces, which may drive them into a different material phase – the processes known as nonthermal phase transitions, the most famous example of which is nonthermal melting in covalently bonded solids [20,21]. Furthermore, in the time window when the electronic temperature (or, more generally, parameters of the nonequilibrium electronic distribution) is different from the atomic one, this so-called two-temperature state departs from the equilibrium material phase diagram and may create states inaccessible under equilibrium conditions [22,23].

[2] https://github.com/N-Medvedev/XTANT-3
[3] https://xm.cfel.de/research/scientific_software/xtant_amp_xtant/

The variety of processes, taking place after ultrafast irradiation, may create synergy leading to nonlinear behavior [24]. For example, the nonthermal phase transitions were studied [25–27], in a few cases with nonadiabatic coupling included [1,28]. Effects of the electronic nonequilibrium were also analyzed separately [6,18,29,30]. The mutual effect of the electronic nonequilibrium and the electron-phonon coupling has also been demonstrated [31,32]. A combined effect of the interplay of the electronic nonequilibrium, nonadiabatic coupling, and nonthermal effects, was studied in Ref.[33] using XTANT-3 code.

The interplay of all these effects needs to be modeled in a unified framework – this is the main point of the XTANT-3 code. It combines on-the-fly a few appropriate models to address these most important effects taking place in the matter under ultrafast irradiation.

## I.2. Multiscale (hybrid, combined) simulation methods

To model the abovementioned effects, the concept of multiscale (or hybrid, or combined) modeling is employed in XTANT-3 (here, we use these terms interchangeably, although one may say that, in general, 'hybrid' and 'combined' approaches are more general terms than the 'multiscale') [5]. The philosophy of this type of approach is to combine a few models into one interconnected, executed together, and exchange the data on the fly. This can be achieved best in the case when the parameter space is separable: e.g., different processes take place at different timescales (models divisible in time [34,35]), particles of notably different masses are involved (models divisible by mass, such as quantum-classical combinations [36], or atomistic-continuum ones [37]), difference in the momentum space (separating close and distant collisions [38]), etc.

XTANT-3 combines four major models (and a countless number of minor ones) under one umbrella, as described below.

## I.3. XTANT-3 concept outline

XTANT-3 combines the following models to trace the essential effects in both, the electronic and the atomic systems of the target:

(i) The electrons with energy above a certain chosen energy cutoff are modeled with the Monte Carlo (MC) simulation.

(ii) The fractional populations of low-energy electrons on the valence and conduction band energy levels are traced with the Boltzmann equation (BE).

(iii) The interatomic forces are calculated from the transferable tight-binding (TB) formalism, which also traces the evolution of the electronic energy levels (molecular orbitals).

(iv) The motion of atoms is traced with help of the molecular dynamics (MD). Below, we will discuss the relevant details of each model, and their interconnection on-the-fly, enabling to model laser irradiation of matter.

The interconnection of these models is schematically shown in **Figure I.1** [39]. On top of that, there are a few models used for data analysis on-the-fly (as well as in post-processing), and minor models used within each of these general modules. Below we describe each model in detail.

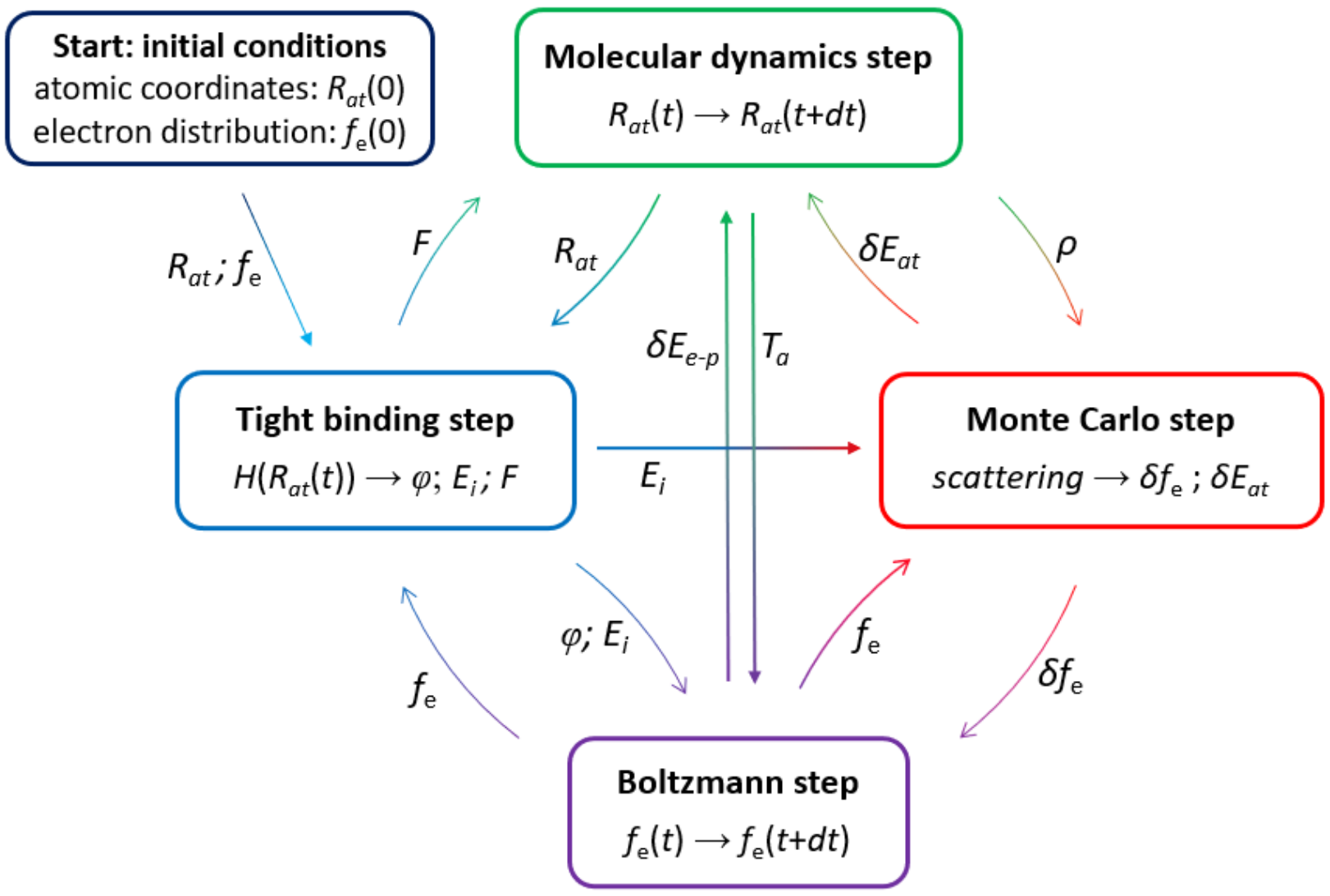

**Figure I.1. Schematics of the XTANT-3 algorithms, interconnecting the main modules of the program. The initial conditions specify the coordinates of all the atoms in the simulation box $R_{at}(t=0)$ and electronic distribution function $f_e$. The molecular dynamics (MD) module propagates the atomic coordinates, using the forces, F, calculated with the help of the tight-binding (TB) module. It also uses the electron transfer from the electronic system, calculated in two steps: elastic scattering of a high-energy fraction of electrons within the Monte Carlo (MC) step, $\delta E_{at}$, and electron-ion (electron-phonon) coupling of low-energy electrons calculated with the Boltzmann equation (BE), $\delta E_{el\text{-}phon}$. TB also provides the BE module with the transient electronic energy levels (eigenstates of the Hamiltonian), $E_i$, and wave functions (eigenfunctions) φ. $E_i$ are also used in MC. MC calculates the source terms for BE associated with the photoabsorption, Auger decays, and electronic scattering: $\delta f_e$. MD step supplies the BE with the atomic temperature Ta, and the MC with the transient material density ρ (if the supercell size is allowed to change in the simulation). Figure reproduced from Ref. [39].**

# II. Transport Monte Carlo simulation

## II.1. Basic concepts

A method that uses (pseudo-)random numbers belongs to the class of Monte Carlo (MC) methods [5]. Generally, in an MC method, any probability appearing in the governing equations may be replaced with a sampled random number uniform in the interval [0,1). If the same procedure is repeated multiple times, the averaged results would reproduce a statistically significant distribution of the desired quantity [38].

For the problem of fast particle transport in the matter, the MC method is used to sample probabilities of scattering [38,40–42]. It involves a few random numbers to sample various effects and processes. Although we denote all the random numbers in this book with the same symbol $\gamma$ (always uniformly distributed in the interval [0,1)), each time a different number is assumed, sampled with a random number generator. XTANT-3 uses standard FORTRAN random number generator. Even though there exist random number generators with proven randomness (e.g., [43]), since many different values are sampled in XTANT-3, possible artificial correlations introduced by the pseudo-randomness of the standard generators do not play a role here.

The following values are sampled randomly in XTANT-3: free flight distance of a mobile particle (photon, fast electron) or decay time of a core hole, type of scattering event (inelastic vs. elastic), the transferred energy in a scattering event, scattering partner in a collision. An MC schematic algorithm can be found, e.g., in Chapters 7 and 15 of Ref. [5].

The free flight time, used to calculate the time of the next scattering event of a particle, is evaluated from its free flight distance, λ, and particle velocity, $v$:

$$t = \lambda / v, \tag{1}$$

where the free flight distance of a particle is sampled from the Poisson distribution (which assumes uniform homogeneous media for the target) [38,41,42]:

$$\lambda = -\lambda_0 \ln(\gamma), \tag{2}$$

where the mean free path – the distance a particle travels between successive collisions, $\lambda_0$, – is defined via the total cross-section of scattering, $\sigma$:

$$\lambda^{-1} = \sigma n_{at} = n_{at} \int_{W_-}^{W_+} \int_{Q_-}^{Q_+} \frac{d^2\sigma}{dW dQ} dW dQ, \tag{3}$$

here $n_{at}$ is the density of target atoms; $W_\pm$ are the minimal and maximal transferred energy ($W = \hbar\omega$); the recoil energy $Q$ is defined by the transferred momentum in a collision as $Q(1 + Q/2m_t c^2) = \hbar^2 q^2/(2m_t)$ (in the nonrelativistic limit $Q = \hbar^2 q^2/(2m_t)$, with $\hbar q$ being the transferred momentum and $m_t$ being the scattering center (atom) mass). For the non-relativistic incident particle energy ($E \ll min(m_{in}, m_t)c^2$), the limits of the transferred momentum are [15,44]:

$$Q_\pm = \frac{m_{in}}{m_t}\left(\sqrt{E} \pm \sqrt{E - W}\right)^2, \tag{4}$$

with $m_{in}$ being the mass of the incident particle (e.g., an electron, $m_{in} = m_e$ free-electron rest mass). The lower limit of the transferred energy ($W_-$) and the upper limit ($W_+$) for scattering on a particle (a target atom or an electron) are defined by the following formulae:

$$\begin{cases} W_- = I_p \\ W_+ = \dfrac{4 m_{in} m_t}{(m_{in} + m_t)^2} E \end{cases}, \tag{5}$$

where $I_p$ is the ionization potential of the atomic shell an electron is being ionized from. The upper limit is written here for a free particle. For scattering on a bound particle with a given ionization potential, the expressions are more complicated, see Ref. [45], but the free-particle approximation works very well for $W_+ >> I_p$. In the case of ionization from the valence band $I_p = E_{gap}$. Here, $E_{gap}$ is the material band gap, which is equal to zero in the case of metals. For the relativistic ones, the reader may refer, e.g., to Refs. [15,38].

An *event-by-event* (or analog) MC simulation traces each scattering event in detail, which is sufficient for tracing material response [46]. Transferred energy in a scattering act is sampled according to the following expression:

$$\gamma\sigma = \int_{W_-}^{W} \int_{Q_-}^{Q_+} \frac{d^2\sigma}{dW dQ} dW dQ, \tag{6}$$

here $\gamma \in [0,1)$ is another random number; Eq.(6) must be solved for the transferred energy *W*. In some cases, the differential equation allows for analytical integration and a closed solution for *W* [38,45]. In general, a numerical solution is required [47], or further approximations to the electron spectrum should be employed [38].

In the first-order Born approximation (linear response), the cross-section of scattering may be represented as a cross-section of scattering on the individual scattering center (Coulombic) and the factor responsible for the collective dynamics of the target particles [15,48]. It then may be represented in terms of the thermal factor and the inverse of the complex dielectric function (CDF, $\varepsilon(\omega,q)$) [15,48]:

$$\frac{d^2\sigma}{d(\hbar\omega)d(\hbar q)} = \frac{2(Z_e e)^2}{n_{at}\pi\hbar^2 v^2}\frac{1}{\hbar q}\left(1 - e^{-\frac{\hbar\omega}{k_B T}}\right)^{-1} \mathrm{Im}\left[\frac{-1}{\varepsilon(\omega,q)}\right] \tag{7}$$

where *e* is the electron charge; $\hbar\omega$ is the transferred energy and $\hbar$ is Planck's constant; $k_B$ is the Boltzmann constant, and *T* is the temperature of the sample; $Z_e$ is the effective charge of the projectile penetrating through a solid (for an incident electron $Z_e = 1$); the particle velocity, *v*, and a transferred momentum, *q*. Relativistic cross sections may be found, e.g., in Refs. [15,38,48].

In XTANT-3, the following convention is used: $n_{sc}$ is the atomic density for scattering on core electrons and is equal to the molecular density for the valence (or conduction) band electrons.

## II.2. Photons

Photons in XTANT-3 experience only one type of interaction: absorption by target electrons in the linear regime (single-photon absorption). This limits the applicability of the code to the photon energies above the band gap of the material (since no multiphoton absorption via virtual levels is not included; note also that it limits the laser intensity at low photon energy to exclude non-linear effects). At high photon energy, it is limited by another approximation for decays of core holes (see below) by the energy of ~30 keV.

The number of photons per pulse is estimated from the user-given absorbed dose. The temporal profile of the pulse may be chosen as one of three options: (i) Gaussian pulse; (ii) flat-top pulse; (iii) spiky pulse mimicking SASE operation of free-electron lasers [49].

A photon, arriving at the supercell at a time sampled according to the chosen pulse profile, will be absorbed by an electron from the shell $N_{sh}$, chosen randomly according to the partial photoabsorption cross-sections for different atomic shells of all elements of the target from the condition:

$$\gamma = \sum_{i}^{N_{sh}} \sigma_i \Big/ \sigma_{tot} \tag{8}$$

where the total cross-section of photon absorption is summed over all shells of all elements: $\sigma_{tot} = \sum_i^{N_{all}} \sigma_i$. The cross sections are taken from the EPICS2017 database[4] (unless CDF-function is provided by the user, see below, in which case the CDF-base cross section is used instead of the atomic ones from this database [47]).

An electron is emitted with the kinetic energy $E_e = \hbar\omega - I_p$ (photon energy of $\hbar\omega$, the ionization potential of the chosen shell $I_p$). A hole in the corresponding shell is created, unless the absorbing shell is the valence/conduction band of the material (in which case its particular energy level is sampled according to the transient electron distribution function traced within the Boltzmann equation as described in Section III), and its decay time is sampled, see below.

[4] https://www-nds.iaea.org/epics/

### II.3. High-energy electrons

Electrons, excited by photons (as well as by impact-ionization by fast electrons or *via* Auger-decays of core holes) to the energy states above a chosen cutoff, $E_{cut}$, then may experience two kinds of scattering events: elastic (in which the total kinetic energy is conserved between the incident particle and the scattering center – meaning, no ionization), and inelastic (in which the kinetic energy is not conserved – a part of the energy is spent to ionization; an impact ionization event). Additionally, if allowed by the user, an electron may be emitted from the target and lost from the MC simulation.

#### II.3.a. Inelastic scattering (impact ionization)

Inelastic scattering (impact ionization) is described by one of the two possible cross-sections: CDF-based or binary-encounter Bethe (BEB) atomic cross-section (if CDF is not available). The CDF function (entering Eq.((7)) is approximated with the set of artificial oscillators according to the Ritchie-Howie formalism [44,50]:

$$Im\left(\frac{-1}{\varepsilon(\omega,0)}\right) \approx \sum_i \frac{A_i \gamma_i \hbar\omega_i}{((\hbar\omega_i)^2 - E_i^2)^2 + (\gamma_i \hbar\omega_i)^2} \qquad (9)$$

This simple analytical representation depends on a set of parameters $(A_i, E_i, \gamma_i)$ determined from the fitting procedure of the optical data [51]. The parameters represent the amplitude $A_i$, the position $E_i$, and the width $\gamma_i$ of the $i^{th}$ oscillator, and may be interpreted as intensity, energy, and an inverse lifetime of collective excitation (plasmon or phonon) associated with the peaks [15]. CDF parameters may be reconstructed from the optical coefficients, as described in detail, e.g., in Refs. [15,44,51]. Many materials' CDF-files may be found in TREKIS-3 code[5]; alternatively, a single-pole approximation may be used for a rough estimation of the CDF coefficients [52]. The position of the oscillator may be chosen according to the position of a collective mode of the particles oscillations (the plasmon mode for inelastic scattering on the valence/conduction band electronic system; ionization potential for core shells):

$$\begin{aligned} E_{0_{sp}} &= \hbar\Omega_p \\ \gamma_{sp} &= E_{0_{sp}} \\ A_{sp} &= (\hbar\Omega_p)^2 \Big/ \int_0^\infty W Im\left(\frac{-1}{\varepsilon(W, Q=0, A=1)}\right)_D dW \end{aligned} \qquad (10)$$

Here, $\Omega_P^2 = 4\pi e^2 n_e / m_e$, the plasmon frequency, with $n_e$ being the valence/conduction band electron density (or core-shell electron density). The $A_{sp}$, the normalization coefficient, is defined by the *k*-sum rule [52].

For materials, for which CDF coefficients are unavailable, as an alternative to the single-pole CDF approximation, the cross-sections of scattering on an independent atom (instead of a solid) may be used. XTANT-3 allows using of BEB cross-sections of impact ionization [53]:

[5] https://github.com/N-Medvedev/TREKIS-3

$$
\begin{gathered}
\frac{d\sigma}{dw} = S\sum_{n=1}^{3} F_n\left(\frac{1}{(1+w)^n} + \frac{1}{(K-w)^n}\right), \\
F_1 = -\frac{F_2}{K+1}, F_2 = \frac{1}{K+u+1}, F_3 = \frac{\ln(K)}{K+u+1}, \\
S = 4\pi a_0^2 N\left(\frac{Ry}{I_p}\right)^2; \\
\sigma = \frac{S}{K+u+1}\left[\frac{\ln(K)}{2}(1-K^{-2}) + (1-K^{-1}) - \frac{\ln(K)}{K+1}\right].
\end{gathered}
\tag{11}
$$

Here $w = dE/I_p$ is the energy transferred to the electron being ionized in this collision; $I_p$ is the ionization potential of the shell being ionized; $u = U/I_p$ is the normalized mean kinetic energy of the shell from which the electron is ejected; $K = E_e/I_p$ is the kinetic energy of the incident electron normalized to the binding energy of the shell it is scattering on; $N$ is the number of electrons in the corresponding shell, $a_0$=0.53 Å is the Bohr radius, and $Ry$=13.6 eV is the Rydberg constant.

Ionization potential and shell-kinetic energies used are taken from the EPICS2017 database [54]. Note that in the case of BEB cross sections, atomic ionization potentials are used for the valence shells (instead of the bandgap of the material), since the parameters of the BEB cross sections were fitted to the atomic data. This may lead to certain problems in the mismatch of the parameters; these problems and solutions used in XTANT-3 will be discussed below in Section II.6.

The electron's energy is then updated, and the next collision is then sampled for this kinetic energy. The sampled energy lost by the incident electron is then transferred to a new electron, and a new hole is created (either in a core-shell or in the valence/conduction band). The new electron's next collision is sampled in the same manner as the primary electron.

If the electron loses energy below the predefined cutoff, it is eliminated from the MC simulation, and its energy is stored for further formation of the source term for the Boltzmann equation, see below. If the hole is created in the valence/conduction band, its energy level is then also used to form the source-term for BE.

### II.3.b. Elastic scattering

At the relativistic energies, the cross-section of elastic (nuclear) scattering, and the corresponding energy loss, can be obtained, e.g., from the Mott or Wentzel-Moliere screened scattering cross sections [55–57]:

$$
\frac{d\sigma}{d\Omega} = \left(\frac{Z_t e^2}{p_e v_e}\right)^2 \frac{1}{(2\eta + 1 - \cos(\theta))^2}, \tag{12}
$$

here $\Omega$ is the solid angle, $p_e$ is the incident electron momentum and $v_e$ is its velocity, $\theta$ is the scattering angle, and $\eta$ is a screening parameter [41]. The screening parameter is used in the modified Moliere form [41]:

$$
\eta = 1.7\times 10^{-5} Z_t^{2/3}\left(\frac{1-\beta^2}{\beta^2}\right)\times\left(1.13 + 3.76\left(\frac{\alpha Z_t}{\beta}\right)^2\sqrt{\frac{E_e}{E_e + m_e c^2}}\right), \tag{13}
$$

Where $c$ is the speed of light in vacuum, $m_e$ is the free electron mass, $\beta^2 = 2E_e/m_e c^2$, and $\alpha = 1/137$ is the fine structure constant.

The corresponding total cross-section used to calculate the elastic mean free path (Eq.(3)) is then:

$$\sigma = \pi a_0^2 \frac{Z_t(Z_t+1)}{\eta(\eta+1)} \left(\frac{Ry}{E_e}\right)^2, \tag{14}$$

Where $E_e$ is the incident electron energy, $a_0$=0.53 Å is the Bohr radius, and $Ry$=13.6 eV is the Rydberg constant (used for normalization of energy here).

The scattering angle can be sampled according to the differential and total cross sections (Eqs. (12)-(14)) as:

$$\cos(\theta) = \frac{\gamma(2\eta+1)-\eta}{\eta+\gamma}, \tag{15}$$

Transferred energy in a collision (connected to the recoil angle $\theta_r$ *via* $dE = \frac{4 m_{in} m_t}{(m_{in}+m_t)^2} E_e \cos^2(\theta_r)$, where $M_{at}$ is the mass of a target atom) can be then calculated from the energy and momentum conservation laws [15,38].

The electron loses the defined amount of energy in the collision. Its next time of collision is then calculated for the updated energy. The energy loss $dE$ is stored, to be later averaged over MC iterations and delivered to atoms, see Section V.5.

If the electron loses energy below the predefined cutoff, it is eliminated from the MC simulation, and its energy is stored for further formation of the source term for the Boltzmann equation, see below.

## II.4. Core holes

Considering the typical parameters XTANT-3 was developed for, the Auger of Koster-Kronig) decay is the dominant process of core-shell decay (not too high photon energies produce not too deep holes, for which radiative decays would be dominant). Thus, all decay in MC models is characterized by the atomic Auger (or Koster-Kronig) times [38,54,58–60]. In the MC simulation, the decay time is sampled with the exponential (Poisson) distribution:

$$t = -t_0 \ln(\gamma), \tag{16}$$

where $t_0$ is the characteristic Auger-decay time, which is taken from the EPICS2017 database [52].

In such a case, the hole jumps into the closest higher shell. The energy release in such a transition is approximated as the difference between the ionization potentials of these shells involved. In the XTANT-3 algorithm, it is assumed that this energy is released as a (virtual) photon, which is then instantly absorbed by one of the atoms. Thus, the shell, absorbing this "photon" is sampled according to the photoabsorption cross-section – such a simulation allows for an interatomic Auger (Knotek-Feibelman) decay [61] since the "absorbing" shell may belong to a different element of a compound target. Its absorption then creates a free electron – the Auger electron – and the second holes, thereby completing the Auger decay.

The new excited electron then undergoes the same sampling procedure as a photo-electron, as well as both newly created holes.

If a hole is produced in the valence/conduction band, the hole is excided from the MC simulations. The level in which this hole appeared (sampled according to the transient electron distribution function defining populations on the transient valence energy levels) and its corresponding energy is then recorded to form the source term for the Boltzmann equation, as described below.

### II.5. Electron emission

In simulations where electrons may be emitted from the surface of the material (such as thin layers, nano samples, near-surface regions), this process must be taken into account. It leads to a few effects to consider.

Two simplistic conditions for electron emission are available in XTANT-3 (set by the user):

i) Emission after a given number of collisions

ii) Emission of all electrons with energy above a defined threshold (e.g., work function)

When one of these conditions (chosen by the user) is satisfied, an electron is eliminated from the MC simulation, and its energy is essentially lost (tracked only for energy balance check, but no processes involving this electron are performed). The number of lost electrons is traced to account for the accumulation of the positive charge to be added to the Coulomb potential of atoms traced in MD, see Section V.4.

### II.6. Exceptions, patches, numerical tricks

If the cutoff energy used is larger than the width of the conduction band of the material (which may be the case of small tight-binding bases in certain parameterizations, see details in Section IV), an electron falling below the cutoff may appear above the highest available conduction band level ($E_{CB} < E < E_{cutoff}$) and will not find the level to merge in the Boltzmann equation (see the next section). Such cases must be excluded from simulation, which is done by an automatic adjustment of the cutoff energy to be not larger than the width of the conduction band.

Another potential problem that may arise is if the cutoff energy is smaller than the material bandgap ($E_{cutoff} < E < E_{gap}$). In such a case, electrons may be stuck in limbo, as they would not be able to lose energy and join the low-energy fraction modeled with the Boltzmann equation. Such cases are excluded in XTANT-3 by automatic adjustment of the cut-off energy to be larger or equal to the transient bandgap.

## III. Boltzmann equation

### III.1. Basic concepts, collision integrals

The methodology, first introduced in XTANT-3 in Ref. [33], uses the following Boltzmann collision integral formulation:

$$\frac{df_e(\varepsilon_i,t)}{dt} = I_{e-e} + I_{e-a} + I_{MC}. \quad (17)$$

Where the distribution function of electrons, $f_e$, is defined on the transient energy levels (the transient molecular orbitals defined in the tight binding module, see Section IV.2), and the source terms here are responsible for the electron-electron scattering $I_{e-e}$, electron-atom (or electron-phonon) $I_{e-a}$, and the source term or article and energy coming in or out to the valence/conduction energy levels from the MC module (high-energy electrons falling below cutoff, holes created in scattering events).

An explicit finite time difference scheme is used for solving the Boltzmann equation (15).

## III.2. Electron thermalization

For the electron-electron scattering [33], the relaxation time approximation (RTA) is used in the current implementation [62]:

$$I_{e-e} = -\frac{f_e(\varepsilon_i,t) - f_{eq}(\varepsilon_i,\mu,T_e,t)}{\tau_{e-e}}. \qquad (18)$$

Here $\tau_{e-e}$ is the characteristic electron-electron relaxation time (defined by the user); $f_{eq}(\varepsilon_i,\mu,T_e,t)$ is the equivalent equilibrium Fermi-Dirac distribution with the same total number of (low-energy) electrons ($n_e$) and energy content ($E_e$) as in the transient nonequilibrium distribution:

$$\begin{cases} n_e = \sum f_e(\varepsilon_i,t) = \sum f_{eq}(\varepsilon_i,\mu,T_e,t) \\ E_e = \sum \varepsilon_i f_e(\varepsilon_i,t) = \sum \varepsilon_i f_{eq}(\varepsilon_i,\mu,T_e,t) \end{cases}. \qquad (19)$$

Eqs. (19) define the equivalent electronic temperature (also called the kinetic temperature, $T_e$ [63]) and the equivalent chemical potential (μ). Within this ansatz, the total number of low-energy electrons and the total energy (in electrons and atoms) are conserved within an MD timestep (changes in the number of electrons may only occur *via* the $I_{MC}$ term).

Eqs. (17)-(18) naturally unify various widely used approaches to quantum-classical dynamics [33]. It recovers the limiting cases of the Born-Oppenheimer (BO) molecular dynamics (in the limit of infinite electronic thermalization time, $\tau_{e-e} \to \infty$, and no nonadiabatic electron-atom coupling, $I_{e-a} = 0$) in which the electronic populations are fixed; the Ehrenfest dynamics which includes average electron-atom energy exchange but no electron thermalization ($\tau_{e-e} \to \infty$, $I_{e-a} \neq 0$); instantaneous thermalization in the adiabatic microcanonical ensemble ($\tau_{e-e} = 0$, $I_{e-a} = 0$, used e.g. in Refs.[3,4]); and nonadiabatic dynamics with instantaneous electron thermalization ($\tau_{e-e} = 0$, $I_{e-a} \neq 0$, used in the previous versions of XTANT-3 [1,2]; the same assumptions are used in the two-temperature-based models, TTM-MD [37,64,65]). It was also noted that the BO approximation only assumes the decoupling of the electronic and atomic wave functions, but not necessarily the ground state – excited adiabatic states may also be calculated in many cases, where potential energy surfaces are far apart, thus suppressing electronic transitions between them (far from such situations as conical intersections or avoided crossings) [66]. The ground-state BO molecular dynamics is a separate additional approximation, which can be reproduced within the relaxation-time formalism as the zero-temperature instantaneous thermalization with no coupling ($\tau_{e-e} \to \infty$, $I_{e-a} = 0$, and $T_e = 0$) [33].

The RTA is commonly used for description of metals, however in semiconductors or insulators, the electron scattering is often divided into two different channels: the intraband (within the valence or the conduction band) and the interband (between the valence and the conductions band, such as three-body recombintation and impact ionization). The intraband electron thermalization is usually faster than the interband one. The first channel leads to partial equilibration for electrons, separately within the valence band (often called holes thermalization) and the conduction band, each with their own tempereatures and chemical potentials. Then, the interband thermalization may take place at longer timescales.

To accommodate the possible different thermalization rates in the valence band, the conduction band, and the total thermalization between the bands, it is possible to introduce partial electronic temperatures and chemical potentials for the two bands, with the same Eqs.(18)-(19) with separate parameters: $f_{eq}^{b}(\varepsilon_i, \mu_b, T_{e,b}, t)$ (where $b$ stands for the band index: valence or conduction) [67]. In this case, the Eq. (19) needs to be solved three times: for the valence band, conduction band, and total electronic ensemble, to defined the three equivalent temperatures and chemical potentials. Then, the separate equilibration rates can be used for each band ($\tau_{e-e}^{b}$), and for the total equilibration (between the bands).

Note that an instantaneous global thermalization produces the conditions of the Two-Temperature Model: separate temperatures of the electrons and the atoms, whereas separate instantaneous thermalizations in the valence band and the conduction band (but not between the bands) reduces to the Three-Temperature Model: different temperatures for excited electrons (in the conduction band), holes (valence-band electrons), and atoms [67].

### III.3. Electron-ion (electron-phonon) coupling

For the nonadiabatic electron-atom (electron-ion, often called electron-phonon) coupling the scattering integral $I_{e-a}^{ij}$ is defined as the time derivative of the electron distribution function,

$$\frac{df(\varepsilon_i, t)}{dt} = \sum_j I_{e-a}^{ij}, \qquad (20)$$

For solids, the method based on the ideas of Tully's surface hoping is used, modified for efficient treatment of a large number of electrons in the modeled ensemble: it can be used to obtain matrix elements (or probabilities $W_{ij} = \left|\langle\psi_j(t)|\psi_i(t_0)\rangle\right|^2$ with the wave-function defined via the TB Hamiltonian, see Section IV) entering the scattering integral $I_{e-a}^{ij}$ [2]:

$$I_{e-a}^{ij} = w_{ij} \begin{cases} f(\varepsilon_j)[2 - f(\varepsilon_i)] - f(\varepsilon_i)[2 - f(\varepsilon_j)]e^{-\varepsilon_{ij}/T_a}, \text{for i} > j \\ f(\varepsilon_j)[2 - f(\varepsilon_i)]e^{-\varepsilon_{ij}/T_a} - f(\varepsilon_i)[2 - f(\varepsilon_j)], \text{otherwise} \end{cases}$$
$$w_{ij} = \frac{dW_{ij}}{dt} \approx 2\frac{\left|\langle\psi_j(t)|\psi_i(t_0)\rangle\right|^2}{\delta t} \approx \frac{4e}{\hbar\delta t^2}\sum_{\alpha,\beta}\left|c_{i,\alpha}(t)c_{j,\beta}(t_0)S_{i,j}\right|^2 \qquad (21)$$

Here again $f(\varepsilon_i)$ is the electron distribution function, normalized to 2 due to spin degeneracy; $\varepsilon_{ij} = \varepsilon_i - \varepsilon_j$; the time derivative is approximated with the finite difference method for the molecular dynamics time step $\delta t$, and the wave functions are calculated correspondingly on two consecutive steps: $t_0$ and $t=t_0+\delta t$; using the linear combination of atomic orbitals (LCAO) basis set within the tight binding Hamiltonian, $\psi_i = \sum_\alpha c_{i,\alpha}\varphi_\alpha$, and $S_{\alpha,\beta}$ is the overlap matrix (in the case of an orthogonal Hamiltonian $S_{\alpha,\beta} = \delta_{\alpha,\beta}$). The exponential terms result from the Maxwellian distribution of the atomic ensemble, and, in general, may be replaced with an integral of the transient nonequilibrium atomic distribution. The factor of $2e/(\hbar\delta t)$ ($e$ is the electron charge and $\hbar$ is the Planck's constant providing the dimensionality of time consistent with that of the MD timestep to render the multiplier dimensionless) is an *ad hoc* correction to eliminate the dependence on the time-step, see [2].

Such a method of combining surface hopping with the Boltzmann collision integral (instead of a random sampling of electronic hops) enables direct calculation of energy flows between the quantum mechanical electrons and classical atoms in the simulation, and, by extension, the coupling parameter [2]. Note that the method is applicable for arbitrary atomic displacements, without assuming harmonic potential and crystalline solids (phonons) – it also works for amorphous materials, liquids, and rapidly changing structures.

### III.4. Source terms

The influx or outflux of electrons from the valence and conduction band (low-energy electrons) *via* such processes as photoabsorption, scattering of high-energy electrons, and Auger decays, are traced in the MC as described above. In each scattering act involving the valence and conduction electrons, the energy level involved is sampled and recorded into the $I_{MC}$. The levels are sampled according to probability following the Pauli blocking term in the Boltzmann collision integral: $w_i \sim f(\varepsilon_i)(2 - f(\varepsilon_i + \Delta E))$, where $\Delta E$ stands for the energy transferred in the collision under consideration (factor 2 is due to spin degeneracy). All the changes in populations in each energy level within the given timestep are then averaged over the MC iterations [33].

Since an incoming electron generally comes with energy somewhere in between the predefined discrete energy levels, the influx of the particles and energy is distributed between the two closest levels under the condition of conservation of the number of particles and energy: increasing the total number of low-energy electrons by one out of $N_{MC}$ iterations and bringing energy $\Delta E$ in the scattering event [33]:

$$\begin{cases} \Delta f_e(\varepsilon_i, t) = \Delta n \dfrac{\varepsilon_j - \Delta E}{\varepsilon_j - \varepsilon_i} \\ \Delta f_e(\varepsilon_j, t) = \Delta n \dfrac{\Delta E - \varepsilon_i}{\varepsilon_j - \varepsilon_i} \end{cases} \tag{22}$$

Where for one incoming electron out of $N_{MC}$, the change in the number is $\Delta n = 1/N_{MC}$, and the corresponding energy change is $\Delta E = E_{el}/N_{MC}$ ($E_{el}$ is the energy brought by this electron). This way, the total number of electrons and energy, in the low- *and* high-energy fractions (MC and BE modules), are conserved. In the case when one of the levels ($i$ or $j=i+1$) is fully occupied, another set of levels can be chosen – Eqs. (22) hold for arbitrary levels $i$ and $j$. The total change of the population on each level is then summed over all scattering acts within the given timestep: $I_{MC} = \sum \Delta f_e(\varepsilon_i, t)$ [33].

### III.5. Exceptions, patches, numerical tricks

Since the source term is gathered from all the independent MC iterations to calculate average quantities, a situation may occur that the source term could lead to changes in the distribution function beyond the physically allowed limits (dropping below 0 or rising above 2). In such a case, the extra number of electrons (the part below zero or above 2) is then redistributed between the closest allowed levels under the conditions Eqs. (22).

In a rare case (typically occurring under extremely high photon fluences – high doses and short pulses), this may not be possible: changes in the distribution function may be too large to be accommodated by such a redistribution. In this case, extra thermalization steps are used (Eq.(18) with artificial time-steps introduced) until the distribution function smoothens closer to the equilibrium Fermi to be within the physical limits (between 0 and 2).

# IV. Tight binding model

## IV.1. Basic concepts, Slater-Koster approximation

XTANT-3 uses the classic Slater-Koster (SK) approximation [68] for evaluation of the Hamiltonian in the tight-binding approximation (using solid harmonics), see educational overview in Ref. [69]. XTANT-3 supports s, p, and d rotational subroutines, which currently allows for the construction of the LCAO basis sets s, $sp^3$, $sp^3s^*$, $sp^3d^5$ (and potentially $sp^3d^5s^*$). At present, basis sets containing f-orbitals are not supported.

Within this method, the Hamiltonian is written in the matrix form via pairwise interaction terms between the atoms (hopping integrals $h_{ij}$). It is generally written in the nonorthogonal basis set, thus overlap matrix *S* is also often necessary (that depends on the particular parameterization; XTANT-3 can deal with both, orthogonal and nonorthogonal parameterizations). The hopping integrals and overlap matrix in TB are split to the radial function depending only on the distance between the two atoms in the simulation, and the angular SK term $Y(l,m)$ (defined by the angular quantum numbers [68,69]):

$$h(R_{ij}) = V(R_{ij})Y(l,m) \tag{23}$$

Such a representation allows for analytical calculation of the derivatives of the potential energy surface, which is convenient for the MD simulations, see below.

## IV.2. Electron energy levels (band structure; molecular orbitals)

Hamiltonian constructed from the on-site terms and hopping integrals is then diagonalized (either directly, or via the Löwdin method for nonorthogonal parameterizations [70]):

$$\varepsilon_i = \langle i|\hat{H}(\{R_{ij}(t)\})|i\rangle \tag{24}$$

Here $\varepsilon_i$ are the electronic energy levels, or molecular orbitals, or electronic band structure – the eigenstates of the electronic Hamiltonian $\hat{H}$ that depends on all atomic positions in the simulation box, and $|i\rangle$ is an eigenvector of this Hamiltonian [69].

## IV.3. Transferable tight binding parameterizations

To make the method transferable to multiple atomic structures, the radial part in the hopping integrals is a function depending on the interatomic distance, not merely constants fitted to reproduce a particular structure of the particular material [69]. XTANT-3 has a few modules to deal with several TB parameterizations: for elemental metals, NRL TB parameterization [71,72] is introduced, and for C and Si the Pettifor-type parameterization can be used, Refs. [73,74], or DFTB parameterization [75].

The potential energy of an atom in the second quantization tight binding formalism can be approximated as a contribution of the ionic repulsion and attraction to electrons [27,73]:

$$V(\{R_{ij}(t)\},t) = E_{rep}(\{R_{ij}(t)\}) + \sum_i f\left(\varepsilon_i(\{R_{ij}(t)\},t)\right)\varepsilon_i, \tag{25}$$

where the potential *V* depends on distances between all the atoms in the simulation box $\{R_{ij}(t)\}$, $E_{rep}$ is effective ion-ion repulsion term (containing all contributions apart from the electronic band energies), and $f_i$ is fractional electron occupation numbers (distribution function) on the transient molecular orbitals.

This is the so-called non-self-consistent TB or zero-order TB. The self-consistent-charge (SCC [76]) iterative subroutine is also included in XTANT-3, but only for Born-Oppenheimer simulation – it currently cannot be used for other types of simulation, since it is a nontrivial task to combine the SCC method with NVE ensemble of simulation or nonequilibrium evolution of the electronic distribution function. Note that the SCC typically assumes Mulliken charges [77] on various elements, and thus high-order TB terms exactly vanish for elemental materials.

Interatomic forces are calculated by analytical derivatives of the given potential energy surface (see e.g. Ref. [62]).

# V. Molecular dynamics

## V.1. Initial conditions

Maxwellian distribution for all atoms in the simulation box for the desired temperature $T_{at}$ can be set with the help of three independent random numbers $\gamma_1$, $\gamma_2$, $\gamma_3$ (all uniformly distributed in the interval (0,1]):

$$E_i = 2T_{at}(-\ln(\gamma_1) - \ln(\gamma_2)\cos^2(\gamma_3\pi/2)) . \tag{26}$$

Where $E_i$ is the kinetic energy of the $i^{th}$ atom, from where its velocity can be calculated; the direction of the velocity (or momentum) is then set uniformly in the solid angle.

This procedure creates a Maxwellian distribution of atomic energies/velocities with the temperature of $2T_{at}$, considering that energy will then be quickly redistributed equally between the kinetic and potential energies (the equipartition theorem), producing the kinetic temperature of $T_{at}$. This also assumes that the initial atomic coordinates must be set in the potential minimum, producing no additional configurational temperature (associated with the potential energy of atoms) [63,78]. Atoms located not in the potential minimum will accelerate, thereby producing a higher temperature than required.

In practice, equilibration of the atomic ensemble to the temperature $T_{at}$, starting from these initial conditions, takes only a few atomic oscillations (typically, times of a few tens to a few hundred femtoseconds).

## V.2. Atomic motion

As is typical for molecular dynamics simulations, we use Lagrange formalism for deriving equations of motion of atoms. Its advantage is in the fact that extended variables can be introduced straightforwardly in the equations of motion, defining, for example, the evolution of the supercell (simulation box) in the case of used barostats, as will be described below in Section V.7. The forces acting on the atoms are calculated as the gradients of the potential energy defined within the TB formalism (see Section IV.3) and possibly from additional classical potentials (see Section V.4):

$$M_i \frac{d^2\boldsymbol{R}_i}{dt^2} = -\frac{\partial V(\{R_{ij}\})}{\partial \boldsymbol{R}_i} . \tag{27}$$

Here $M_i$ is the mass of a simulated particle (an ion or a target atom), $\mathbf{R}_i$ is its coordinate vector and $R_{ij}$ is the distance between a pair of atoms *i* and *j*, $V(\{R_{ij}\})$ is interaction potential (collective potential energy surface or a pairwise potential, depending on all atoms *j* involved).

Equations of motion (27) for the ion-matter scattering may be solved using molecular dynamics (MD) simulations [79]. The set of Eqs. (27) is solved for all the atoms in the simulation box by numerical discretization of time into time steps, and propagating coordinates of all atoms accounting for the dynamical change of the potential of interaction among them.

XTANT-3 supports three integrators of atomic trajectories: Verlet (2nd order) [80], Yoshida (4th order) [81], and Martyna-Tuckerman (4th order) [82].

Additionally, the nonadiabatic energy transfer from the Boltzmann collision integral and the energy transfer from high-energy electrons elastic scattering are fed to atoms *via* velocity scaling at each timestep, which ensures the energy conservation in the entire system: all electrons and atoms (microcanonical ensemble), see Section V.5.

Using Eq.(17) allows directly tracing an effect of the nonequilibrium electronic distribution function on the interatomic forces in Eq.(25) (as well as on the electron-atom coupling, $I_{e-a}$ [2]), and thus the dynamics and stability of the material under such conditions. As mentioned above, various approximations for the relaxation time also enable comparisons of different standard methods (e.g., BO, instantaneous thermalization approximation).

Note that the formalism of coupled BE-TBMD, at its core, relies on the Ehrenfest dynamics, and not on the finite-temperature extension to *ab initio* simulations [83]. The difference is, the underlying physical picture of the finite-temperature models is that the constant electronic temperature is enforced by the interaction with the bath. In our case, however, even in the limit of the instantaneous electron thermalization ($\tau_{e-e} \rightarrow 0$), no interaction with the bath is assumed, and the electronic thermalization is an intrinsic process (non-mean-field effect added). The electronic temperature and chemical potential are not variables in this formulation but parameters. As was noted in Ref. [33], Eq.(18) could be replaced with the electron-electron Boltzmann collision integral, which does not employ any electronic temperature or chemical potential, even as parameters of the equivalent equilibrium distribution [18].

## V.3. Artificial dynamics: coordinate path, zero-temperature MD

Instead of the MD simulation, XTANT-3 allows to calculate all the properties of the system over a given coordinates path between two chosen states of matter (by setting the initial and final states in the initial data). This may be helpful for the coordinate-path calculations, allowing to identify potential barriers along predefined paths.

XTANT-3 also allows nullifying of atomic velocities (and velocities of the supercell vectors in Parrinello-Rahman MD) in every given number of MD steps (quenching). This method reduces to the common zero-temperature (steepest descent) MD in the case if atomic velocities are being nullified at each time step of the simulation. Choosing a larger step of nullifying the velocities (once in *N* simulation steps) allows to speed up the search of the potential minimum, especially in case a barostat is used and the evolution of the supercell vectors needs to be modeled (which is typically significantly slower than the atomic motion and thus requires nullifying its “velocities” not as often as the atomic ones).

### V.4. Interatomic potentials

TB-calculated potential energy surface in some cases needs to be augmented with additional classical potentials. It is very common to have additional short-range repulsive potential, which is typically included within the TB parameterization. In some cases, such short-range potential may be needed to be added externally. XTANT-3 supports a range of such potentials, in the form of polynomial, exponential, Coulomb, and universal ZBL [84] potentials (see details in Part II, Section VI. 4(e)).

Coulomb potential is softly truncated in the calculations [85]. In the case of electron emission included, XTANT-3 counts the average charge accumulated due to the emission and distributes it evenly among all the atoms in the simulation, creating additional Coulomb potential, which thus evolves in time depending on the number of emitted electrons.

Long-range potentials may also be used in XTANT-3, e.g. to include dispersion corrections (London corrections, etc. [86]). Currently, only (improved) Lennard-Jones-type (LJ) potentials approximating van der Waals (vdW) forces are included [85]. Long-rage potential must be softly cut at short distances not to overlap with the TB-potential (a subject of further development to include models such as, e.g., d2, d3, and d4 dispersion corrections [87]). The vdW potential in the LJ form may be represented in three most common forms: 12-6 form (ε-σ form), AB form, and n-exp form:

The ε-σ form:

$$V_{LJ}(r) = 4\varepsilon\left(\left(\frac{\sigma}{r}\right)^{12} - \left(\frac{\sigma}{r}\right)^{6}\right) \tag{28}$$

The AB form:

$$V_{LJ}(r) = \frac{A}{r^{12}} - \frac{B}{r^{6}} \tag{29}$$

The n-exp form:

$$V_{LJ}(r) = \varepsilon\left(\left(\frac{r_0}{r}\right)^{2n} - 2\left(\frac{r_0}{r}\right)^{n}\right) \tag{30}$$

Which is the most convenient form since the coefficients have a clear physical meaning: $r_0$ is the equilibrium interatomic distance, and ε is the depth of the potential at this equilibrium distance.

It is also possible to use so-called improved Lennard-Jones (ILJ) potential [88]:

$$V_{ILJ}(r) = \varepsilon\left(\frac{m}{n-m}\left(\frac{r_0}{r}\right)^{n} - \frac{n}{n-m}\left(\frac{r_0}{r}\right)^{m}\right) \tag{31}$$

which reduces to the standard Lennard-Jones for $n$=12, $m$=6.

### V.5. Source terms, electron-ion coupling (two-temperature MD)

Energy, transferred to atoms from high-energy electrons *via* elastic scattering (from MC module) and from the nonadiabatic coupling of low-energy electrons (from BE module), is distributed to atoms at each time-step of MD simulation *via* velocity scaling:

$$v_i' = v_i\sqrt{1 + \frac{dE}{E}}, \tag{32}$$

Where $E = \sum_i E_i$ is the total kinetic energy of all the atoms in the simulation box, $dE$ is the total energy transferred from electrons at this time step, $v_i$ is the velocity of the $i^{th}$ atom before the scaling, and $v_i'$ is its velocity after the scaling (corresponding to the total energy increased by $dE$). This Eq.(32) ensures that the total kinetic energy in the simulation box increases (decreases) by exactly the amount of $dE$ while conserving the total momentum in the simulation box.

### V.6. Thermostats

If high gradients in the energy/temperature in the target are present, the effects of the transport of energy may be important. Simulation box sizes that can be modeled with ab-initio-like MD are too small to capture such transport effect. To account for energy transport out of the simulation box, thermostats may be used. They allow the temperature or the energy in the supercell to change towards a defined temperature over a given characteristic time. XTANT-3 allows to use generalized Berendsen thermostat [89] for both, electronic and atomic systems (with independent parameters in the two systems). It is generalized in the sense that instead of a linear rate equation for the temperature, it uses a full exponential solution [52]:

$$T(t + \delta t) = T_0 + (T(t) - T_0)\exp(-\delta t/\tau), \tag{33}$$

Where $T(t + \delta t)$ and $T(t)$ are the atomic temperature on the current and previous time steps (before and after cooling), $T_0$ is the bath temperature (typically, room temperature, 300 K), $\tau$ is the characteristic cooling time, and $\delta t$ is the MD time step. Note that Eq.(33) is an exact solution of the equation for the temperature rate, in contrast to the linearized one in the original work [89], and thus it does not require the condition $\tau \gg \delta t$ [52].

For electrons, in the case of instantaneous thermalization, the same Eq.(33) is used for the electronic temperature; in the case of nonequilibrium electron simulation, the same type of equation is used but for the distribution function instead of the temperature:

$$f_e(\varepsilon_i, t + \delta t) = f_{eq}(\varepsilon_i, \mu, T_0, t) + (f_e(\varepsilon_i, t) - f_{eq}(\varepsilon_i, \mu, T_0, t))\exp(-\delta t/\tau), \tag{34}$$

With the equilibrium distribution function to relax towards.

### V.7. Barostats: variable supercell

For simulations, where the system needs to change the size and shape of the supercell, ensembles of constant volume do not do the job. Such situations arise, for example, when different phases of material are required to be modeled. A phase transition between phases with different densities and symmetry may be modeled with MD using a barostat allowing the simulation box to change adjusting to the changes in the pressure tensor. XTANT-3 uses the Parrinello-Rahman barostat [90], in which case the supercell vectors are introduced in the Lagrangian of motion as additional variables. Their equation of motion is then defined by the pressure tensor.

# VI. Data analysis

## VI.1. Excited electrons and holes

XTANT-3 traces the number of high-energy electrons and the number of core-shell holes within the MC module. Additionally, it defines the number of the conduction band electrons (per atom in the simulation) as the number of electrons on the TB energy levels above the LUMO (lowest unoccupied molecular orbital) state at $T_e$=0.

## VI.2. Electronic structure

The electronic structure is calculated at the Gamma point according to Eq.(24) unless the calculation of DOS is specified by the user. In the calculation of the forces, however, only Gamma-point calculations are used in all cases. It limits the simulation size to a sufficiently large number of atoms (typically, 64 atoms is sufficient).

## VI.3. Mulliken charge

Mulliken charge analysis [77] is performed by XTANT-3 for different elements in the target. Charges on individual atoms may be printed out, as well as the average charges on various chemical elements in the simulation.

## VI.4. Electron distribution function, thermalization and entropy

The electronic entropy $S_e = -k_B 2 \sum[(f_e/2)\ln(f_e/2) + (1 - f_e/2)\ln(1 - f_e/2)]$, with $k_B$ being the Boltzmann constant, and factors of 2 due to the normalization of the distribution function according to the spin degeneracy. XTANT-3 prints out the transient entropy according to the transient distribution function, and according to the equivalent equilibrium distribution function (corresponding to the maximal entropy, to compare and identify the degree of the deviation of the electronic state from the equilibrium) [33].

## VI.5. Temperatures, electronic and atomic

Electron temperature is used as the kinetic or equivalent temperature defined by Eqs. (19).

Atomic temperature is defined as the classical kinetic temperature in the periodic system (hence, factor of $\frac{2}{(3N_{at}-6)}$ accounting for the reduced number of degrees of freedom, instead of 2/3 for free particles) [15,63]:

$$T_{kin} = \frac{2}{k_B(3N_{at} - 6)} \sum_{i=1}^{N_{at}} \frac{M_i v_i^2}{2} \tag{35}$$

## VI.6. Energy balance and conservation

XTANT-3 traces the evolution of the total energy in the system, which must conserve in NVE simulations (without quenching, thermostats, or barostats), except for the laser pulse. At the time of arrival of the laser pulse, the energy increases by the value, defined by the deposited dose.

### VI.7. Pressure and stress tensor

The pressure tensor is defined according to the Parrinello-Rahman method (Section V.7), which is calculated even in NVE simulations (without calculating the equation of motion for the supercell vectors) [90].

### VI.8. Electronic heat capacity

The electronic heat capacity is calculated via the derivative of the electron entropy with respect to the electron temperature at a constant volume: $C_e(T_e, T_a) = T_e(\partial S_e/\partial T_e)_V$ [91]. Under the assumption of thermodynamic equilibrium, the electronic distribution function adheres to the Fermi-Dirac distribution, so the heat capacity (per area of material) reduces to the following expression:

$$C_e(T_e, T_a) = \frac{1}{V_0}\sum_i \frac{\partial f_e(E_i)}{\partial T_e}(E_i - \mu(T_e)), \qquad (36)$$

where the derivative of the Fermi-Dirac function by the (kinetic) electronic temperature is:

$$\frac{\partial f_e(E_i)}{\partial T_e} = \frac{2\exp((E_i - \mu(T_e))/k_B T_e)}{k_B T_e (1 + \exp((E_i - \mu(T_e))/k_B T_e))^2}\left(\frac{E_i - \mu(T_e)}{T_e} + \frac{\partial \mu(T_e)}{\partial T_e}\right) \qquad (37)$$

Where the equivalent (kinetic) temperature and chemical potentials are used, and the derivative $\partial\mu(T_e)/\partial T_e$ is calculated numerically; the factor of two is due to spin degeneracy.

### VI.9. Electron-ion (electron-phonon) coupling parameter

As described in Section III.3, in the one-particle approximation, the electron-ion coupling parameter *G* is defined *via* [18]:

$$G(T_e, T_a) = \frac{1}{V_0(T_e - T_a)}\sum_{i,j} \varepsilon_j I_{e-a}^{ij} \qquad (38)$$

Where summation runs over all electronic states *i* and *j*. See an example of the coupling parameter, calculated and averaged over 10 simulation runs, in Figure VI.1, according to the methodology described in [92].

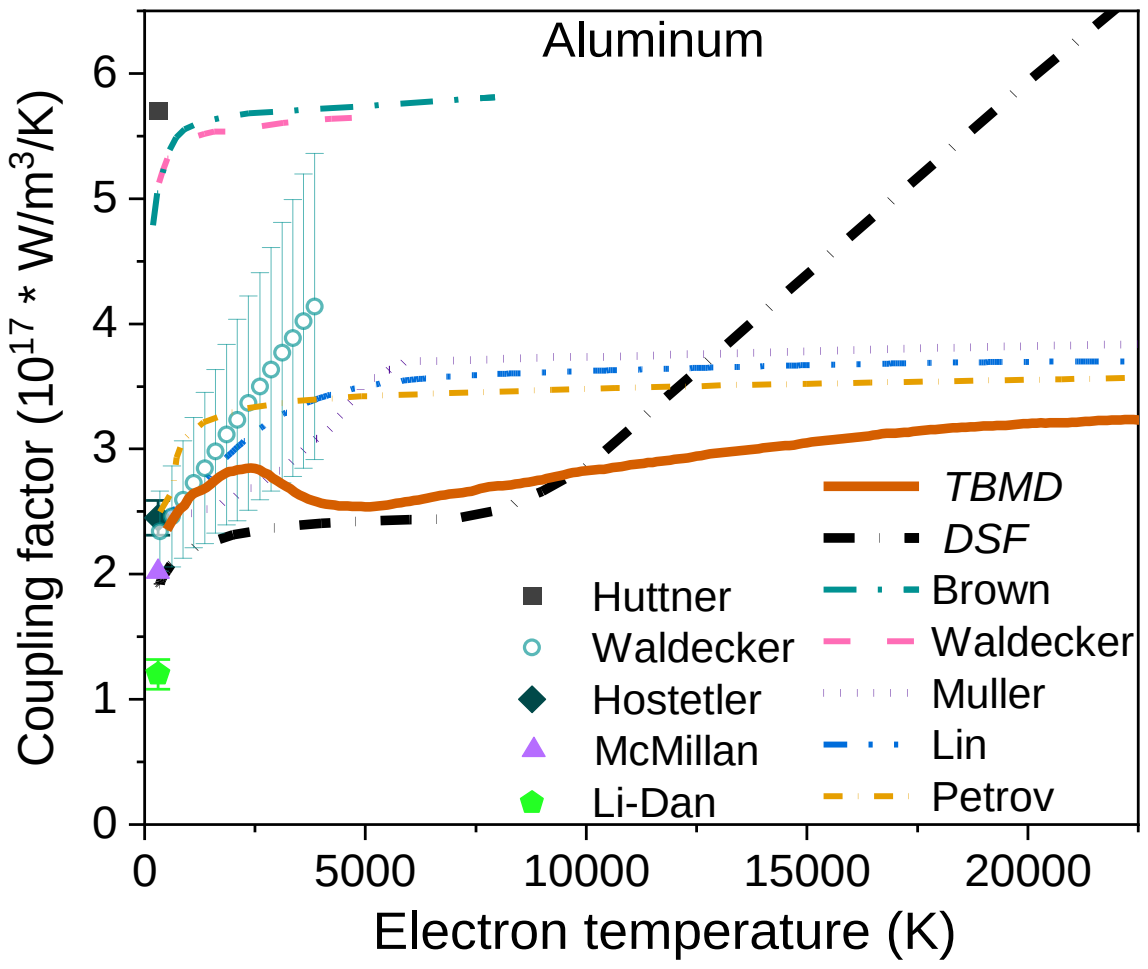

**Figure VI.1. Electron-ion coupling parameter as a function of electron temperature in aluminum calculated within the XTANT-3 TBMD approach (Eq.(31)) [2]; DSF-based approach by Gorbunov et al.** [93]**; compared with other estimates by Lin et al.** [94]**, McMillan** [95]**, Allen et al.** [96]**, Brown et al.** [97]**, Petrov et al.** [98]**, Muller et al.** [99]**, Waldecker et al.** [100]**. Experimental data for comparison are from Huttner et al.** [101]**, Waldecker et al.** [100]**, Hostetler et al.** [102]**, and Li-Dan et al.** [103]**. Reproduced from Ref.** [2]

## VI.10. Atomic structure and displacement

Atomic structure is printed out in various formats and then can be analyzed with the standard MD visualization tools such as OVITO [104], VMD [105], Mercury [106], etc. Mean atomic displacement (in arbitrary degree, e.g. linear or square displacement) may be printed out for all chemical elements of the target. This, e.g., allows to identity melting into a liquid phase, for which the average square displacement is proportional to time ($d \sim \sqrt{t}$).

## VI.11. Pair correlation function

The pair correlation function (or radial distribution function) is defined as:

$$g(\boldsymbol{r}) = \frac{1}{\rho} \langle \sum \delta(\boldsymbol{r} - \boldsymbol{r_i}) \rangle \tag{39}$$

Similarly, the phonon spectrum can be calculated with help of the Fourier transform of the velocity autocorrelation function (with the damping factor $a$) [107]:

$$I(\omega) = \int \langle v(t) v(0) \rangle \exp{(-a t^2)} \exp{(-i \omega t)} dt \tag{40}$$

## VI.12. Neighbors' analysis and fragmentation

XTANT-3 simulation package allows us to evaluate the number of nearest neighbors within the given radius, and count the number of fragments in case the system may break up in the simulation (e.g., ablation). Those fragments are then sorted by mass to create a mass spectrum.

To create a spectrum in units of m/z (mass over charge of the fragments), as observed in experiments, an average charge of the fragments can be used. In XTANT-3, fractional Mulliken charge on each atom may be calculated, and then summed for each fragment. In this case, the integer charge is calculated by splitting the total fragment charge between the two neighboring states to calculate the weights and add two "effective fragments" to the spectrum: One with the mass/charge ratio as M/($\lfloor Z \rfloor$) and one with M/($\lfloor Z \rfloor$+1), where $\lfloor Z \rfloor$ is the nearest integer smaller than the fractional charge Z. The weight of the first fragment is then (Z-$\lfloor Z \rfloor$), and of the second one is ($\lfloor Z \rfloor$-Z+1).

## VI.13. Optical coefficients

XTANT-3 can evaluate transient optical properties: complex dielectric function (CDF), complex index of refraction, and optical coefficients (transmittance, reflectance, and absorbance), within the random phase approximation (RPA, or, equivalently, Kubo-Greenwood approach [108]). The complex dielectric function within RPA is defined as follows [109]:

$$\varepsilon^{\alpha\beta}(\omega) = \delta_{\alpha\beta} + \frac{e^2\hbar^2}{\Omega m_e^2 e_0} \sum_{nn'} \frac{F_{nn'}}{E_{nn'}^2} \frac{f_{n'} - f_n}{\hbar\omega - E_{nn'} + i\gamma} \tag{41}$$

Here Ω is the volume of the simulation box, $E_{nn'} = E_{n'} - E_n$ is a transition energy between two eigenstates $|n'\rangle$ and $|n\rangle$; $f_{n'}$ and $f_n$ are the corresponding transient occupation numbers (electron distribution function) normalized to two accounting for the spin degeneracy; $m_e$ is the free-electron mass; $e$ is the electron charge; ħ is the Planck constant; and $e_0$ is the vacuum permittivity (in SI units). The parameter $\gamma$ is an inverse electron relaxation time (in XTANT-3 set as $\gamma = 1.5\times10^{14}$ 1/s; a particular choice of $\gamma$ does not affect the results beyond the broadening of peaks in the CDF [109]). $F_{nn'}$ are the oscillator strength corresponding to the energy transition [110]:

$$F_{nn'} = |\langle n'|p|n\rangle|^2 = \left| \sum_{R,R',\sigma,\sigma'} B_{\sigma n}(\boldsymbol{R}) P(\boldsymbol{R},\boldsymbol{R}') B_{\sigma' n'}(\boldsymbol{R}') \right|^2 \tag{42}$$

where $\boldsymbol{R}$ are the coordinates of atoms; $\sigma$ are the atomic orbitals; and $B_{\sigma n}(\boldsymbol{R})$ and $B_{\sigma' n'}(\boldsymbol{R}')$ are the corresponding eigenvectors of the TB Hamiltonian matrix $H(\boldsymbol{R},\boldsymbol{R}')$. There are two approaches to *approximately* calculate the momentum operator with the TB formalism:

1) directly using nonorthogonal Hamiltonian and overlap matrix (see [111]), in which case:

$$P(\boldsymbol{R},\boldsymbol{R}') = \frac{m_e}{\hbar}\left( \frac{\partial H(\boldsymbol{R},\boldsymbol{R}')}{\partial k} - E_n \frac{\partial S(\boldsymbol{R},\boldsymbol{R}')}{\partial k} \right) \tag{43}$$

2) Or, first orthogonalizing it (e.g., with Löwdin method [70]); then, using the operator identity, the momentum operator matrix elements can be calculated within the TB as [110]:

$$P(\boldsymbol{R},\boldsymbol{R}') = i\frac{m_e}{\hbar}[\boldsymbol{R} - \boldsymbol{R}']H_{orth}(\boldsymbol{R},\boldsymbol{R}')$$
$$|\langle n'|p|n\rangle|^2 = \left| \sum_{R,R',\sigma,\sigma'} B_{\sigma n}^{orth}(\boldsymbol{R}) P(\boldsymbol{R},\boldsymbol{R}') B_{\sigma' n'}^{orth}(\boldsymbol{R}) \right|^2 \tag{44}$$

This simple representation allows to calculate the CDF within the TB method. Knowing the CDF, optical coefficients can be obtained *via* the complex refractive index: $n(\omega) = \sqrt{\varepsilon(\omega)}$. The optical coefficients depend on the polarization of light (of the probe pulse), its angle of incidence, and the thickness of the layer, according to the standard theory of light propagation in multilayer systems (see [112]).

Currently, XTANT-3 can only calculate either single-ray coefficients, or a coherently summed infinite number of reflections. The first case describes well the limit of thick material (if the absorption is high enough so that the rays reflected from the back surface do not contribute significantly). The latter case corresponds to the limit of ultrathin layer, so that many reflections are possible within the duration of the probe-pulse to create interference.

### VI.14. Electronic heat conductivity

The total electronic conductivity consists of two terms: the electron-atom (or electron-phonon, $\kappa_{e-a}$) and electron-electron ($\kappa_{e-e}$) scattering contributions combined using the Matthiessen's rule [113]:

$$\kappa_{tot} = \left(\frac{1}{\kappa_{e-a}} + \frac{1}{\kappa_{e-e}}\right)^{-1} \quad (45)$$

The electron-phonon part of the electronic heat conductivity dependent on the electronic temperature ($T_e$) may be approximately calculated within the Kubo-Greenwood formalism [114]:

$$\kappa_{e-a} = L_{22} - T_e \frac{L_{12}^2}{L_{11}} \quad (46)$$

Where the Onsager coefficients are [114]:

$$L_{ij} = -\frac{(-1)^{i+j}}{\Omega m_e} \sum \frac{df}{dE_k} (E_k - \mu)^{i+j+2} \, |\langle k|p|k'\rangle|^2 \quad (47)$$

The momentum matrix elements in the Onsager coefficients used for calculation of the conductivity may be approximated with Eqs.(44). Since only the atomic displacements are used for evaluation of the momentum matrix elements with this representation, this term only describes the electron-atom heat conductivity.

The electron-electron scattering probabilities are not accessible within the TB formalism, but its contribution to the conductivity can be evaluated with the electron-electron scattering cross sections instead. We use the same formalism with the Onsager coefficients replaced with the following classical expression:

$$L_{ij}^{e-e} = -\frac{(-1)^{i+j}}{\Omega m_e^2} \sum \frac{df}{dE_k} (E_k - \mu)^{i+j-2} \frac{v\lambda}{3} \quad (48)$$

For the electron-electron scattering, the electron velocity, $v$, may be approximated as a free-particle velocity (here counted from the bottom of the valence/conduction band), and the inelastic mean free path, $\lambda$, can be obtained with the cross section of electron-electron scattering. We use the same cross-sections as in the MC module of XTANT-3, namely, the complex dielectric function formalism (Eqs.(7)-(10)) [50].

# PART II: User Manual

## I. Limits of applicability of XTANT-3

The code has the following limitations due to approximations of the model:

- The photon absorption is modeled in the linear regime only. Since no simultaneous multi-photon absorption is included, the lowest allowed photon energy is limited by the material bandgap (for insulators). When modeling photon energy that is smaller than the energy cut-off (see description below), nonequilibrium electron kinetics may be included via the relaxation time approximation.
- The upper limit for the photon energy is ~20-30 keV (no radiative decays of core holes are included, which becomes important for heavy elements at higher photon energies; no relativistic effects for electrons are included).
- The deposited dose is limited to ~5 eV/atom, depending on the TB parameterization of the material. The code *may be* used at higher doses, if you understand the risks, limitations, and shortcomings – such as, e.g., the results at ultrashort timescales may still be reliable, or ultrafast cooling or electron emission that may quickly reduce the dose, etc.
- At low deposited doses, the electron-phonon coupling model is unreliable, it is limited to electronic temperatures above some ~2-3 kK.
- The number of atoms in the simulation box must be larger than some 200-300, defined by the two factors: 1) the TB cut-off radius must be smaller than half of the simulation box with periodic boundaries; 2) the electron-phonon coupling, calculated at the gamma-point only, requires a few hundred atoms for convergence (must be checked for each material).
- Duration of simulation is typically limited to a few picoseconds, due to possible accumulating MD instabilities – they must be checked by the energy conservation in the simulation. Poor convergence (non-conserved energy) should be corrected by reducing the MD time step. Note: the energy drift may be mediated by using Berendsen thermostats.

## II. Compiling XTANT-3

If the executable (e.g. XTANT.x, or XTANT.exe) does not exist, compile the source files. With the help of the makefile, the compilation under Linux is as follows: go into the directory with the whole code (all necessary files that must be present are described below), and execute the command

```
make
```

in a terminal. It will compile the code and create an executable named XTANT.x in the same folder. It might take a few moments. It automatically compiles the code with OpenMP[6].

Then execute the corresponding shell-script, which specifies paths to MKL libraries on your computer/cluster. Examples of the shell scripts prepared for running XTANT.x on some clusters it was used before include:

[6] More on OpenMP with fortran: http://www.openmp.org/presentations/miguel/F95_OpenMPv1_v2.pdf

./XTANT_DESY.sh - to run XTANT-3 at the cluster in DESY (Hamburg)

./XTANT_Metacentrum.sh – to run XTANT-3 on Luna cluster as part of the Metacentrum (Prague)

Alternatively, if you are using Windows (with Intel Fortran compiler, OpenMP, and MKL-library installed, and paths set in the environmental variables), rename the file Make.bat.txt into Make.bat, and execute it. It will create an executable XTANT.exe.

Run this XTANT.exe in case you are using Windows.

In case the program cannot find libraries required for OpenMP parallelization, you might have to specify the paths to them manually by executing commands similar to these (insert your current paths accordingly):

```
export LD_LIBRARY_PATH=/opt/intel/2011/lib/intel64:$LD_LIBRARY_PATH
export LD_LIBRARY_PATH=/opt/products/mkl/11.0/mkl/lib/em64t:$LD_LIBRARY_PATH
```

Some other libraries might be missing on your workstation. In this case, find the paths to them and export them analogously.

In case of having problems with a lack of memory (which XTANT-3 usually does), you must set it unlimited by executing

```
ulimit -s unlimited
limit stacksize unlimited
```

(For Windows, use the compiler option /F9999999999)

These four commands usually are not necessary since they are already inside of the XTANT.sh

If you wish to recompile the code without OpenMP, first you have to clean up all the old compiled files with the command

make clean

After that, you can compile the code without OpenMP by specifying

make OMP=no

Again, new recompilation with OpenMP must be preceded by making clean, because there are pre-processing options included. So, each recompilation that changes the involvement of OpenMP must be done *only after cleaning up* the files. This option is mainly for code-developing, since executing is much faster with parallelization with OpenMP.

Another option for debugging during the development of the code is compiling with:

make DEBUG=yes

Under Windows a user may compile it in the following way: Make.bat X or make X, where X is one of the following options (no '*make clean*' needed for recompiling):

DEBUG - for debugging during the developing of the code (such as checking arrays boundaries, undeclared variables, etc.), compiles XTANT_DEBUG.exe.

DEBUGOMP or db - to compile parallel version with all debug options on, compiles XTANT_DEBUG_OMP.exe.

FAST or slow – to compile parallelized code with no optimizations and no debug (fast compilation, not-the-fastest execution), compiles XTANT_OMP.exe.

For a compilation for release, simply use Make.bat (or make), with no additional option, which compiles XTANT.exe (or XTANT.x, correspondingly).

## III. Compiling additional post-processing programs

XTANT-3 package has additional useful programs for data analysis, saved in the folder !XTANT_ANALYSIS_SUBROUTINES. They can be compiled all together by using Make.bat (under Windows) file *in this* directory.

## IV. Running XTANT-3 with additional options

To include additional options in the code, you can run it with some additional options as follows:

./XTANT.sh X (or XTANT.x X, or XTANT.exe X)

where X is an option available. At the moment, there are only a few options:

1) help – If you need some short info on the XTANT-3 when starting the code, you can also call the following help-commands: help (so, run it as ./XTANT.sh help, or correspondingly ./XTANT.x help or XTANT.exe help). It will printout the numbers and meaning of possible errors in the error-file (see below), how to communicate with the program on-the-fly (see below), etc. This flag will only print some info and stop the execution of the code; no calculations will be performed.
2) info – Before running the code will print out some basic information about the XTANT. Running with this option will tell you some information about references, disclaimers, how to cite the code, and similar info. This flag will only print some info and stop the execution of the code; no calculations will be performed.
3) size – If you want to create the file with cohesive energy of the material as a function of the nearest neighbor distance (to compare with other works), run XTANT-3 with this option will create an output file named OUTPUT_Energy.dat with the energy as a function of the nearest neighbor distance, see a description below. Note: this file is overwritten every time you run the program with the flag “size”.

*Note:* if the size option is specified in the input file (INPUT.txt or INPUT_MATERIAL.txt, see below), then the user may (optionally) specify the parameters of the grid for the variable size change in the next line in the following format:

min_a, max_a, points

where min_a and max_a are real(8) parameters setting the minimal and maximal sizes in the units of the supercell, and points is the number of grid points to be used.

The default options are: min_a=0.7, max_a=2.1, points=300. They are used if there is no line with alternative parameters provided.

4) allow_rotation – By default, initializing MD removes the total angular momentum of the system. If you want to start MD simulation without removing it, which will allow the whole system to rotate (e.g. might be useful for modeling individual molecules), use this flag.

5) matter or material – To list all available materials in the directory INPUT_DATA. XTANT-3 will read out the existing directories with the corresponding material parameters, create the file List_of_materials.txt and also print it on the screen. After that, the code terminates; no calculations will be performed.

6) verbose – If you want to see a lot of information on the screen during the simulation run (such as, when which subroutine was called, etc.), which may be useful for debugging and testing, use this option.

7) For a regular simulation run without additional options simply call:

./XTANT.sh or XTANT.exe for Windows.

Note that XTANT-3 supports as many flags at a simulation run as you need, they can be combined.

Also note that all these additional options may be listed at the end of the input file INPUT_MATERIAL.txt, as described below.

# V. Files of the code

This code XTANT-3 is contained in several files that have to be compiled together, plus additional files for post-processing of the data if needed. All the files listed below must be in the same directory for XTANT-3 code compilation and execution:

- Makefile – this file uses a standard Linux program make[7] to create links between the compiler and source files, compile all the modules, and the final program: XTANT.x. Use it for Linux-based systems. For Windows, alternatively, rename Make.bat.txt to Make.bat and execute the file Make.bat in the command line (assuming all the paths to needed libraries such as OpenMP are provided in your system, or the libraries themselves are in the same folder).

By default, it uses the Intel Fortran-2013 compiler (ifort2013; or ifort.exe for Windows); this can be changed in the Makefile. Note that the code uses some of the intel-features, that would need to be corrected, should you want to compile the code with gnu-fortran (gfortran).

- XTANT.sh – this is the shell script to be executed for running the XTANT-3 (executable XTANT.x) under Linux. Alternatively, under Windows run the created XTANT.exe.
- XTANT_MAIN_FILE.f90 – this file contains the main part of the XTANT-3 code. It assembles the program into one piece and performs the dynamics of atoms and electrons by calling all necessary subroutines. Also, all initialization of variables, reading input files, and creating output files are called from here.
- Algebra_tools.f90 – this file contains linear algebra necessary subroutines (also with references to the LAPACK library[8]).
- Atomic_tools.f90 – this file contains subroutines used for the atomic subsystem.

[7] Details on what is ‘make’: http://linux.about.com/library/cmd/blcmdl1_make.htm
[8] http://www.netlib.org/lapack/

- BS_Basis_sets.f90 – this file contains subroutines to deal with Gaussian basis sets that are required for xTB calculations (*unfinished*)
- BS_Cartesian_Gaussians.f90 – this file contains subroutines to deal with Cartesian Gaussian basis sets that are required for xTB calculations (*unfinished*)
- BS_Spherical_Gaussians.f90– this file contains subroutines to deal with Spherical Gaussian basis sets that are required for xTB calculations (*unfinished*)
- Coulomb.f90 – this file contains Coulomb potential and forces for modeling the Coulomb explosion of finite-size systems.
- Dealing_with_3TB.f90 – this module contains subroutines for reading and interpreting files in 3TB format [115]. Note that currently XTANT-3 only supports the 2-body part of the parameterization, the 3-body part is unfinished.
- Dealing_with_BOP.f90 – this module contains subroutines for reading and interpreting files in BOP format [116]. Note that this parameterization only supports dimer molecules, not solids.
- Dealing_with_DFTB.f90 – this module contains subroutines for reading and interpreting files in the Slater-Koster format[9], as provided by DFTB [117].
- Dealing_with_EADL.f90 – this file contains subroutines to read from the EPICS2017 (former EADL and EPDL) databases[10], needed for naming atomic shells, extracting information on Auger-decay rates, ionization potential, photoabsorption cross sections (used in one of the options of MC, see below).
- Dealing_with_eXYZ.f90 – this file contains subroutines to read from extended XYZ format[11].
- Dealing_with_files.f90 – this file contains useful subroutines to deal with files, such as counting lines and columns, reading, checking for errors, etc.
- Dealing_with_mol2.f90 – this file contains subroutines deal with mol2 format[12].
- Dealing_with_output_files.f90 – this file contains all subroutines to create and prepare output directories and files, communicate with the program, and interpret the user's commands.
- Dealing_with_POSCAR.f90 – this file contains subroutines to read input atomic coordinates in POSCAR format[13].
- Dealing_with_xTB.f90 – this module contains subroutines for reading and interpreting files in the extended tight binding, xTB, format [87] (*unfinished*)
- Electron_tools.f90 – this file contains subroutines to deal with the Fermi function of low-energy electrons.
- Exponential_wall.f90 – this module contains short-range exponential repulsive potential and forces needed in case TB parameterization provides too low barrier for atoms at short distances.
- Gnuplotting.f90 – this module contains subroutines to create gnuplot shell scripts.

---

[9] http://www.dftb.org/parameters/introduction/

[10] See details of this database here: https://www-nds.iaea.org/epdl97/libsall.htm, physical details and references for the database are here: https://www-nds.iaea.org/epdl97/document/epdl97.pdf

[11] Original XYZ format: https://en.wikipedia.org/wiki/XYZ_file_format, its extension used in XTANT is described below

[12] https://www.structbio.vanderbilt.edu/archives/amber-archive/2007/att-1568/01-mol2_2pg_113.pdf

[13] https://www.vasp.at/wiki/index.php/POSCAR

- Initial_configuration.f90 – this file sets up the initial conditions, such as constructing initial atomic positions and velocities and so on.
- Little_subroutines.f90 – this file contains useful subroutines, such as for approximations, search in arrays, resizing arrays, etc.
- MC_cross_sections.f90 – this file contains subroutines for the calculation of electron cross-sections and the mean free path used in the Monte Carlo part. It uses complex dielectric function formalism [50,51], or BEB cross sections [53,118].
- Monte_Carlo.f90 – this file contains all Monte Carlo model subroutines for photons, high-energy electrons, and core holes Auger decays.
- Nonadiabatic.f90 – this file contains subroutines for Boltzmann collision integrals and nonadiabatic electron-ion energy exchange [119,120].
- Objects.f90 – this file contains all the introduced objects in the framework of object-oriented programming[14]and some subroutines to deal with these objects.
- Optical_parameters.f90 – this file contains subroutines for the calculation of the optical part of the complex dielectric function within the tight binding and RPA [110], or within the Drude model [121].
- Periodic_table.f90 – this file contains subroutines to extract information about each elements from the periodic table (must be attached as one of the input files, see below).
- Read_input_data.f90 – this file contains subroutines to read all necessary input files (see below).
- TB.f90 – this file contains general subroutines to deal with tight binding (TB) formalism.
- TB_complex.f90 – this file contains subroutines that assemble subroutines requiring complex tight binding Hamiltonian, such as DOS, CDF and electronic heat conductivity calculations over multiple *k*-points (parallelized *via* OpenMP).
- TB_3TB.f90 – contains subroutines to calculate the TB Hamiltonian within one of the following basis sets: s, $sp^3$, $sp^3d^5$, and corresponding forces, according to the 3TB model [115].
- TB_BOP.f90 – contains subroutines to calculate the TB Hamiltonian within one of the following basis sets: s, $sp^3$, $sp^3d^5$, and corresponding forces, according to the BOP method [116]. *Note that this parameterization only supports dimer molecules, not solids (unfinished).*
- TB_DFTB.f90 – contains subroutines to calculate the TB Hamiltonian and the repulsive term within one of the following basis sets: s, $sp^3$, $sp^3d^5$, and corresponding forces, according to the DFTB method [117].
- TB_Fu.f90 – contains subroutines to calculate TB Hamiltonian within the $sp^3$-basis set and repulsive energy, and corresponding forces, as a combination of Pettifor's parameters, according to Fu *et al.*[122]. Note that the tests showed unstable systems, even though with correct band structure; *not recommended for using until solved.*
- TB_Koster_Slater.f90 – contains some subroutines for the Koster-Slater angular parameterizations up to *d*-orbital [68].

[14] Quick introduction into object-oriented programming in FORTRAN: http://www.pgroup.com/lit/articles/insider/v3n1a3.htm and http://www.pgroup.com/lit/articles/insider/v3n2a2.htm

- TB_Molteni.f90 – contains subroutines to calculate TB Hamiltonian within the $sp^3s^*$-basis set and repulsive energy, and corresponding forces, according to Molteni *et al*.[123].
- TB_NRL.f90 – contains subroutines to calculate TB Hamiltonian within the $sp^3d^5$-basis set and corresponding forces, according to NRL format [72].
- TB_Pettifor.f90 – contains subroutines to calculate TB Hamiltonian within the $sp^3$-basis set and repulsive energy, and corresponding forces, according to Pettifor *et al*.[62].
- TB_xTB.f90 – contains first attempts to introduce xTB parameterization [87] (*unfinished, cannot be used*)
- Transport.f90 – contains simple rate equations mimicking heat transport out of the system using a Berendsen thermostat [89].
- Universal_constants.f90 – this file contains all universal constants.
- Use_statements.f90 – contains 'use' statements included in the main file.
- Van_der_Waals.f90 – this file contains all subroutines to deal with van der Waals potential.
- Variables.f90 – this file contains all global variables (mainly as defined objects) used throughout the code.
- ZBL_potential.f90 – core-core repulsion in the form of the Universal ZBL potential [84].

Additionally, the following modules for post-processing of the data can be compiled (stored in the directory !XTANT_ANALYSIS_SUBROUTINES, require separate compilation as described above):

1) XTANT_atomic_data_analysis.f90 – program to analyze atomic data: pair correlation function
2) XTANT_autocorrelators.f90 – program to analyze atomic motion: velocity autocorrelation and phonon spectrum
3) XTANT_coupling_parameter.f90 – program to extract the coupling parameter (and electronic chemical potential, heat capacity, and pressure) as a function of the electronic temperature from a series of XTANT-3 calculations
4) XTANT_dielectric_function_analysis.f90 – program to extract the optical coefficients from the calculated CDF for the given incident pulse and layer parameters
5) XTANT_el_distribution_analysis.f90 – program to analyze electronic distribution function, plot its evolution
6) XTANT_entropy.f90 – program to analyze electronic entropy
7) XTANT_fragmentation.f90 – program to analyze atomic fragmentation for given separation distance in the simulation

## VI. INPUT FILES

The code requires input files stored in the directory: INPUT_DATA. This name cannot be changed. The directory contains the following files and directories:

- INPUT_MATERIAL.txt or INPUT.txt – input file with all the parameters of the material and laser pulse (and, optionally, numerical parameters).

- NUMERICAL_PARAMETERS.txt – optional input file with all the numerical parameters of the calculations (can be included in the INPUT.txt, or set in this separate file).

These files cannot be renamed, the program has to be able to find them by these exact names.

The directories with the following names must also be present:

- Atomic_parameters must contain the following databases:
  - EADL2017.all – Electronic atomic database (ionization potentials, Auger- and radiative decay rates, kinetic energies of atomic electrons, etc.)
  - EPDL2017.all – Photoionization cross sections database for all elements.
  - INPUT_atomic_data.dat – periodic table of elements.
  - INPUT_Hubbard_U.dat – table with Hubbard U parameters for (chemical hardness) for selected elements, according to the ThreeBodyTB model[15] [115].
- DFTB containing directories with Slater-Koster files within DFTB format, e.g. matsci-0-3, and others. Inside the directories .skf files must be present, named [*El*]-[*El*].skf, where [*El*] stands for the element which overlap parameters with the second listed element this file contains[16].
- DFTB_no_repulsion optional directory with Slater-Koster files within DFTB format without the repulsive potential fitted, such as, e.g., can be found in [17]. Inside this folder must be a directory with the parameterization name specified in the TB_Parameters file (e.g., "2elements"). In the directory, .skf files must be present, named [*El*]-[*El*].skf, where [*El*] stands for the element which overlap parameters.
- 3TB_PARAMETERS – files containing ThreeBodyTB parameterizations for elemental solids and binary compounds in xml format
- BASIS_SETS – files with Gaussian basis sets in the gbs format[18] (*currently unused, since xTB or ab-initio model is unfinished*)
- BOP_data – all the dimer parameters in BOP format in the file models.bx [116]

A few other folders with the names of the materials must be there. The material name given in the INPUT_MATERIAL.txt (see below) must exactly coincide with the name of the folder, such as e.g.:

- Diamond, Silicon, Gold, etc. Each of the folders contains a few files describing the material properties, used in the code as described below.

## 1. File INPUT_MATERIAL.txt or INPUT.txt

File INPUT_MATERIAL.txt or INPUT.txt contains the following mandatory lines, which must be exactly in this order, with exactly as many numbers inside each line, as described below.

Note that the name INPUT_MATERIAL.txt takes precedence over the name INPUT.txt, so if both files are present, the INPUT_MATERIAL.txt will be read and INPUT.txt will be ignored.

---

[15] The ThreeBodyTB code and its parameters can be found here: https://github.com/usnistgov/ThreeBodyTB.jl

[16] Detailed description of the files format is provided here: http://www.dftb.org/parameters/introduction/ The skf files can also be downloaded from there with the format that can be read by XTANT, no alterations needed.

[17] Download files from https://github.com/by-student-2017/Slater-Koster-parameters-no-repulsion_v1, place in the INPUT_DATA folder and rename the directory Slater-Koster-parameters-no-repulsion_v1-main into DFTB_no_repulsion

[18] Basis set files can be downloaded from: https://www.basissetexchange.org/

```
1  PMMA                  ! material name
2  C5O2H8                ! chemical formula of the compound
3  300.0                 ! initial electron temperature [K]
4  300.0                 ! initial atomic temperature [K]
5  0.0                   ! start of simulation [fs]
6  1000.0                ! end of simulation [fs]
7  1                     ! number of FEL-pulses
8  1.0d0                 ! 1.0d0   0.1e0 ! absorbed dose per this pulse [eV/atom] (min, max, step)
9  1000.0  100.0         ! hw, mean photon energy [eV], FWHM of the distirbution of hw [eV]
10 10.0d0                ! pulse FWHM-duration [fs]
11 1                     ! type of pulse to be analysed: 0 = rectangular, 1 = Gaussian, 2 = SASE
12 0.0d0                 ! position of the maximum of the laser pulse [fs]
13
14 ! probe
15 0   1   1.0d-1  25.0d0  1.0d-1  ! optic coef: 0=no, 1=Drude, 2=Trani-k, 3=gamma; spectr (1) w/ KK (2), or no (0); min, max, d_hw (eV)
16 2   800.0   -70.0     ! how many rays (0=exclude, 1=1st ray, (>1)=sum all); probe-pulse wavelength [nm]; FWHM probe pulse [fs]
17 89.0    50.0d0        ! angle of prob-pulse with respect to normal [degrees]; material thickness [nm]
18
19 ! water
20 100
21
22 ! coupling
```

Figure VI.1 INPUT_ MATERIAL.txt or INPUT.txt example.

- Line 1: One or two char-variables must be present in this line:

  First variable (mandatory): material name, must exactly coincide with the name of the folder, where the material parameters are stored (mentioned above, described below).

  Second variable (optional): name of the file with atomic coordinates and supercell vectors (the file with this name must be present in the directory, specified by the first variable, the material name; note the formats supported, as specified in Section 4.f)

- Line 2: chemical composition or element, which material consists of. Each element **must** start with a capital letter, followed by the small letters and/or a number corresponding to the contribution of this element to the compound – that is how the program parses the names into separate chemical elements to be used for the periodic table reading. For example, diamond or graphite must be set as C; silicon – Si; gallium arsenide – GaAs (e.g., “Gaas”, “Ga As”, “GAAS”, “Ga_As” or “gaas” will not interpret correctly); quartz – SiO2, etc.

- Line 3: setting the initial electron distribution function can be done in two ways:

  - If a number is provided in this line, it sets the initial electron temperature in [K]. Then, the initial electron distribution is assumed to be the Fermi function, defined by the given temperature and the number of electrons in the valence band of the material (calculated from the atomic valence electrons, given in the database INPUT_atomic_data.dat, see above).

  - If the name of the file is given, XTANT-3 will attempt to read the electron distribution function from this file. The file must be located in the directory, containing all the material parameters (as provided in the first line). If the file with the given name cannot be found in the directory, then the Fermi distribution will be assumed instead, with the electronic temperature equal to the atomic one (see next line). The file with the distribution function must be in the following format:

    The first line is the comment line that will be skipped while reading.

    Second and all the other lines will be read. The last column in the file is assumed to contain the distribution function (compatible with the output file with the distribution function created by XTANT: OUTPUT_electron_distribution.dat, see below).

    This option can be used for setting nonequilibrium electron distributions that can be run in BO approximation (see flag setting it below, file NUMERICAL_PARAMETERS.txt);

currently, no other simulation option supports nonequilibrium distributions – the initial one will be thermalized in all but BO simulations.

- Line 4: initial atomic temperature in [K].
- Line 5: starting time of simulation in [fs]. The starting time in the simulation will be chosen as the minimum between the user-provided value here, and [-50+FWHM*2.35], where FWHM is the full width at half maximum of the laser pulse (see line 12 below).
- Line 6: total duration of the simulation in [fs]. *Can be later changed during the simulation, see below description of the Communication file.*
- Line 7: number of FEL-pulses to be simulated (multiple-pulses allowed). A number of next lines depends on this. In this example of Figure VI.1, there is only 1 pulse to be modeled. In case you want to model two pulses, set here 2.
- The next lines specify the parameters of each FEL-pulse:
    - Line 8: specifies the incoming fluence or the absorbed dose of the pulse

- Column #1: marker (text), may optionally specify if the numbers given are the incoming fluence or the absorbed dose. The marker may be:

  d or D: for absorbed dose; then the following numbers are interpreted in [eV/atom]

  f or F: for incoming fluence; then the following numbers are interpreted in [J/cm$^2$]

  *Note #1*: XTANT-3 uses the absorbed dose internally, so if the incoming fluence is specified, it is converted into the absorbed dose prior to simulation, according to the assumption of the normal incidence of the laser pulse along the Z axis [1]:

$$D = F\,(1 - exp(-d/L)\,)/(e\,n_{at}\,(d\cdot 10^{-8})) \tag{49}$$

  Here $D$ is the absorbed dose in [eV/atom]; $F$ is the incoming fluence in [J/cm$^2$]; $d$ is the thickness of the sample along Z axis in [Å] (simulation box, or a layer of material); $L$ is the photon attenuation length for the given photon energy in [Å]; $n_{at}$ is the atomic concentration in [atoms/cm$^3$]; $e$ is the electron charge for conversion between [J] and [eV]; and the factor $10^{-8}$ is for conversion between [Å] and [cm]. This expression assumes linear absorption, and the program will not work in the nonlinear regime (too low photon energy, or too high fluence).

  *Note #2:* if there is no marker given but a number instead, it is assumed by default that the number corresponds to the absorbed dose (to be back-compatible with the legacy format).

  *Note #3:* the output directory name contains the absorbed dose value, not the incoming fluence (if needed, converted according to Eq.(42); so, don't be surprised by the folder name containing the dose value).

- Column #2: the absorbed dose in [eV/atom] (or the incoming fluence in [J/cm$^2$] depending on the marker in the Column #1) used for energy deposition from this pulse.
  Setting it equal to 0 gives NO laser pulse, modeling the dynamics of the unirradiated system (for example, for relaxation of the system, or for electron-ion thermalization, if the nonadiabatic coupling is included).
  There are two options to set the absorbed dose:

(i) if a single number is given in this line, the single absorbed dose is set in [eV/atom] (or incoming fluence in [J/cm$^2$])

(ii) if there are three (real) numbers in this line they are interpreted as a grid in dose (or fluence): *first dose (or fluence), last dose (or fluence), step* (all in [eV/atom] for dose, or in [J/cm$^2$] for incoming fluence). In this case, the program will create a set of input files with all identical parameters, except for the dose, which will be varied between the first and the last number, by the *step* given in this line. The program then will run a sequence of simulations one after another automatically for all set doses, see below section VI.3.

- Line 9: parameters of the photon spectrum. It can be set by one or two numbers:

  First (real) number: $\hbar\omega$, mean photon energy of the incoming FEL in [eV]

  Second (real) number: FWHM of the photon spectrum in [eV], assuming a Gaussian distribution of the photon energies around the mean.

  If the second number is not given, then zero-FWHM is assumed, and all the photon energies will be equal to the mean (ideal monochromatic pulse).

- Line 10: duration of the pulse, $\tau$, in [fs] (FWHM for Gaussian pulse, total duration for flat-top or SASE [124]).
- Line 11: type of the pulse to be used: 0 means flat-top pulse, 1 gives Gaussian, and 2 mimics SASE-like spiky pulse [124].
- Line 12: position of the center, $t_0$ (Gaussian maximum) of the laser pulse [fs]. The simulation will start at ($t_0 - \tau - 50$) fs if this value is smaller than the starting time set by the user (line 5).
- If you want to set a second pulse, repeat the same lines 8-12 (with different pulse parameters) in the same order.

- Optional lines:

To provide additional options for calculations, one may use optional lines at the end of the file.

- Any of the additional options that may be passed to XTANT-3 *via* the command line (see Section IV) may also be specified here (e.g., size, info, help, etc.).
- Probe: specifies that calculation of the optical parameters of a probe pulse is required. Optional line 1: probe. Below this marker, the following three lines in the given order (and no empty lines in between) must be provided:

  Optional line 2: this line contains 5 numbers (2 integers + 3 reals):

- first one sets whether you want to calculate the evolution of the optical properties (set 0 if not), and within which model:

a) 1 for the Drude model. If the Drude model is used, it requires additional parameters to be set in the file NUMERICAL_PARAMETERS using the optional block of data "DRUDE", see below. If this block is not specified, the default values are used: $n$=1, $k$=0, $m_{e_eff}$=$m_{h_eff}$=$m_e$, and $t_e$=$t_h$=1 fs.

b) 2 for the RPA model (in Trani *et al.*'s approximation) at many k-points, distributed according to the user-defined grid or to the Monkhorst-Pack grid [125] (this option requires many diagonalizations of TH Hamiltonian which are currently not parallelized, and thus is very slow). It uses nonorthogonal Hamiltonian representation, see Eq.(43).

c) 3 for the RPA model (in Trani *et al.*'s approximation) [110] at the Gamma-point only.

d) 4 for Kubo-Greenwood (KG) model (equivalent to RPA) using orthogonalization for momentum operator calculations, Eq.(44).

e) 5 for KG model using non-orthogonal momentum operator calculations, Eq.(43). *This should be the default option.*

*Note #1*: The two formalisms – KG in a non-orthogonal representation and RPA – produce nearly identical results. However, KG implementation is parallelized with OpenMP, while Trani's RPA is not; thus, option 5 is recommended as the default choice, while option 2 is obsolete. Option 4 produces worse results than 5.

- The second number in this line indicates whether you want to calculate the complex dielectric function only for a given (probe-photon) energy (set 0), or for the whole spectrum (set 1). The default choice is 0.
- The next three numbers define the interval of the spectrum you'd like to calculate, in case the previous number is set to 1: The third number is the starting point in [eV], the fourth is the ending point in [eV], and the fifth is the energy step in [eV] to make a uniform grid. If any of these three numbers is set to negative, then default values for the interval are used, which are: from 0 to 50 eV with the step of 0.05 eV.

Optional line 3: contains three numbers:

- First one sets for how many rays propagation you want to calculate the optical parameters (transmission, reflection, and absorption of the probe-pulse): set 1 for the first ray, or a value larger than 1 for summing up all rays. For femtosecond probe pulse, the default choice is 1 (however, for very thin samples, thinner than ~50 nm, sum up all).
- The second number sets the wavelength of the probe pulse in [nm].
- The third one is the duration of the probe pulse in [fs]. If the number if set to a positive value, the output files will be additionally convolved with the Gaussian probe pulse of the given duration. A set of additional convolved output data will be created with the tag 'CONVOLVED' (see below). To exclude this option, set the duration to zero or a negative value.

Optional line 4: contains two numbers:

- The first one sets the angle of incidence of the probe-pulse [in degrees] *with respect to the normal*.
- The second one sets the thickness of the material layer through which the probe pulse absorption and reflection are calculated in [nm]. Must be equal to the experimental target thickness, if it is thinner than the FEL photon attenuation length; or may be to the FEL-photon attenuation length otherwise.

- Kappa: option to calculate the electronic heat conductivity (and together with it, the electronic chemical potential and the electronic heat capacity) vs. the electronic temperature.

Optional line 1: Kappa (or kappa, or do_kappa, or get_kappa) – the marker to include this calculation.

*Note #1*: the calculations are performed for multiple k-points, grid for which is defined in the last line in the NUMERICAL_PARAMETERS block described below; for single gamma-point calculations, set there 1x1x1 points (3*1).

Optional line 2: three real numbers, specifying the grid in the electronic temperature:

Te_min Te_max step

Where Te_min is the start of the grid in [K]; Te_max is the end of the grid in [K]; step is the grid step in [K]. This line is optional. If it is not defined, the default values are used: Te_min=300 K; Te_max=30000 K; step = 100 K.

- Save_CDF (or print_CDF, or get_CDF): option to print out the cdf-file with the Ritchie-Howie parameters used for the specified material. This option does not require any additional specifications.
  *Note #1*: this option is particularly useful if you use single-pole approximation for Ritchie-Howie CDF to see the automatically fitted coefficients.

- Coupling: option for automatic preparation of XTANT-3 input for calculations of the average electron-phonon coupling (see details in Section (VIII.3)) may be passed here. To set it, use the following command (2 lines):

Optional line 1: Coupling – the marker, identifying that input files for electron-phonon coupling calculations should be prepared.

Optional line 2: *N* is the number of simulations to be used for average electron-phonon coupling calculations. If *N* is not specified, the default value of *N*=10 is used.

If the marker "Coupling" is found in the file, the code will create automatically *N* copies of the input files (INPUT_DATA_*i*.txt and NUMERICAL_PARAMETERS_*i*.txt, for *i*=1..*N*), in which the following parameters will be set:

- the start of simulation $t_0$ (1 ± 0.1*RN*), where *RN* is a random number in the interval [0,1]

- end of simulation $t_f$ (1 ± 0.25*RN*), where *RN* is a different random number in the interval [0,1]

- absorbed dose *D* (1 ± 0.1*RN*), where *RN* is a different random number in the interval [0,1]

- pulse FWHM duration will be equal to the end of simulation time.

The values of $t_0$, $t_f$, and *D* are taken from the existing input file, and in each successive file their own characteristic values are written, sampled around these ones. This way, slightly different initial conditions, and the absorbed dose will be used in each calculation, which then allows us to average the data for reliable calculation of the electron-phonon coupling parameter, see Section (VIII.3) for details.

This option allows for more convenient calculations of the electron-phonon coupling parameter, instead of the manual creation of input files.

- Water: option for the automatic embedding of the material or molecule into water. It will set randomly water molecules around the given material. The following lines must be specified:

  Optional line 1: water, the keyword specifying that the target will be embedded in water.

  Optional line 2: number of water molecules to be placed around the target material (integer).

  This option performs the following procedure: sets the target material as specified above (by the unit cell or supercell parameters). Then, extends the size of the supercell, and places the specified number of water molecules around it. The molecules are randomly placed and randomly oriented but checked not to be placed too close to each other or the given target atoms.

  Note, however, that there is no guarantee that thusly constructed water environment will make the supercell relaxed (in fact, it never does). If the code would be unable to place all the water molecules at the first attempt (it performs a certain number of iterations), then it will increase the size of the supercell and try again. A message about it will be displayed on the screen. Randomly placed water molecules will not be in an equilibrium state, and will need a two-step relaxation: (1) with quenching, attempting to find the equilibrium positions minimizing the potential energy (note that some molecules may break apart at this stage and the code may need to be run again!); (2) thermalization with Berendsen thermostat (to reach thermal equilibrium). See below options for quenching and thermostat.

  This option is convenient for setting bio-molecules in water, without manually specifying all the water molecules.

- NUMERICAL_PARAMETERS (or NUMERICS, or NUMPAR): the option to specify that all the numerical parameters will be provided in this file, just under this line, instead of the separate file NUMERICAL_PARAMETERS.txt described below. After this flag, the entire set of parameters in exactly the same order must be provided, as described below. This is just an option for the user to choose where to set numerical parameters for convenience: in a separate file, or all in one file, see an example in Figure VI.2.

  Note that this option takes precedence over the separate file NUMERICAL_PARAMETERS.txt: if both are present, the parameters will be read from here, and the separate file will be ignored.

- All optional parameters at the end of the NUMERICAL_PARAMETERS.txt (described below) may also be used here (e.g., MASS, AUGER, etc.).

```
PbI2                ! material name
PbI2                ! chemical formula
300.0               ! initial electron temperature [K]
300.0               ! initial atomic temperature [K]
0.0                 ! start of simulation [fs]
-1000.0             ! end of simulation [fs]
1                   ! number of FEL-pulses
0.0e0   !1.0d0  0.5e0   ! absorbed dose per this pulse [eV/atom] (min, max, step)
2.0e0  0.0          ! hw, mean photon energy [eV], FWHM of the distirbution of hw [eV]
15.0e0              ! pulse FWHM-duration [fs]
1                   ! type of pulse to be analysed: 0 = rectangular, 1 = Gaussian, 2 = SASE
0.0e0               ! position of the maximum of the laser pulse [fs]

NUMERICS
3   3   1       ! number of unit-cells in X,Y,Z
1   1   1       ! periodicity along X,Y,Z directions (1=yes, 0=no)
BEB             ! where to take material parameters from (EADL, CDF)
-1.0d0          ! [g/cm^3] density of the material (used in MC in case of EADL parameters), if <0 uses data from MD supercell
1000            ! number of MC iterations
4               ! number of threads for OPENMP
2               ! MD algorithm: 0=Verlet (2d order); 1=Yoshida (4th order, slow); 2=Martyna (4th order, fast)
1               ! frozen atoms (=0), or moving normally (=1)
25.5d0          ! Parinello-Rahman super-cell mass coefficient
dt_grid.txt     ! time step for MD [fs] (or a file name with the variable time-grid)
1.0d0           ! printout data into files every 'dt_save_time' [fs]
1               ! it's = 1 if P=const, or = 0 if V=const
0.0d0           ! external pressure [Pa]
```

Figure VI.2 INPUT_ MATERIAL.txt or INPUT.txt example with option NUMERICS included.

## 2. File NUMERICAL_PARAMETERS.txt

File NUMERICAL_PARAMETERS.txt (or the block “NUMERICAL_PARAMETERS” in the INPUT.txt file) contains the following lines, which must be exactly in this order, with exactly as many numbers inside each line, as described below:

```
1  2   2   2        ! number of unit-cells in X,Y,Z
2  1   1   1        ! periodicity along X,Y,Z directions (1=yes, 0=no)
3  BEB              ! where to take material parameters from (EADL, CDF)
4  -1.0d0           ! [g/cm^3] density of the material (used in MC in case of EADL parameters), if <0 uses data from MD supercell
5  10000            ! number of MC iterations
6  -1               ! number of threads for OPENMP (<1 = max threads available)
7  2                ! MD algorithm: 0=Verlet (2d order); 1=Yoshida (4th order, slow); 2=Martyna (4th order, fast)
8  1                ! frozen atoms (=0), or moving normally (=1)
9  25.5d0           ! Parinello-Rahman super-cell mass coefficient
10 0.1 dt_grid.txt    ! time step for MD [fs] (or a file name with the variable time-grid)
11 1.0d0            ! printout data into files every 'dt_save_time' [fs]
12 0                ! it's = 1 if P=const, or = 0 if V=const
13 0.0d0            ! external pressure [Pa]
14 F   0   0.35     ! include SCC (True, False), model of gamma (default 0); mixing factor (default 0.35)ss
15 4   1.0e10   10.0e0    50.0e0    ! scheme (0=decoupled; 1=E_tot=const; 2=T=const; 3=BO; 4=Relaxation time); full tau [fs]; tau_CB; tau_VB
16 -1               ! -1=nonperturbative (default), 0=no coupling, 1=dynamical coupling, 2=Fermi golden rule
17 -1.0d+3    4.0d0    ! [fs] when to switch on the nonadiabatic coupling; scaling factor (if needed; default=4.0)
18 5.0d0    0.001d0    ! [eV] acceptance window for nonadiabatic coupling; [eV] tolerance for quasidegenerate levels
19 0    -50.0    5.0    ! quenching (0=no, 1=yes); starting from when [fs]; how often [fs]
20 0    150.0    10.0   ! Berendsen thermostat for atoms (0=no, 1=yes); bath temperature [K]; cooling time [fs]
21 0    150.0    10.0   ! Berendsen thermostat for electrons (0=no, 1=yes); bath temperature [K]; cooling time [fs]
22 15.0d0            ! [eV] cut-off energy in MC (<0 means Ecut=top of CB)
23 1.0d30            ! [eV] work function, for electron emission (>0 eV; <0 - number of collisions)
24 0                ! printout electron energy levels (1) or not (0)
25 0    0.1    1    ! printout DOS (0=no,1=Gamma,2=k-points); smearing of gaussian; print PDOS yes (1) or no (0)
26 1       ! printout Mulliken charges for types of atoms
27 1       ! printout electron distribution (1) or not (0)
28 0       ! printout atomic pair correlation function (1) or not (0)
29 1       ! printout atomic positions in XYZ (1) or not (0)
30 0       ! printout atomic positions in CIF (1) or not (0)
31 0       ! printout raw data file OUTPUT_coordinates_and_velosities.dat (1) or not (0)
32 1       ! power of mean displacement to print out (set integer N: <[R(t)-R(t=0)]^N>)
33 1.75e0  ! printout numbers of nearest neighbors within the given radius (<=0 No, >0 = radius in [A])
34 png     ! which format to use to plot figures: eps, jpeg, gif, png, pdf
35 9   9   9    ! number of k-points in each direction (used only for Trani-k!)
```

Figure VI.3. NUMERICAL_PARAMETERS.txt (or optional lines in INPUT.txt after the flag NUMERICS) example.

Line 1: the number of unit-cells used in the code along each direction X, Y, Z. If one unit cell contains Nat atoms, the total number of atoms in the supercell will be **Ntot**=Nat*Nx*Ny*Nz.

a. The numbers in a raw must be separated by TAB, not SPACE.

b. *Setting here 0 0 0 should in principle exclude the atomic dynamics and run only electronic MC simulations (analogous to XCASCADE code*[118]*); this option, however, has not yet been tested!*

Line 2: three numbers specify conditions at surfaces along X, Y, and Z axes: setting here 0 creates an open surface along the axis (by adding empty space around the sample), whereas setting here 1 means periodic boundaries. For example: 1 1 0 means periodic boundaries along X and Y, but free boundary along Z (thin layer of material).

*Note #1: Non-periodic simulation uses a periodic boundary in a supercell, in which the sample is surrounded by empty space. The code increases the size of the simulation box by 50 times its given value and places the atoms in the middle. That means, all the values that include normalization to the supercell volume (i.e. electron-ion coupling parameter, electron heat capacity, pressure) will include the empty space volume and must be rescaled manually for interpretation of the results.*

Line 3: May use one or three columns (one mandatory, and two optional):

a. Column 1: flag (character), specifying which cross sections to use in the MC module: set here 'CDF' (a cdf-file should be provided, see below, or a single-pole approximation could be used) to use cross sections based on the complex dielectric function formalism [47]; or 'BEB' (or 'EADL') to use atomic BEB-cross-sections (default choice) [53,118].
b. Column 2 (optional): the materials bandgap in [eV]; this value is used if single-pole approximation for CDF is specified. In this case, the bandgap value is required for using CDF cross sections – without it, XTANT-3 will use an atomic energy level, which is a very poor approximation for the bandgap of a material.
c. Column 3 (optional): name of the cdf-file (character). This column may specify a full path to the file with cdf, or a name of the cdf-file to be found in the directory with the input material data (if this file name does not have an extension, the default extension '.cdf' will be assumed). If this column is not provided, or the file with the given name is not found, the default name of the cdf-file is checked: [*material*].cdf (see below, Section 8)1)f).
   If no file with CDF is found, the single-pole approximation will be used, with the coefficients automatically fitted, following the procedure described in Ref. [52].

*Note #1*: specifying 'CDF', but no file and no bandgap (using atomic ionization potential) results in cross sections close to BEB, which means it has no advantages, but much slower calculations (because CDF cross section employ numerical integration, whereas BEB has analytical solution). So, this option is not recommended for use – instead, provide a correct band gap, this will provide significant advantage over the BEB cross-sections, since CDF-formalism accounts for collective effects, and thus describes solids much better than the atomic approximation.

Line 4: the density of the material in [g/cm$^3$] to be used in the MC simulations. This value overwrites the default value given in the cdf-file (see below) if set positive. If you wish to use the default value (defined by the number of atoms and size of the supercell), set here any negative number (this must be the default choice).

Line 5: number of iterations to be performed within the Monte Carlo module. A small number of iterations gives not smooth curves. Too large numbers give too long computation times. The optimal value empirically determined is ~2,000,000/(Dose * Ntot).

*Note #1: large values here result in large arrays taking a lot of memory. In case you don't need MC simulations but only TBMD, set here 0 or 1.*

Line 6: number of threads used for parallel calculation via OpenMP. Set 1 for nonparallel calculations. Set any non-positive number to make it automatically equal to the number of available threads on your machine.

Line 7: which MD integrator to use: 0 = velocity Verlet algorithm (2$^{d}$ order) [80], 1=Yoshida algorithm (4$^{th}$ order; *it is ~4 times slower than Verlet*) [81], 2 = Martyna predictor-corrector algorithm (4$^{th}$ order, *about as fast as Verlet*) [82]. The default option is 2.

*Note #1: Martyna's algorithm is included only for atomic coordinates, while for the supercell vectors, Verlet is used (2d order).*

Line 8: exclude the MD module, freezing the atoms in their equilibrium positions (if set 0), or allow the atoms to move (set 1). The default option is 1.

Line 9: the effective mass of the super-cell in [atomic mass] used in the framework of the Parrinello-Rahman MD [90] (only used in case of constant-pressure simulation, see below).

Line 10: time-step for the MD calculations.

It can be set in two ways:

a. Constant time-step: write any real(8) number of dt in [fs]. The default value is 0.1 fs (or smaller for P=const simulations, see below), but in some simulations can be as large as 1 fs or even larger (especially for V=const simulations). Larger steps can lead to instabilities, smaller steps conserve energy better but run the program slower.

b. Variable time-step: write here a name of a file where the array of timesteps is provided. The file must be present in the same folder INPUT_DATA. The file must contain two columns. First column: time instant in [fs], when to change the timestep to the one given in the second column.

   E.g., if the file contains the following lines:

```
-1.0e10    1.0
-50        0.2
100 0.5
```

   At the first timestep of the simulation, the timestep is set to 1.0 fs. At the time instant of -50 fs, the timestep is changed to 0.2 fs. At the time instant of 100 fs, it is changed to 0.5 fs.

Line 11: time step how often the output data files must be saved in [fs] of the simulation time. Does not have to be equal to the MD time step, and can be larger to sparse the output data (but cannot be smaller). The default choice is 1 fs.

Line 12: here 1 means constant pressure simulations (P=const; Parrinello-Rahman scheme of the super-cell motion, NPH ensemble); 0 means constant volume (V=const, NVE ensemble) [126]. For femtosecond dynamics, V=const (0) is the default choice.

Line 13: external pressure applied (set 0 to use for normal atmospheric pressure). Used only in case of P=const simulation.

Line 14: contains 3 numbers to set self-consistent-charge (SCC) calculations parameters:

a. Column 1: to use or not the SCC (*T* or *F* - for True or False)

*this option works only with the Born-Oppenheimer approximation (option 3 in the next line), but not with any other simulation scheme, so cannot be combined with irradiation!*

b. Column 2: which model for gamma to use: -1 is bare Coulomb (*do not use, only for testing*)

0 = Wolf's method of softly truncated Coulomb (this is the default choice) [127]

1 = Klopman-Ohno [76]

2 = Mataga-Nishimoto [76]

c. Column 3: mixing factor for self-consistent calculations: weight of the new charge in the next iteration. Recommended values are between 0.2 and 0.7. Smaller values lead to too slow convergence, whereas higher values may not converge at all. The empirically found optimal value is 0.35, which should be the default choice.

*Note #1*: the default option must be *F* (no SCC). It should only be used with TB parameterizations that are specifically fitted to account for SCC (such as 3TB, and some of DFTB parameterizations) and with BO simulation *only*, while in most cases, the parameterizations cannot account for the SCC effects, and including this option may lead to qualitatively incorrect potentials.

Line 15: contains 4 numbers:

a. The first one describes which scheme of simulation to use:

0 sets a scheme of decoupled electrons and ions, with instant electron thermalization (something like Two-Temperature Model).

1 sets enforced total energy conservation. *Obsolete, do not use!*

2 sets here fixed electron temperature instead of total energy (*does not work well with a pulse on or with an electron-ion coupling on, use only for tests of unirradiated material*).

3 uses the true Born-Oppenheimer (BO) scheme – constant electron populations.

*Note #1:* if the laser pulse is on, the populations will change; also, if the electron-phonon coupling is on, it will affect the populations, so the simulation instead of BO will be Ehrenfest-like dynamics.

4 uses the relaxation time approximation to trace the evolution of the electron distribution function. *This should be the default choice.*

b. The second number sets the characteristic relaxation time in [fs], used in the relaxation-time approximation only.

*Note #1*: To turn the simulation into instantaneous thermalization (equivalent to option 0 in "a", the Two-Temperature-Model conditions), set here time = 0.0e0. To turn the simulation into Ehrenfest-like (or BO, if the electron-ion coupling is off), set here infinite time, e.g., 1.0e20.

c. The third number sets the characteristic relaxation time of electrons in the conduction band (CB) in [fs], used in the relaxation-time approximation only.

d. The fourth number sets the characteristic relaxation time of electrons in the valence band (VB) in [fs], used in the relaxation-time approximation only.

*Note #1*: the separate band-resolved thermalization times are only meaningful in bandgap materials (semiconductors or insulators); to switch the separate thermalization off, set any negative number in the third or fourth number.

*Note #2*: to leave only the separate band-resolved thermalization, but not the interband thermalization, set the second number to a very large number, and use only the 3$^{d}$ and 4$^{th}$ numbers to set separate thermalizations.

*Note #3*: one may model instantaneous thermalization in each band separately, by setting 3$^{d}$ and 4$^{th}$ numbers equal to zero, but finite time (or infinite) by setting a positive number in the 2$^{d}$ position (interband thermalization) – this setup results in the Three-Temperature Model: different temperatures for excited electrons (in the conduction band), holes (valence-band electrons), and atoms.

Line 16: which electron-phonon coupling to use:

a. 0 means no coupling included;

-1 generalization of dynamical coupling as described in Ref.[2] (this must be the default choice);

1 means first-order dynamical coupling as described in Ref.[28];

2 uses Fermi's Golden Rule, which might overestimate the coupling rate (*do not use*).

Line 17: Coupling model; has two numbers:

a. Sets the time, when the nonadiabatic coupling switches on (for test purposes, one can first thermalize the system, and only later let it exchange energy). The default choice for real simulations: 1d-3.

b. Set the scaling factor for coupling calculations. For numerical reasons, must be equal to 4.0 for producing correct results with dynamical coupling. Can be smaller to artificially reduce coupling in the calculations of the coupling parameter (as described in section Calculation of electron-ion coupling parameter g(Te), Ce(Te), μ(Te)), but the results must then be rescaled back manually!

Line 18: The coupling model (continuation); contains two numbers:

a. acceptance window for nonadiabatic coupling in [eV]. It excludes electron transitions between the levels separated by more than this specified value. E.g. set 5 eV by default to separate over-band-gap nonadiabatic transitions in diamond.

b. tolerance for quasi-degenerate levels in [eV]. It excludes transitions between too-close levels, separated by smaller energy than this given number, to exclude degenerate states. The default value is 0.001 eV.

Line 19: Quenching; three numbers here specify:

The first number defines whether to include artificial quenching, (0=no, 1=yes), the 'yes'-option must be used *only* for relaxation or the construction of amorphous materials. Any 'real' simulation must have 0 here. 'Yes' here means that once in a time-step specified by the next numbers of the line, atomic velocities will be set to zero. Similar to the method known as "zero-temperature molecular dynamics".

The second number in this line is defining when to start cooling from in [fs].

The third number means how often set the atomic velocities to zero (in [fs]).

Line 20: Berendsen thermostat for atoms; it can be set in two ways:

Option #1: If three numbers are given here, they define a simple model (rate equation) for artificial cooling mimicking transport effects [128], using the Berendsen thermostat [89]:

First number: include electron heat transport out of the atomic system (1), or not (0).

Second number: in case there is transport, sets the atomic bath temperature towards which the cooling/heating will be made until temperatures equilibration in [K].

The third one is the characteristic time of cooling/heating of atoms in [fs].

Option #2: If a file name is given here, the file with the parameters of the Berendsen thermostat must be provided in the directory INPUT_DATA. The file may contain an arbitrary number of lines. Each line must specify 3 parameters:

First column: time of simulation in [fs], when to switch to the Berendsen thermostat parameters given in the next two numbers.

Second column: the atomic thermostat temperature [K].

Third column: the characteristic time of the Berendsen thermostat for atoms in [fs].

E.g., if the file contains the following lines:

```
-1.0e10   300   1.0e15
0.0  500   100.0
100 1000   700.0
```

It will be interpreted as follows: at the beginning of the simulation (at a time larger than -1.0e10 fs), the thermostat is off (essentially infinite characteristic time, 1.0e15 fs). Then, at the time instant of 0.0 fs, it switches on with the bath temperature of 500 K and the characteristic time of 100 fs. At the time instant of 100 fs, it changes to the bath temperature of 1000 K and characteristic time of 700 fs.

Line 21: Berendsen thermostat for electrons; it can be set in two ways, analogous to the atomic thermostat:

Option #1: If three numbers are given here, they define a simple model (rate equation) for artificial cooling mimicking transport effects [128], using the Berendsen thermostat [89]:

First number: include electron heat transport out of the electronic system (1), or not (0).

Second number: in case there is transport, sets the electronic bath temperature towards which the cooling/heating will be made until temperatures equilibration in [K].

The third one is the characteristic time of electronic cooling/heating in [fs].

Option #2: If a file name is given here, the file with the parameters of the Berendsen thermostat must be provided in the directory INPUT_DATA. The file may contain an arbitrary number of lines. Each line must specify 3 parameters:

First column: time of simulation in [fs], when to switch to the Berendsen thermostat parameters given in the next two numbers.

Second column: the electronic thermostat temperature [K].

Third column: the characteristic time of the Berendsen thermostat for electrons in [fs].

*Note* that if an electronic thermostat is used, it will equilibrate the electronic distribution function towards the given temperature, and electronic nonequilibrium simulation is thus affected.

Line 22: energy cut-off in [eV] that separates the low-energy and high-energy subspaces for electrons within MC and Boltzmann-equation [4]. The default value is 10 eV. If set negative, it uses dynamical evolution of the cut-off, adjusting it to the transient top-most CB level at each time-step (*only meaningful for small basis sets that do not include a lot of CB orbitals*).

Line 23: the work function, setting whether we want to allow for electron emission and build-up of an unbalanced charge in the system, which may lead to a Coulomb explosion. If the work function here is set higher than 1.0d25 [eV], no emission will take place. This must be the default choice.

a. If the work function is set smaller, an electron with energy above the set number will be considered emitted from the sample and will disappear from the calculations (will forever stay in the high-energy domain, making no collisions).

b. If the work function is set to a negative value, another model for electron emission is used: an electron is considered to be emitted after a certain number of collisions that is specified by the absolute value of the number set in this line. E.g. if -2.0 is set here, an electron will be emitted after performing 2 collisions (unless is falls below the cut-off energy and joins the low-energy fraction).

c. If electrons are emitted, it builds up an uncompensated charge for the atomic system, inducing additional Coulomb repulsion of atoms (ions), if a file with Coulomb parameterization is present (see below) [85].

Line 24: to print out electron energy levels (eigenvalues of the TB Hamiltonian) as output (set 1) or not (set 0) at each saving time step. *Produces large files.*

Line 25: the three numbers here specify parameters of the density of states, DOS, and calculations at each saving time step:

First column: First number sets: to calculate total DOS or not. Use the following parameters: (0) exclude DOS calculations; (1) calculate DOS on the *k*-points grid set below (in line 35).

Second column: spreading to use for constructing the DOS out of the discrete energy levels in [eV].

Third column: to printout partial DOS (PDOS) for the atomic shells of each element in the compound (1), or not (0).

Line 26: to save Mulliken charges for types of atoms (set 1), or not (set 0)

Line 27: Three numbers defining printout of the electron distribution function at each saving timestep.

- First is the flag that defines what to print out:

0 means no printing out of electron distribution (this should be the default choice because printing out the distribution produces large files, which are usually unnecessary);

1 means to printout the electron populations on the energy levels only (below $E_{cutoff}$);

2 means to also printout the electron distribution on a grid with the parameters set in the other two numbers in this line (produces twice the large files; not recommended to be used);

-2 means to printout only the distribution on the given grid, with the parameters defined by the two other numbers in this line;

- The second number in the line defines the step in the energy grid in [eV], on which the distribution is printed out if flag 2 or -2 is set.
- The third number defines the maximal energy for the grid in [eV], used if the flag 2 or -2 is set.

Most of the time, it is not necessary, thus use the default value 0. Option 1 is useful in case of nonequilibrium electron distributions, such as BO or relaxation-time approximation simulation. Option -2 is useful for modeling the electron emission. Option 2 may be useful for code development and debugging.

Line 28: to save the atomic pair correlation function (set 1), or not (set 0) at each saving timestep. Produces large files. Most of the time, it is not necessary, thus the default value is 0.

Line 29: Line defining the printout of XYZ format. May contain two columns:

Column #1: (integer) save atomic positions additionally in extended XYZ-format[19] (set 1), or not (0). This format is used for plotting and making movies of the atomic positions e.g. with VMD[20], OVITO[21] , or a similar program.

Column #2: (character) a flag defining if additional atomic parameters need to be printed out. May include the following flags:

m: to printout atomic mass.

q: to printout atomic charge (Mulliken charge for each atom).

e: to printout atomic kinetic energy.

For example, if the user wants to save XYZ file and additionally include kinetic energy and charge in it, the line 29 may look like:

---

[19] See description here: http://en.wikipedia.org/wiki/XYZ_file_format (in XTANT-3, the extended format defined in OVITO is used, in which the comment line is used to specify the supercell size and parameters for coordinates, see https://www.ovito.org/docs/current/reference/file_formats/input/xyz.html)

[20] VMD: http://www.ks.uiuc.edu/Research/vmd/

[21] OVITO: https://www.ovito.org/

1 qe ! comment: printout XYZ-file with the charge and the kinetic energy included

*Note #1*: the second column is optional. By default, no additional output is included, with only atomic coordinates in the XYZ file.

Line 30: save atomic positions additionally in CIF-format[22] (set 1), or not (0). This format is used for powder diffraction patterns calculations e.g. with Mercury[23] software.

Line 31: to save raw data file OUTPUT_coordinates_and_velosities.dat (1) or not (0). This file is necessary for post-analysis calculations of atomic velocity autocorrelations and phonon spectra (see below). If not needed, do not save it, as it produces very large files.

Line 32: sets *N*, the power of mean displacement to print out (set integer *N*: <u^*N*>-<u0^*N*>). For example, for mean square displacement, set 2.

Line 33: save numbers of nearest neighbors within the given radius: to exclude optional set a number <0, a number >0 means the radius within which the atoms are considered to be neighbors in the units of [Å].

Line 34: which format to use to plot output figures: eps, jpeg, gif, png, pdf

Line 35: contains three numbers (integers): numbers of k-points in each direction x, y, z; used only with the 'Trani-k' or 'Kubo-Greenwood' options (numbers 2, 4 or 5 in line 12 in the file INPUT_MATERIAL.txt), and ignored with other options.

*Note #1:* all numbers <u>must</u> be odd for more reliable results, even numbers do not work well.

*Note #2:* FORTRAN allows specifying repeating numbers in the input by using "*" (times) symbol; e.g., the line

3*5

Is equivalent to

5 5 5

Line 36: This and the next lines are optional, each block may be written in arbitrary order or skipped altogether (recommended). The optional blocks allow the user to set model-specific data, and to overwrite the default atomic data, such as the atomic mass, Auger decay times, kinetic energies of shell electrons, electronic populations, and the name of the element if needed. The default values are taken from the EPICS-2017 database, and recommended for use, so only replace them if you are absolutely sure of what you are doing.

- DRUDE: to set the parameters of the Drude optical model (if in the file INPUT_DATA above, option "Probe" is set to use the Drude model for the calculation of the optical coefficients of the probe pulse). The following lines must be exactly in the following order and format:

Line Op1: DRUDE – the keyword, indicating that the following three lines set the parameters of the Drude model

Line Op2: two real numbers: initial values on unexcited materials optical coefficients *n* and *k*.

[22] See description here: http://www.iucr.org/resources/cif
[23] Mercury: https://www.ccdc.cam.ac.uk/solutions/csd-system/components/mercury/

Line Op3: two real numbers: effective mass of the conduction band electrons and valence band holes [in units of the free-electron mass].

Line Op4: two real numbers: mean scattering times of electrons and holes in [fs].

- MASS: To replace an atomic mass, the following block of data must be used:

Line Op1: MASS – the keyword, indicating that the following line defines the atomic mass

Line Op2: must contain two numbers: *atom* (integer), and *mass* (real).

The *atom* must correspond to the number of the element in the used compound, as defined by the chemical formula in line 2 of the file 'INPUT_MATERIAL.txt'. For elemental targets, the number 1 still must be present.

The *mass* sets the mass of the element in the atomic mass units. See the example of a compound SiAu, setting the mass of Au (element #2) to "infinity" (1.0d30), in **Figure VI.4**. Note that such a mass will essentially freeze all the atoms of the chosen element (which will not speed up the calculations, since the force is still calculated for them, just they are too heavy to move).

```
MASS
2   1.0d30

NO_AUGER

AUGER
1       1       1d25
```

**Figure VI.4. Example of optional lines in the file NUMERICAL_PARAMETERS.txt**

- NAME: to replace the name of the element, the following block of data must be used:

Line Op1: NAME – the keyword, indicating that the following line defines the name of the element

Line Op2: must contain two numbers: *atom* (integer), *name* (character(3)).
The *atom* is analogous to the one from the block MASS above.
The *name* sets the new name of the element.

- NO_AUGER: to switch off all Auger decays in the MC module, use the one-line option NO_AUGER, see an example in **Figure VI.4**.

- AUGER: to replace Auger decay times of selected shells of selected elements, use the optional block (see an example in **Figure VI.4**):

Line Op1: AUGER – the keyword indicating overwriting of Auger decay times in this block.

Line Op2: three numbers: *atom* (integer), *shell* (integer), *time* (real) [fs].

The number *atom* sets the number of the element in the compound, analogously to the block MASS above, following the numbers set in the chemical formula in the input file.

The number *shell* sets the atomic shell, for which the Auger decay time must be replaced. The shell numbers are printed out in the output file !OUTPUT_[*material*]_Parameters.txt, see below. The number of the shell must coincide with the number printed out in this file.

The number *time* sets the Auger decay time for the given shell of the given element in [fs].

- Ip: to replace ionization potentials of selected shells of selected elements, use the optional block:

  Line Op1: Ip – the keyword indicating overwriting of ionization potentials in this block.

  Line Op2: three numbers: *atom* (integer), *shell* (integer), *ionization_potential* (real).

  The numbers *atom* and *shell* are analogous to those in the block Auger (see above).

  The *ionization_potential* sets the ionization potential in the shell in [eV].

- Ek: to replace the kinetic energy of electrons in selected shells of selected elements (this value is only used in the BEB cross sections in the MC module), use the optional block:

  Line Op1: Ek – the keyword indicating overwriting of kinetic energies in this block.

  Line Op2: three numbers: *atom* (integer), *shell* (integer), *kinetic_energy* (real).

  The numbers *atom* and *shell* are analogous to those in the block Auger (see above).

  The *kinetic_energy* sets the kinetic energy of electrons in the shell in [eV].

- Ne: to replace the number of electrons in selected shells of selected elements (electron population of the atomic shell), use the optional block:

  Line Op1: Ne – the keyword indicating overwriting of electronic populations in this block.

  Line Op2: three numbers: *atom* (integer), *shell* (integer), *population* (real).

  The numbers *atom* and *shell* are analogous to those in the block Auger (see above).

  The *population* sets the number of electrons in the shell of a given element.

## 3. Executing consecutive runs of the program automatically

If you want to run XTANT-3 program several times in a raw with different parameters (useful, e.g., for finding damage threshold by varying only the pulse fluence while keeping all others parameters the same, or for calculations of electron-phonon coupling parameter vs. electronic temperature), you can (apart from setting corresponding optional parameters in the INPUT file) create several input files – either automatically (see below), or manually in the following manner:

1) Manual creation of multiple input files:

The first simulation run will use as input files INPUT_MATERIAL.txt (or INPUT.txt) and possibly NUMERICAL_PARAMETERS.txt

After the end of the simulation, the program will check the presence of the next input files in the same folder named with consecutive integer numbers at the end of file names:

INPUT_MATERIAL_1.txt (or INPUT_1.txt) and possibly NUMERICAL_PARAMETERS_1.txt

If they are present, XTANT-3 will read the data from these files and start the simulation over from the beginning automatically. The next automatic simulation run must have the next integer number at the end of the files (…_2), and so on.

*Note #1*: that if the file NUMERICAL_PARAMETERS_1.txt with the numerical parameters for the second run is absent, XTANT-3 will use the original file NUMERICAL_PARAMETERS.txt instead.

So, if the numerical parameters are identical in the simulation runs, there is no need to copy the same file – the original file will be reused.

*Note #2*: if the numerical parameters are provided in the INPUT_PARAMETER.txt (or INPUT.txt), the file NUMERICAL_PARAMETERS.txt is unnecessary (as well as its numbered copies).

*Note #3*: if in the INPUT_PARAMETER.txt (or INPUT.txt) file you set a grid for the absorbed dose, several input files will be created automatically with the varying doses, corresponding to the grid given. In this case, all existing numbered files (INPUT_MATERIAL_*i*.txt (or INPUT_*i*.txt) and possibly NUMERICAL_PARAMETERS_*i*.txt) will be overwritten, so make sure there is no conflict between the pre-existing files and the given dose or coupling grid.

2) Automatic creation of multiple input files:

To create copies of the INPUT.txt with only a few lines modified, you can specify how many copies to create, and what lines should be replaced, in an optional file Copy_input.txt, placed in the same directory as INPUT.txt.

The file Copy_input.txt must contain blocks of data with the following lines:

Line #1: ‘new’ or ‘copy’, which specifies the beginning of the block with the lines to be replaced in a copy of the INPUT.txt file. XTANT-3 will create as many copies of the input files, as there are lines containing the flag ‘new’ (or ‘copy’), each file named INPUT_*i*.txt, as described above.

Line #2 and all the other lines within the block: must contain two columns:

Column #1 (integer) sets the number of the line to be replaced in the file INPUT.txt that you are using (which may differ, depending on the optional words and blocks used, see above).

Column #2 (character) specifies the line to be used as the replacement instead of the corresponding line in the file INPUT.txt.

For example, in file shown in Figure VI.5, two copies of the input file will be created. In the first one (INPUT_1.txt), the first line will be replaced with ‘Aluminum’, the second one with ‘Al’, the line #26 (which in this example is responcible for setting number of unit-cells in the block NUMERICAL_PARAMETERS) will be replaced with ‘3 2 2’, and the line 60 will be replaced with ‘3*1’, setting the number of k-points. The second file (INPUT_1.txt) will be created with only first two lines replaced, setting Gold as the material to be modelled.

```
new
1   Aluminum
2   Al
26  3   2   2       ! number of unit-cells in X,Y,Z
60  3*1        ! number of k-points in each direction (used only for Trani-k!)

new
1   Gold
2   Au
```

**Figure VI.5. Example of file Copy_input.txt**

*Note #1*: It is allowed to replace as many lines as necessary. Up to 100 copies of INPUT.txt may be created *via* this method of automatic copying *via* the file Copy_input.txt.

## 4. Folder [*material name*]

Folders with material parameters, named exactly as the material name given above in the file INPUT_MATERIAL.txt (or INPUT.txt), contain several files describing necessary material parameters for the simulation.

*Do not change these files, unless you want to change the properties of the material!*

Folders with the files for already created materials are typically already set, and the user does not need to worry about them. If you want to create a new material, create it by example of already existing folders.

### a) [*A*]_[*A*]_TB_Hamiltonian_parameters.txt and [*A*]_[*A*]_TB_Repulsive_parameters.txt

These files contain all the parameters used in the tight-binding Hamiltonian for each pair-wise interaction of atoms [*A*]. These files contain attractive and repulsive parts, correspondingly. For instance, in the case of a diamond, we only have carbon atoms, thus only files C_C_TB_Hamiltonian_parameters.txt and C_C_TB_Repulsive_parameters.txt should be in the folder. In the case of GaAs, the combination of each interaction should be present in the files: Ga-Ga, Ga-As, and As-As.

In case identical parameters are used for multiple combinations of elements (this is possible within DFTB parameterization files as they only contain links to databases), one can use files: TB_Hamiltonian_parameters.txt and TB_Repulsive_parameters.txt. In a case XTANT-3 cannot find files [*A*]_[*A*]_TB_Hamiltonian_parameters.txt or [*A*]_[*A*]_TB_Repulsive_parameters.txt, it will look up the files without “[*A*]_[*A*]_” prefix.

This also means that one can set default parameterizations in the files without prefixes, and for specific elements use the files with the prefix to specify exceptions.

The first line of the files defines which TB model will be used. The models supported can be specified by the following code words:

- 3TB : ThreeBodyTB model for elemental solids and binary compounds (*only standard 2-body TB is implemented currently, the three-body contribution is unfinished, see below*)
- BOP : a bond-order potential model for dimers [116] (*unfinished, no forces, do not use!*)
- DFTB : one of the s-, $sp^3$- or $sp^3d^5$-basis set according to DFTB [117,129]
- DFTB_no_repulsion: one of the s-, $sp^3$- or $sp^3d^5$-basis set according to DFTB [117,129], but without the repulsive potential in the file
- Fu : $sp^3$-basis set according to Fu [122] (*unfinished, do not use!*)
- Mehl : $sp^3d^5$-basis set according to NRL model [130]
- Molteni : $sp^3s^*$-basis set according to Molteni [123]
- Pettifor : $sp^3$-basis set according to Pettifor [62]
- xTB : extended tight binding [87] (*unfinished, do not use!*)

Thus, the first line in the file must contain one of the possible names of the parameterization. Depending on that, the further lines will define one or another parameterization:

### 1) 3TB parameterization

```
3TB
9.0     0.1      ! cut off [A], smoothing [A]
1.06             ! rc, rescaling coefficient
F                ! include 3-body terms or not (true, false)
T                ! exclude diagonal part of crystal field
```

Figure VI.6. 3TB TB_Hamiltonian_parameters.txt

```
3TB
```

Figure VI.7 3TB TB_Repulsive_parameters.txt

The files contain the parameters, according to ThreeBodyTB[24] parameterization, which may be s, $sp^{3,}$ or $sp^3d^5$, depending on the element.

Line 1 specifies the model name, which must be “3TB”

Line 2 defines embedding cut-off function parameters, ensuring the interaction is short-ranged

Line 3 defines the rescaling coefficient *rc* in the Laguerre polynomials for the TB radial function: *L*(*d*rc*), which shifts the location of the potential minimum, allowing adjustment of the minimum to a desired value. A typical value here is between 1 and 1.1.

*Important note*: In certain cases, it is important to adjust it before productive calculations. Additionally, rigorous tests must be performed for each material: the shape of the cohesive energy curve and the stability of the lattice must be checked since the parameterization was not designed for MD runs (especially in highly excited systems), and not all materials are stable and behaving well dynamically.

Line 4 currently must have option *F*, meaning three-body interactions must be excluded because this option is not yet fully implemented in XTANT.

Line 5 must have the value *T* since the diagonal part of the crystal field should be excluded in calculations.

### 2) BOP parameterization

```
BOP
40.0e0   0.5e0        ! rcut [A], dcut [A]
```

Figure VI.8. BOP TB_Hamiltonian_parameters.txt

```
BOP
```

Figure VI.9 BOP TB_Repulsive_parameters.txt

The files contain the following lines, see for example Figure VI.8 and Figure VI.9.

### 3) DFTB parameterization

```
DFTB
matsci-0-3        ! DFTB parameterization
3.5d0    0.1d0           ! cut-off radius and smoothing dist
```

Figure VI.10. TB_Hamiltonian_parameters.txt

```
DFTB
matsci-0-3        ! DFTB parameterization
1                 ! Type of repulsive: 0=polinomial, 1=spline
```

Figure VI.11 TB_Repulsive_parameters.txt in DFTB format

[24] https://pages.nist.gov/ThreeBodyTB.jl/

The files contain the following lines, see for example Figure VI.10 and Figure VI.11 of C interaction with C (e.g. in diamond). There are two ways to set the path to the skf-files, *via* specifying the parameter in the second line. In the Hamiltonian file, the following lines must be specified (Figure VI.10):

- Line 1: The first line sets the parameterization type (DFTB).
- Line 2: The second line sets the path to the skf-files.
    a) If in this line only the name of the SK parameters is provided, the default directory is assumed ("DFTB" inside of the directory "INPUT_DATA": INPUT_DATA/DFTB under Unix system, or INPUT_DATA\DFTB under Windows).
    b) If the marker "Path" (acceptable forms: PATH, Path, path) is listed in this line, the next line must specify the full path to the skf file or a directory with the skf file (note that the user can use either slash or backslash, the program will make sure it is the correct path separator for the given OS). XTANT-3 first checks if there is a file with the given name in this line and if such a file is not found, it assumes that it is a directory, constructs the default filename (X_Y.skf, where X and Y are the elements), and looks for it in the provided directory.
- Line 3: The third (or fourth, if "PATH" is specified in the second line and the full path in the third) line defines the soft cut-off radius in [Å] and its smoothing in [Å]. Those can be adjusted empirically, but should not be larger than ~10 Å, as this is typically the limit in the SK files of DFTB. It usually makes sense to set the cut-off radius somewhere after the second or third nearest neighbor.

DFTB SK parameter files for various materials can be found, e.g., here:

1. https://dftb.org/parameters/download - official DFTB parameterizations distribution, contains most of the well-tested parameters (but does not have newly developed ones)
2. http://kiff.vfab.org/dftb - alternative distribution of parameter files
3. https://github.com/by-student-2017/Slater-Koster-parameters-no-repulsion_v1 - untested parameters across the periodic table *without* repulsive potential fitted (with no documentation but apparently came from here https://www5.hp-ez.com/hp/calculations/page441). It may (and in most cases does) require the manual fitting of the repulsive potential, for which one may use the function of TB_short below. In the case of compounds, it may also require manual adjustment of the on-site energies of elements to match those expected in compounds.

Apart from the SK-parameters from the official DFTB website, XTANT-3 has the following parameterizations included:

1) PtRu containing parameters for Ru-Pt compounds from [131]
2) TransMet containing Ag-Ag, Au-Au, Cu-Cu, Ni-Ni, Pd-Pd, Pt-Pt (for nanoclusters) from [132]
3) trans3d-LANLFeC for Fe-C-O from [133]
4) Al2O3 containing files Al-Al, Al-O, O-O, optimized for solid $Al_2O_3$ from [134]
5) Ga2O3 containing files Ga-Ga, Ga-O, O-O, optimized for solid $Ga_2O_3$ from [135]
6) BC containing B-C from https://github.com/tlyoon/BC_SKfiles
7) 3ob_cnoh containing parameters for C-H-N-O-P-S interactions from https://bitbucket.org/solccp/adpt_core/src/master/erepfit/example/3ob_cnoh/

8) Verners_2023 containing parameters for a range of materials (similar to pbc parameterization) from http://dx.doi.org/10.13140/RG.2.2.28507.54564
9) Li-C containing data for Li-graphite interaction from [136]
10) Al-C containing improved Al-C interaction based on matsci-parameters from https://apps.dtic.mil/sti/tr/pdf/AD1026685.pdf
11) dataset_Ru_RuO for Ru-O from https://doi.org/10.17617/3.CRSJQV (requires specifying full path in the parameters file, because it contains separate subdirectories for Ru and Ru-O)
12) MoS containing Mo-S from https://github.com/hyllios/dftb_auto (need to be converted from svn format to skf!)

- List of tools that can be used to create your own SK-parameters:

https://github.com/dftbplus/skprogs
https://github.com/pekkosk/hotbit (Instruction for creation of parameterizations: https://github.com/pekkosk/hotbit/wiki/Parameters-and-parametrization)
https://www.dftbaby.chemie.uni-wuerzburg.de/DFTBaby/mdwiki.html#!WIKI/main_page.md
https://bitbucket.org/solccp/adpt_core/src/master/
https://gitlab.com/mvdb/tango
https://github.com/v2quan89/DFTBparaopt

In the Repulsive parameters file (Figure VI.11):

Line 1: The first line sets the parameterization (DFTB). Must be the same as in the Hamiltonian file.

Line 2: The second line sets the set of SK parameters to be used according to DFTB, which must coincide exactly with the directory name existing within the DFTB directory. Must be the same as in the Hamiltonian file.

Line 3: The third line defines which form of the Repulsive term to use according to DFTB format: polynomial (0) or spline (1). It is recommended to use spline (since it contains exponential repulsion as short distances, which polynomial does not; also, apparently not in all sk-files polynomial form is given), thus 1 is the default choice.

All the parameters within the skf-files are described on the dftb-website[25]. Do not change those files (unless you know what exactly you want to change in the parameterization of TB Hamiltonian)!

Also note that currently XTANT-3 only supports zero-level DFTB, non-self-consistent calculations for productive runs (only for BO simulations with DFTB specifically).

### 4) *DFTB_no_repulsion parameterization*

This parameterization is identical to the DFTB described above, but without the repulsive potential, e.g. found in https://github.com/by-student-2017/Slater-Koster-parameters-no-repulsion_v1. In this case, there may be an additional line present in the Hamiltonian file, specifying the default inner directory: in the “2elements” directory, there are other directories with pair-wise elements, e.g. Al-X, etc.), meaning if the second line specifies the parameterization “2elements”, there must be the third line specifying which subfolder to take the parameterization from (see example in Figure VI.12).

[25] http://www.dftb.org/parameters/introduction/

```
DFTB_no_repulsion
2elements                    ! DFTB parameterization without repulsion
Al-X                             ! inner folder with parameters
5.0d0    0.1d0               ! cut-off radius and smoothing distance [A]
```

**Figure VI.12. TB_Hamiltonian_parameters.txt**

```
DFTB_no_repulsion
```

**Figure VI.13 TB_Repulsive_parameters.txt in DFTB_no_repulsion format**

In this case, the file TB_Repulsive_parameters.txt needs only one line with the name. Don't forget to use TB_short additional potential as a separate file.

### *5) Fu parameterization*

The files contain parameters in exactly the same format as Pettifor, where only the first line should read "Fu" instead of "Pettifor". See the description below. *Unfinished, not recommended for use!*

### *6) Mehl parameterization*

The Hamiltonian files should contain 3 introductory lines described below, and then a set of 97 parameters as defined in the NRL format. The Repulsive potential files should contain only one line "Mehl", identifying the parameterization. They are all already prepared, so no need to change anything. To set new parameters for new material, use existing files as an example (starting from the third line, only the first column of parameters is used in the code, the others are comments explaining the meaning of the parameters).

```
NUMERICAL_PARAMETERS.txt | INPUT_MATERIAL.txt | C_C_TB_Repulsive_parameters.txt | C_C_TB_Hamiltonian_parameters.txt | C_(
1   Mehl
2   1    0    1              ! parameterization index (1=with split between t2g and e2, or 0=without); (Sll'm 1=with
3   10.5    0.5                          (RCUT and SCREENL for 1-1 interactions)
4     0.159901905594E+01   0   1     lambda  (equation 7)
5     -.102789972814E+00   0   2     a_s     (equation 9)
6     -.162604640052E+01   0   3     b_s     (equation 9)
7     -.178884826119E+03   0   4     c_s     (equation 9)
8      .451611342028E+04   0   5     d_s     (equation 9)
9      .542619178314E+00   0   6     a_p     (equation 9)
10     .273454062799E+01   0   7     b_p     (equation 9)
11    -.671397098834E+02   0   8     c_p     (equation 9)
12     .438528831452E+03   0   9     d_p     (equation 9)
13     .100000000000E+02   1  10     a_t2g   (equation 9)
14     .000000000000E+00   1  11     b_t2g   (equation 9)
15     .000000000000E+00   1  12     c_t2g   (equation 9)
16     .000000000000E+00   1  13     d_t2g   (equation 9)
17     .100000000000E+02   1  14     a_eg    (equation 9)
18     .000000000000E+00   1  15     b_eg    (equation 9)
19     .000000000000E+00   1  16     c_eg    (equation 9)
20     .000000000000E+00   1  17     d_eg    (equation 9)
```

**Figure VI.14 File TB_Repulsive_parameters.txt in NRL format**

Line 1: must be "Mehl" identifying the parameterization name

Line 3: contains three numbers:

- First: included split between t2g (1) or excluded (0), as specified in an original NRL parameters file[26][72]
- Second: includes the terms Sll'm with delta (1) or without (0)
- set f_bar = 0 or not (=1)

Line 3: sets the cutoff distance in [A] and its smoothing in [A].

[26] The NRL parameters files used to be available at http://cst-www.nrl.navy.mil/bind/, and later at http://esd.spacs.gmu.edu/tb/tbp.html, but currently can be extracted by means of internet archives such as Wayback Machine: https://archive.org/web/

All these parameters must be extracted from the original NRL .par files. The rest of the lines must be just copied from NRL .par files without any changes. The available NRL .par files are saved in the directory NRL_TB_database and can be used to construct required parameterizations for chosen materials.

*Note #1*: that NRL parameterization often needs additional short-range repulsive potential [2], as may be defined in a short-range (or "wall") file, see below.

*Note #2*: NRL parameterization has only been tested in XTANT-3 with elemental materials, no compounds have been tested with this parameterizations, so may not work.

### *7) Molteni parameterization for sp$^3$s* basis set*

The files contain the following lines, see Figure VI.15 and Figure VI.16 for examples of Ga interaction with As (GaAs). All the parameters are described in [123,137]. Do not change these files (unless you know what exactly you want to change in the parameterization of TB Hamiltonian).

```
Molteni
-2.657d0     ! Es, on-site energy of s-orbital [eV]
3.669d0      ! Ep, on-site energy of p-orbital [eV]
6.739d0      ! Es*, on-site energy of s*-orbital [eV]
-1.613d0     ! V0 (s s sigma)
2.504d0      ! V0 (s p sigma)
3.028d0      ! V0 (p p sigma)
-0.781d0     ! V0 (p p pi)
2.082d0      ! V0 (s* p)
2.46d0       ! r0
2.0d0        ! n
3.511d0          ! rc
13.0d0       ! nc
1.3d0        ! N; rcut = N*r0
0.1d0        ! d
```

Figure VI.15. File TB_Hamiltonian_parameters.txt in Molteni forma

```
Molteni
1        ! NP, type of potential
2.3906d0     ! phi1
1.2347d0     ! phi2
2.46d0       ! r0
1.3d0        ! N; rcut = N*r0
0.1d0        ! d
0.3555d0     ! alpha (NP=1) or m (NP=2)
```

Figure VI.16 File TB_Repulsive_parameters.txt

### *8) Pettifor parameterization*

The files contain the following lines, see Figure VI.17 and Figure VI.18 for an example of Carbon. All the parameters are described in [4,74].

```
Pettifor
-5.25d0      ! Es, on-site energy of s-orbital of Si [eV]
1.20d0       ! Ep, on-site energy of p-orbital of Si [eV]
-2.038e0     ! V0 (s s sigma)
1.745e0      ! V0 (s p sigma)
2.75e0       ! V0 (p p sigma)
-1.075e0     ! V0 (p p pi)
2.360352e0   ! r0
2.0e0        ! n
4.0e0        ! r1
4.16e0       ! rm
9.5e0        ! nc(1)
8.5e0        ! nc(2)
7.5e0        ! nc(3)
7.5e0        ! nc(4)
3.5e0        ! rc(1)
3.55e0       ! rc(2)
3.7e0        ! rc(3)
3.7e0        ! rc(4)
-6.4651e-5   ! c0(1)
0.0014704        ! c1(1)
-0.010804        ! c2(1)
0.025871     ! c3(1)
0.0026017    ! c0(2)
-0.031794        ! c1(2)
0.092548     ! c2(2)
0.028371     ! c3(2)
0.028336     ! c0(3)
-0.20485     ! c1(3)
-0.76001     ! c2(3)
5.8340       ! c3(3)
-0.011077        ! c0(4)
0.08077      ! c1(4)
0.29709      ! c2(4)
-2.2806      ! c3(4)
```

```
Pettifor
8.7393204    ! E0_TB [eV]
1.0e0        ! phi0
6.8755e0     ! m
13.017e0     ! mc
2.360352e0   ! d0
4.0e0        ! d1
4.16e0       ! dm
3.66995e0    ! dc
1.8790e-11   ! c0
1.3221e-9    ! c1
1.4324e-8    ! c2
-4.2468e-8   ! c3
0.0e0        ! a0
2.1604385    ! a1
-0.1384393   ! a2
5.8398423e-3     ! a3
-8.0263577e-5    ! a4
```

**Figure VI.17. TB_Hamiltonian_parameters.txt in Pettifor format** **Figure VI.18 TB_Repulsive_parameters.txt**

### c) [*A*]_[*A*]_vdW.txt

This optional file contains the parameters used in the van der Waals (vdW) potential for each pair-wise interaction of atoms [*A*]. For instance, in the case of diamond, we only have carbon atoms, thus only files C_C_vdW.txt should be in the folder. At present, the vdW potential is set according to Girifalco's model, in the shape of the (improved) Lennard-Jones potential (see e.g. [138]), smoothly cut at large (for better energy conservation, and to limit it to some reasonable distance in the spirit of TB) as well as at short distances (not to overlap with the TB covalent bonds). The cut-off may be performed in two different ways: *via* Fermi-like cut-off functions, or *via* switching to fitted polynomials as described in Ref.[85].

Line 3: The first line in the file must contain the exact name of the parameterization. The available options are:

1) LJ (or Lennard-Jones, Lennad_Jones), to use Fermi-like cutoffs:

$$V(r) = V_{LJ}(r) f_s(r) f_l(r)$$
$$f_s(r) = 1 - \left(1 + \exp\left((r - \delta d_{short})/d_{0_short}\right)\right)^{-1} \quad (50)$$
$$f_l(r) = \left(1 + \exp\left((r - \delta d_{long})/d_{0_long}\right)\right)^{-1}$$

In this case, one can use three different forms of the LJ potential[27]: 12-6 form (ε-σ form), AB form, and n-exp form (see Section V.4).

Lines 2 and 3 must specify the form and its parameters, coprrespondingly:

- SE for ε-σ form. In this case, the next line must specify two numbers:
  ε σ - in [eV] and in [Å], correspondingly;
- AB for AB form. In this case, the next line must specify two numbers:
  A B - in [eV*$Å^{12}$] and in [eV*$Å^6$], correspondingly;
- n-exp for n-exp form. In this case, the next line must specify three (or two) numbers:
  ε $r_0$ n - in [eV], in [Å], and [-] correspondingly; if the power coefficient *n* is not provided, the default value of *n*=6 is used. Note that in this representation, the coefficients have a straightforward meaning: ε is the depth of the potential well, $r_0$ is the position of the potential minimum, hence this form is the most convenient to use.

Line 4: in all cases must be the parameters of the cutoff at short distances:

Two numbers: $d_{0_short}$, $\delta d_{short}$ – cutoff radius and its width, both in [Å]

Line 5: in all cases must be the parameters of the cutoff at long distances:

Two numbers: $d_{0_long}$, $\delta d_{long}$ – cutoff radius and its width, both in [Å]

2) ILV, to use improved LJ potential (see Section V.4), which has the same general form as in Eq.(50), but with $V_{LJ}(r)$ replaced with $V_{ILJ}(r)$. In this case, three more lines are required:

[27] https://en.wikipedia.org/wiki/Lennard-Jones_potential

Line 2: Four numbers specifying the parameters of $V_{ILJ}(r)$:

ε r0 n m - in [eV], in [Å], and [-], [-] correspondingly.

Line 3: in all cases must be the parameters of the cutoff at short distances:

Two numbers: $d_{0_short}$, $\delta d_{short}$ – cutoff radius and its width, both in [Å].

Line 4: in all cases must be the parameters of the cutoff at long distances:

Two numbers: $d_{0_long}$, $\delta d_{long}$ – cutoff radius and its width, both in [Å].

3) Girifalco, to use polynomial cutoffs.

Examples of the files are shown in Figure VI.19. This type of parameterization assumes LJ potential in AB form, which switches to polynomials at short and at large distance. All polynomial coefficients must be provided. The coefficients must be set such that the potential, its first derivative (and long distances) and second derivatives (at short distances) all coincide with those of the LJ potential at the chosen distances, and goes to zero at the chosen short distance.

Note: since this requires careful fitting, it is more convenient to use LJ-type potential instead.

If the file is absent, the calculations will proceed without vdW forces. So, this module will not affect the materials for which there is no vdW force or parameterization.

```
1  LJ
2  n-exp
3  0.03d0  3.5d0   6.0d0            ! epsylon, r0, n            ! [eV], [A], [-]
4  3.2d0   0.03d0        ! d0_short, dd_short     ! [A] short-range cutoff radius, [A] cutoff width
5  6.0d0   0.1d0         ! d0_cut, dd_cut         ! [A] long-range cutoff radius, [A] cutoff width
```

```
Girifalco
22.5d3          ! C12 [eV*A^12] Lenard-Jones C12
15.4d0          ! C6 [eV*A^6] Lenard-Jones C6
5.0d0           ! dm [A] radius where to switch to polinomial
6.0d0           ! d_cut [A] cut-off radius
-0.8253440000d-3        ! a, fitting polinomial coefficients: a*x^3+b*x^2+c*x+d
0.1313740800d-1         ! b
-0.6851174400d-1        ! c
0.1163980800d0          ! d
3.4d0           ! dsm [A] radius where to switch to polinomial at small distances
2.5d0           ! ds_cut [A] cut-off radius on small distances
0.1286478847d0          ! as, fitting polinomial coefficients: a*x^3+b*x^2+c*x+d
-1.858707955d0          ! bs
10.66718892d0           ! cs
-30.40360058d0          ! ds
43.05091787d0           ! es
-24.23710839d0          ! fs
```

Figure VI.19 Examples of vdW files: with Lennard-Jones-type parameters (top), and with Girifalco-type parameters (bottom).

### d) [A]_[A]_TB_Coulomb.txt

This optional file contains parameters of the softly-cut Coulomb potential, softly cut according to [85]. It is only used in the case we have electron emission included in the simulation (e.g. modeling thin films). An example of the file is shown in Figure VI.20.

If the file is absent, the calculations will proceed without Coulomb forces. So, the Coulomb module will not affect the materials for which there is no Coulomb force or parameterization.

```
Coulomb_cut
20.0d0  ! dm
0.1d0   ! dd
```

**Figure VI.20 File C_C_TB_Coulomb.txt for $C_{60}$.**

### e) [*A*]_[*A*]_TB_short.txt or [*A*]_[*A*]_TB_wall.txt

This optional file contains parameters of the short-range repulsive potential, *in addition* to the repulsive potential included in the used TB parameterization. E.g., it is to be used in case we have kinetic energies of atoms higher than the TB-provided barrier at short distances (in such a case, it is necessary to make the system stable at short distances); or in TB parameterizations where the repulsive part is not provided. The current parameterization may combine a few functional shapes described below (the obsolete version called "exponential wall" only includes the inverse exponential).

Generally, the repulsive potential is expressed in the form:

$$E_{SR} = F_{SR} f_{cut}, \tag{51}$$

Where $F_{SR}$ is the short-range function, and $f_{cut}$ is the cut-off function in the shape of the Fermi-function: $f_{cut} = f_l$ (from Eq.(50)).

This function is always included, independently of the choice of $F_{SR}$, thus, its parameters must be always listed in the file, if additional short-range potential to be included. If unspecified, zero potential will be used (cut off at zero radii).

Line 1: The first line in the file must specify either the path to the file with the parameter, or the parameterization name. This means, it can be either:

"Path", or parameterization name:

"General" – to specify the new format described above, or

"Simple_wall" – to specify the obsolete format which includes only the inverse exponential function.

An example of the new and the obsolete files containing the parameters in the same notations is shown in Figure VI.21.

```
1   PATH
2   INPUT_DATA\!Short_range_potentials\Fe_O_trans3d-0-1.txt
```

```
1   General
2
3   Cut-off
4   2.7d0    0.25d0        ! d0 [A]; dd [A]
5
6   Inv_exp
7   35.0     1.5e0         ! C [eV]; r0 [A]
```

```
1   Simple_wall
2   1.0d2         ! C [eV]
3   0.0d0         ! r0 [A]
4   1.165d0           ! d0 [A]
5   0.01d0            ! dd [A]
```

**Figure VI.21 Equivalent presentation of inverse exponential short-range (exponential wall) function in three possible formats: specifying path to the file with the parameter (top); the new format (left) and in the obsolete format (right).**

If the flag "Path" is specified, the next line (Line 2) must contain the path to the file with the short-range potential parameters. Note that it does not matter what path separator is used in the path (slash or backslash), the program will automatically correct it to the one used in Windows or Linux. The specified file, then, must contain one of the allowed flags: "General" or "Simple_wall", as described below.

If the file is absent or anything else but "General" or "Simple_wall" is specified in the first line, the calculations will proceed without an additional short-range repulsive potential.

The short-range repulsive function, $F_{SR}$, may assume the following shapes, summed up pairwise for all the atoms in the simulation box ( $F_{SR} = \frac{1}{2}\sum_{i,j}^{N_{at}} F_{SR}^{i,j}$ ), Line 2 (and more):

| **Flag** | **Function** | **Parameters to be specified in the file** (must be exactly in the given order!) |
|---|---|---|
| Cutoff, or Cut_off, Cut-off | $\left(1 + exp\left(\frac{r - d_0}{d_a}\right)\right)^{-1}$ | One line:<br>$d_0$ $d_d$ – cut-off radius [A], smoothing distance [A] |
| Exp, Exponential | $\phi \exp\left(-\frac{r - r_0}{a}\right)$ | One line:<br>$\Phi$ $r_0$ $a$ – Energy [eV], characteristic radius [A], characteristic distance [A] |
| InvExp, Inv_Exp | $C \exp\left(\frac{1}{r - r_0}\right)$ | One line:<br>$C$ $r_0$ – Energy [eV], characteristic radius [A] |
| Pow, Power | $F_{SR} = \Phi\left(\frac{r}{r_0}\right)^m$ | Two or more lines:<br>$N$ – number of power functions (integer)<br>$N$ more lines each with three numbers for each power-function:<br>$\Phi$ $r_0$ $m$ – Energy [eV], characteristic radius [A], power |
| ZBL | Ziegler-Biersack-Littmark potential [84] | *No additional lines required, all parameters are fixed and included in the code* |
| TAB, Tab, TABLE, Table,<br>Tabulated | Tabulated potential | Line 1: must specify the number of points, and may (optionally) specify the type of integration of this potential; thus, the line must contain at least one number, but may contain a number and a text variable:<br>$N$ (integer) – number of grid points in the tabulated potential; text marker: “diff” or “spline”. The marker “diff” sets the finite difference method for evaluation of the potential and forces between the tabulated points. The marker “spline” sets the cubic spline method for calculations of potential and forces (similar to the DFTB repulsive potential). Note that the default option is “diff”, which is used if no text marker is specified in this line. Note #2: finite difference method works better for dense grids in the potential, spline works better for spars grids.<br>Lines 2-N+1:<br>Tabulated potential in two columns:<br>Column 1: (real) interatomic distance [Å]<br>Column 2: (real) potential energy [eV] |
| Simple_wall (*obsolete, cannot be combined with* | $C \exp\left(\frac{1}{r - r_0}\right) f_{cut}$ | Four lines:<br>$C$ – Energy [eV], |
| *other functions!*) | | $r_0$ – characteristic radius [A]<br>$d_0$ – cut-off radius [A]<br>$d_d$ – smoothing distance [A] |

It is allowed to use a few power functions to construct a polynomial (or inverse fractions, using negative powers "m", such as those used in [123]).

### f) Setting initial atomic configuration

The following formats may be used in setting the initial atomic configuration, in the following priorities:

1) Files with reaction coordinate path (using internal XTANT-3 SAVE-files format, see below) between two different phases of material (instead of full MD simulation); if the coordinate-path files are absent, the next option is checked.
2) SAVE-files (internal XTANT-3 format for setting the supercell parameters and atomic coordinates); if SAVE-files are absent, the next option is checked.
3) XYZ-file (in extended XYZ format): user-specified name with XYZ-file is allowed (must be a file with the extension ".xyz"). If the user did not provide the name with the file, the default filename is checked: Cell.xyz. If this file is absent, the next option is checked.
4) POSCAR format[28] file: user-specified name with poscar-file is allowed (must be a file with the extension ".poscar"). If the user did not provide the name, the default file is check: Cell.poscar. If this file is absent, the next option is checked.
5) mol2 format[29] file: user-specified name with mol2-file is allowed (the file must have the extension ".mol2"). If the user did not specified the name, the default filename is checked: Cell.mol2. If this file is absent, the next option is checked.
6) Unit-cell parameters and atomic coordinates in it (internal XTANT-3 format).

All the formats are described below.

#### i. Files PHASE_[*i*]_atoms.dat and PHASE_[*i*]_supercell.dat

The user can set the initial and final states of the simulation cell for calculation of the free-electron along a reaction coordinate path. Here [*i*] runs from 1 to 2, the index of the initial and the final phase. It can be done in the following way:

#### Calculation of free-energy along reaction coordinate path

Set initial atomic configuration in the files PHASE_1_atoms.dat and PHASE_1_supercell.dat

Set final atomic configuration in the files PHASE_2_atoms.dat and PHASE_2_supercell.dat

Run XTANT. If these files are present in the folder with the material data, XTANT-3 will linearly interpolate coordinates from the first to the second phase (accounting for periodic boundaries) and save all the output data along this coordinate path. The free energy will be calculated for the electronic temperature provided in the INPUT_MATERIAL.txt (or INPUT.txt) file (see above).

[28] https://www.vasp.at/wiki/index.php/POSCAR
[29] https://www.structbio.vanderbilt.edu/archives/amber-archive/2007/att-1568/01-mol2_2pg_113.pdf

### ii. Files SAVE_supercell.dat, SAVE_atoms.dat and SAVE_el_distribution.dat

An alternative way to set the initial configuration of the atoms and supercell is to have the files SAVE_supercell.dat and SAVE_atoms.dat in the directory. If these files are present, the program will use them instead of the 'Unit_cell_atom_relative_coordinates.txt' and 'Unit_cell_equilibrium.txt' files described above. With these files, you can add any desired atomic configuration, not only perfect periodic crystalline lattice.

The file SAVE_el_distribution.dat sets the electron distribution function. It must be present in the folder if the given electronic distribution needs to be used (e.g., to restart calculations with exactly the same parameters). In the absence of this file, the Fermi function with the given temperature will be used. This file allows to restore a nonequilibrium electronic distribution. The presence of the file supersedes the electronic temperature given in the input file: if the file is present, this distribution is used, and the electronic temperature specified is ignored.

The file 'SAVE_atoms.dat' must contain the data in the same format as the file 'OUTPUT_coordinates_and_velosities.dat' (see below).

The format of the file 'SAVE_supercell.dat' must coincide with the format of the file 'OUTPUT_supercell.dat' (see below).

If these files are present, XTANT-3 will ignore the number of unit-cells specified in the NUMERICAL_PARAMETERS file, and use the atomic coordinates and the supercell provided here.

These files can be used, e.g., for simulation of amorphous materials, and relaxation process before productive runs. They must be constructed separately as follows:

#### Creation of initial configuration of an amorphous material

1) Choose a material you'd like to construct (e.g. carbon or silicon-based)
2) Choose a number of atoms in the simulation box, which will be used for all further simulations of the amorphous state (each number of atoms chosen requires separate preparation of the initial state!)
3) Set the proper density of the desirable amorphous material by adjusting the volume in the file 'Unit_cell_equilibrium.txt' of the material we will start from the melt
4) Set parameters for quenching in the NUMERICAL_PARAMETERS input file as follows:

```
1   1000.0      10.0
```

The first parameter tells to include material cooling (quenching), which is made by setting atomic velocities to zero starting from the time, given in the second number (e.g. 1000.0 fs, after the melting), then propagates the atomic trajectories and repeats the procedure every (~10.0) femtosecond (e.g. for silicon-based material, chose here ~30.0 as an optimal time). Set initial conditions that would create a melted state (e.g., by setting high atomic temperature or irradiation with a high deposited dose). Run this simulation for a few picoseconds, until the total energy stops dropping. This means the material is relaxed into its equilibrium amorphous state.

Don't forget to switch this off (by setting the first parameter to 0) for further real simulations!

5) The files 'SAVE_atoms.dat' and 'SAVE_supercell.dat' are created during the simulation run in the output folder, and are updated at each saving time step; thus, the data in them after the simulation has finished correspond to the last time-step of the simulation and can be just copied into the input file, provided the simulation delivered desired quality of results. Place these files into the folder with the new material name, see next step.
6) Place both files into the folder 'Amorphous_[material_name]'. Copy all other input files from the directory of 'parental' material (ideal material that you just melted) into the same folder, and set the name of the folder the same as the material name in the file 'INPUT_FILE.txt'.
7) Place the files 'SAVE_atoms.dat' and 'SAVE_supercell.dat' into the same directory.
8) Check if there is no artificial void in the new created state, and the density is uniform (and any other properties that are needed to be reproduced well in your amorphous material). If the amorphous material looks good, these files with the relaxed amorphous atomic state can now be used for further simulations of amorphous material. If not, repeat the procedure from the beginning until the quenched state is produced to satisfy your conditions.

### iii. Extended XYZ files

Another alternative way to set the initial conditions is to use a file Cell.xyz (or a name of the file with the extension xyz specified in the input file), which uses extended XYZ format[30,] where the extended format means that instead of the comment-line, line #2 specifies important parameters not included in the simple XYZ format. The format is as follows:

Line 1: number of atoms (integer)

Line 2: Specifiers. For the usual XYZ format, the following specifiers are needed[31]:

Lattice="ax ay az ba by bz cx cy cz" Properties="species:S:1:pos:R:3"

- "Lattice" sets the unit- or super-cell vectors;
- "Properties" define in which format the next lines are specified:
  - "species" indicates that the first column will be defining the element type; if after the column "S" is specified then the symbol of the element from the Periodic Table must be in the first column; if "I" is specified then the index corresponding to the order in the provided chemical formula (in the INPUT file) must be in the first column.
  - "pos" defines that in the next 3 columns, atomic coordinates will be provided: if "R" is defined here, then absolute coordinates in [Å] must be given; if "S" is specified then relative coordinates within the supercell must be provided.
  - "vel" or "velo" specifier identifies that columns 5-7 will be defining atomic velocities; if "V" is defined here, absolute velocities in [Å/fs] must be given; if "S" is specified, then relative velocities within the supercell must be provided.
  - "mass" specifies that the column will contain the atomic mass (in [kg]); this option is only used in the output XYZ file, but not read in case of the input XYZ file.

[30] https://en.wikipedia.org/wiki/XYZ_file_format
[31] https://www.ovito.org/manual/reference/file_formats/input/xyz.html

- "charge" specifies that the column will contain atomic charge (in the units of electron charge; currently only Mulliken charge is supported); this option is only used in the output XYZ file, but not read in case of the input XYZ file.
- "kinetic_energy" specifies that the column will contain atomic kinetic energy (in [eV]); this option is only used in the output XYZ file, but not read in case of the input XYZ file.

Line 3: and further until the number of atoms specified in line #1: Element name or atomic type; Coordinates.

In the second line, instead of the standard specifiers "Lattice" and "Properties", the user may choose to set a random atomic arrangement with the given material density. For that, the following identifier must be used:

"Random" (Note that it is not a part of the standard extended XYZ format, but specific to XTANT-3). If only the flag Random is set, then the unit-/super-cell will be cubic, with the size chosen to reproduce the density specified in the next lines. Alternatively, the user may set the X and Y dimensions of the supercell, while the Z dimension will be chosen to reproduce the density of the material. E.g.:

Random - cubic cell

Random="X 15.0 Y 14.0" – X dimension is set to be 15.0 Å, Y dimension is set to be 14.0 Å.

In this format, inside the quotation marks, 4 values must be set: character "X" (or "Y"), value (real) of the supercell in the dimension specified, character "Y" (or "X"), value (real) of the supercell in the dimension specified.

If this specifier is used, the next lines must define the following parameters:

Element name from the periodic table (character); density (real) [g/cm$^3$]; the number of atoms (integer)

E.g., for a single aluminum cell with 100 atoms, one should write:

Al 2.7 100

For a multilayer system, containing, e.g., Al (100 atoms) and Cu (80 atoms) layers, the following should be set:

Al 2.7 100
Cu 8.96 80

The layers will be stacked along the Z direction.

The elements set in this file *must* be included in the chemical formula set in the INPUT_MATERIAL.txt file (see above).

*Note #1*: that the XYZ file sets a "unit cell", and a supercell will be constructed, according to the number of images set in the NUMERICAL_PARAMETERS.txt file.

*Note #2*: similar to setting a molecular in water, such random placing of atoms requires two-step relaxation: first, quenching to the ground state is required, and after that, setting the required temperature (e.g., room temperature) may be done with help of Berendsen thermostat.

### iv. POSCAR file

A file must have the extension ".poscar" (the default file name is "Cell.poscar"); the format is the standard[32] with no modifications.

### v. mol2 file

A file with the input atomic and supercell data may also be provided in the format mol2, which then must have an extension ".mol2" (the default file name is "Cell.mol2"). The format is standard[33] with no modifications. May be useful for files with molecules in SYBYL format[34].

### vi. Unit_cell_atom_relative_coordinates.txt and Unit_cell_equilibrium.txt

The file Unit_cell_atom_relative_coordinates.txt contains the coordinates of atoms inside of the equilibrium unit-cell in relative coordinates. The number of lines in this file defines how many atoms we have in the unit cell of the material. For example, it is 8 atoms for GaAs; see Figure VI.22. These 8 lines contain the following information:

The first number stands for the kind of atom according to its chemical formula given in the input file. For example, for GaAs number 1 stands for Ga, and number 2 stands for As. *Make sure your order of elements in the chemical formula in the file matches the order of elements in this file!*

The next 3 numbers in each line represent the initial relative coordinates $S_x$, $S_y$, $S_z$ of 8 atoms inside the unit cell (normalized from 0 to 1).

```
Unit_cell_atom_relative_coordinates.txt
1 1 0.0e0  0.0e0  0.0e0
2 1 0.5e0  0.5e0  0.0e0
3 1 0.5e0  0.0e0  0.5e0
4 1 0.0e0  0.5e0  0.5e0
5 2 0.25e0 0.25e0 0.25e0
6 2 0.75e0 0.75e0 0.25e0
7 2 0.75e0 0.25e0 0.75e0
8 2 0.25e0 0.75e0 0.75e0
```

**Figure VI.22 File Unit_cell_atom_relative_coordinates.txt for GaAs.**

The file Unit_cell_equilibrium.txt contains the initial vectors of the unit cell in angstroms [A], to be later evolved according to the Parrinello-Rahman method [90] (for P=const simulation). The file contains three vectors (as columns), see Figure VI.23.

```
Unit_cell_equilibrium.txt - Notepa
File Edit Format View Help
5.431d0 0.0d0   0.0d0
0.0d0   5.431d0 0.0d0
0.0d0   0.0d0   5.431d0
```

**Figure VI.23 File Unit_cell_equilibrium.txt for silicon**

[32] https://www.vasp.at/wiki/index.php/POSCAR
[33] https://www.structbio.vanderbilt.edu/archives/amber-archive/2007/att-1568/01-mol2_2pg_113.pdf
[34] E.g., from the CCDC molecular structures database: https://www.ccdc.cam.ac.uk/structures/

### f) File with Ritchie CDF coefficients: [*Material*].cdf

The file with cdf-coefficients fitted within the Ritchie-Howie formalism [50]. The name (and the path to) of the file may be defined by the user, or the default name may be used: [*material*].cdf. This file contains all parameters needed for Monte Carlo calculations of the electron cross-sections within the Ritchie-Howie complex dielectric function formalism [50]. The file format is compatible with that used in TREKIS-3 code[35]. It contains the following lines, see Figure VI.24, described below. Only used if the cdf-cross sections are chosen, not used for BEB cross sections.

Silicon.cdf

```
1  Silicon
2  Si
3  2.3290  8430.0  ! density [g/cm^3], speed of sound [m/s]
4  4               ! number of shells of the first element: Si
5  1   1   1838.9e0    2.0 1.66e0       ! number of CDF , shell-designator, ionization potential, num
6  1579.84393e0    236.1525e0  1192.8108e0 ! E0, A, Gamma coefficients
7  1   3   148.7e0 2.0 0.375e0          ! number of CDF functions, shell-designator, ionization poten
8  219.31853e0 223.14459e0     199.22160e0 ! E0, A, Gamma coefficients
9  1   4   99.2e0  6.0 16.06e0          ! number of CDF functions, shell-designator, ionization poten
10 100 685 145                              ! E0, A, Gamma coefficients
11 2   63  1.12    4.0 1.0e23           ! number of CDF functions, shell-designator, ionization poten
12 15.9504e0   113.883e0   3.212e0          ! E0, A, Gamma coefficients
13 17.499583e0 135.1805e0  3.0291e0         ! E0, A, Gamma coefficients
```

**Figure VI.24 Example file [*Material*].cdf (Silicon)**

Line 1: material name

Line 2: chemical element name(s), according to the rules described above

Line 3: density of the material in [g/cm$^3$] (this number is the default value, which may be overwritten by the input file, see above), its speed of sound (not used in this version of MC), and the Fermi-level in [eV] (also unused in the current implementation)

Line 4: number of shells of the first element in the compound

Line 5: for the first shell, the next five numbers in the line:

1) Number of complex dielectric function (CDF) oscillators used in the formalism [51]
2) Shell designator according to the EADL[36] database (see a copy in Table VI below), with an additional notation of shell number 63 corresponding to the valence band.
3) The ionization potential of this shell in [eV]
4) Number of electrons in this shell
5) Time of Auger-decay in [fs]; for the valence band set here a huge number such as 1.0d23.

[35] The code TREKIS-3 and cdf-files for many materials may be found at: https://github.com/N-Medvedev/TREKIS-3
[36] See details of this library here: https://www-nds.iaea.org/epdl97/libsall.htm, physical details and references for the library are here: https://www-nds.iaea.org/epdl97/document/epdl97.pdf

**Table VI. Atomic Subshell Designators**

| Desig-nator | Subshell | | Desig-nator | Subshell | | Desig-nator | Subshell | |
|---|---|---|---|---|---|---|---|---|
| 1. | K | (1s1/2) | 21. | N4 | (4d3/2) | 41. | P1 | (6s1/2) |
| 2. | L | (2) | 22. | N5 | (4d5/2) | 42. | P23 | (6p) |
| 3. | L1 | (2s1/2) | 23. | N67 | (4f) | 43. | P2 | (6p1/2) |
| 4. | L23 | (2p) | 24. | N6 | (4f5/2) | 44. | P3 | (6p3/2) |
| 5. | L2 | (2p1/2) | 25. | N7 | (4f7/2) | 45. | P45 | (6d) |
| 6. | L3 | (2p3/2) | 26. | O | (5) | 46. | P4 | (6d3/2) |
| 7. | M | (3) | 27. | O1 | (5s1/2) | 47. | P5 | (6d5/2) |
| 8. | M1 | (3s1/2) | 28. | O23 | (5p) | 48. | P67 | (6f) |
| 9. | M23 | (3p) | 29. | O2 | (5p1/2) | 49. | P6 | (6f5/2) |
| 10. | M2 | (3p1/2) | 30. | O3 | (5p3/2) | 50. | P7 | (6f7/2) |
| 11. | M3 | (3p3/2) | 31. | O45 | (5d) | 51. | P89 | (6g) |
| 12. | M45 | (3d) | 32. | O4 | (5d3/2) | 52. | P8 | (6g7/2) |
| 13. | M4 | (3d3/2) | 33. | O5 | (5d5/2) | 53. | P9 | (6g9/2) |
| 14. | M5 | (3d5/2) | 34. | O67 | (5f) | 54. | P1011 | (6h) |
| 15. | N | (4) | 35. | O6 | (5f5/2) | 55. | P10 | (6h9/2) |
| 16. | N1 | (4s1/2) | 36. | O7 | (5f7/2) | 56. | P11 | (6h11/2) |
| 17. | N23 | (4p) | 37. | O89 | (5g) | 57. | Q | (7) |
| 18. | N2 | (4p1/2) | 38. | O8 | (5g7/2) | 58. | Q1 | (7s1/2) |
| 19. | N3 | (4p3/2) | 39. | O9 | (5g9/2) | 59. | Q23 | (7p) |
| 20. | N45 | (4d) | 40. | P | (6) | 60. | Q2 | (7p1/2) |
| | | | | | | 61. | Q3 | (7p3/2) |

The number of next lines depends on the number of CDF-oscillators specified. For each oscillator, there will be a separate line, containing the following $E_0$, $A$, $\Gamma$ coefficients of the CDF [51].

Then, for each shell, there will be the same set of lines with its own parameters.

If there is more than one element in the compound (e.g. GaAs), the same lines from 5 and further must be present for the second element. However, the last orbital (energy level) must be skipped, because it forms the valence band, and the valence band was already described in the first element. Thus, for all the next elements the number of shells must be one less than for the case of an elemental material (first element).

In case one uses BEB cross sections (by setting the EADL option in the input file NUMERICAL_PARAMETERS.txt), the file [*Material*].cdf is unnecessary. And vice versa, if you do not have cdf data and the corresponding file, switch to the EADL option for the cross sections.

### g) Files with electron mean free paths

[*Matter*]_Total_Electron_IMFP.txt contains the total electron mean free path in the material. The first column is the electron energy in [eV], and the second one is the mean free path in [Å].

Files named as [*A*]_[*CS*]_Electron_IMFP_Ip=[*Ip*]eV.txt specify in the same format electron mean free paths for each shell of each kind of atom.

Here [*A*] represents the atomic species (Si, C, Ga, As…);

[*CS*] represents within which formalism the cross-section (and, correspondingly, mean free path) is calculated: CDF (for CDF read from a file), CDFsp (for single-pole approximation to CDF) or BEB (for atomic cross-section);

[*Ip*] is the ionization potential of the shell, the given value of *Ip* must coincide with the ionization potential specified in the cdf-file.

If these files are not present in the folder, at the first XTANT-3 run it will automatically calculate them for the given choice of the cross-section (CDF or BEB), and save. Next time, it will read from the saved files, instead of recalculating them again. This means, if you modify something in the atomic parameters or the cdf, you have to delete the mean free paths files and let the program recalculate the new ones at the next run.

At the first run with CDF, the code also informs you about the corresponding sum-rules for the cdf you provided or the single-pole approximation [47,52].

### h) Files with photon attenuation lengths

The files are constructed in exactly the same way as the files for electrons described above, named similarly: [*Matter*]_Total_photon_IMFP.txt. In the case of the given cdf, the photon attenuation lengths (mean free paths) are calculated from the cdf; in the case of chosen EADL (or BEB) cross sections, the photoabsorption cross sections are extracted from the EADL (part of EPICS2017) database.

### i) K-points grid

File k_grid.dat may contain several lines with 3 values in each specifying the grid points ($k_x$, $k_y$, $k_z$) in the reciprocal space for calculations of the CDF and/or DOS (if the corresponding options are set in the input file; otherwise, this file is ignored). See an example in Figure VI.25.

If no such a file is present, the Monkhorst-Pack[125] sampling of k-space points is used.

```
0.0d0   0.0d0   0.0d0
0.0d0   0.0d0   0.1d0
0.0d0   0.0d0   0.2d0
0.0d0   0.0d0   0.3d0
0.0d0   0.0d0   0.5d0
0.1d0   0.0d0   0.5d0
0.2d0   0.0d0   0.5d0
0.3d0   0.0d0   0.5d0
0.4d0   0.0d0   0.5d0
0.5d0   0.0d0   0.5d0
```

**Figure VI.25. Example of k_grid.dat file**

# VII. Output Files

XTANT-3 produces several output files.

## 1) OUTPUT_Error_log.dat

file must be empty if there were no errors during the execution of the code. In this case, it is automatically deleted after the execution is finished. If it’s not empty and not deleted at the end, have a look inside for the description of a known error that you would have to find later in the code and figure out why it occurred. Known types of errors and their meaning:

- Error #1: file not found
- Error #2: file could not be opened
- Error #3: file could not be read on the line number given
- Error #4: some problem with databases (EADL, EPDL, periodic table file)
- Error #5: inconsistent TB parameterization (only the same type of parameterization is allowed for all kinds of atoms within the compound)
- Error #6: diagonalization subroutine with LAPACK failed (uses MKL library)

- Error #7: some errors in low-energy electrons (probably in temperature or chemical potential calculation)
- Error #8: error in optical coefficients (probably in complex Hamiltonian)
- Error #9: error in conversion between the fluence and the dose

## 2) OUTPUT_Energy.dat

In case you included an additional option size, the code produces this file with the following information:

Column 1 is the nearest neighbor distance in [Å]

Column 2 is the total energy [eV/atom]

Column 3 is the repulsive part of the energy [eV/atom]

Column 4 is the attractive part of the energy [eV/atom]

Column 5 is the van der Waals contribution in [eV/atom]

Column 6 is the ZBL potential in [eV/atom] (unused in real calculations but may be useful for creating repulsive potential or checking short-range potential)

## 3) Directory OUTPUT_[*material*]_hw=[*hw*]_t=[*t*]_F=[*F*]

Such a directory contains all output files with the results of the code execution. Its name itself contains details of the parameters of the run:

[*material*] is the name of the material used (diamond, silicon, etc.). If the option "water" was specified in the INPUT_MATERIAL file, then the name will be augmented with the words "in_water" (e.g. it will be *material* = "diamond_in_water")

[*hw*] is the photon energy of the FEL pulse used [eV]

[*t*] is the duration of the FEL pulse [fs]

[*F*] is the pulse fluence in terms of the absorbed dose in [eV/atom]

For the case of more than one FEL-pulse modeled, these parameters are shown for the *first* pulse, but the directory-name is appended with the following:

OUTPUT_[*material*]_hw=[*hw*]_t=[*t*]_F=[*F*]_[*N*]_pulses

Where *N* shows the number of pulses specified in the input file.

Alternatively, for no pulse calculations (*F*=0), the name will be OUTPUT_[*material*]_Te=[*Te*]_Ta=[*Ta*]_[*coupling*] where

[*Te*] is the initial electron temperature [K]; in case if electron distribution is read from the file (instead of Fermi distribution with a given temperature), then here "-1.0" will be used, to indicate that the distribution may be out of equilibrium and the electronic temperature is undefined.

[*Ta*] is the initial atomic temperature [K]

[*coupling*] will be either "no_coupling" (if no electron-phonon coupling is included), or "with_coupling" if the nonadiabatic coupling is switched on.

If a run with the same parameters already was performed, and the data file with the same name already exists, the new file will be created with a number at the end, e.g.

OUTPUT_[*material*]_hw=[*hw*]_t=[*t*]_F=[*F*]_v1

An output directory will also contain several files, including a copy of the INPUT_MATERIAL.txt (or INPUT.txt) and, if exists, NUMERICAL_PARAMETERS.txt for your records, and the file

!OUTPUT_[*material*]_Parameters.txt with essentially the same information, plus the atomic data that are extracted from either cdf-file, or EADL database, and duration of execution of the program. Also, if you use a communication file (see the next subsection), its results will be saved here for your information.

### 4) SAVE_[*something*].dat

Files with the atomic coordinates SAVE_atoms.dat, with supercell vectors SAVE_supercell.dat, and SAVE_el_distribution.txt containing electronic distribution. These files are updated at each printout-MD-step and can be used to restart the simulation from the current step (including the last step).

### 5) Communication with the program on-the-fly

In the output folder, XTANT-3 creates a text file named Comunication.txt. This file is checked by the program at each saving-time-step. You can send the following messages to the program that it will interpret and act upon:

- time "number" : to change the total duration of the simulation (type 'time' and the new number in [fs], without quotation marks, e.g. time 10000, or Time 2e3)
- SAVEdt "number" : to change how often outputs are saved (type the new number in [fs], e.g. SAVEdt 2.0 – will make the program save output data with the time step of 2 fs)
- MDdt "number" : to change the timestep of MD simulation (type the new number in [fs], e.g. MDdt 0.01).
- OMP "number" : to change the number of OpenMP threads in the parallel calculations (integer). Setting here a zero or negative number will set the number of threads equal to the maximum number of threads on your machine.
- Thermostat_Ta "number" : to change the temperature of the atomic Berendsen thermostat. The number sets the new atomic temperature in [K]. A negative number switches off the atomic thermostat.
- Thermostat_dt_a "number" : to change the characteristic time of the atomic Berendsen thermostat. The number sets the new characteristic time of the thermostat in [fs]. A negative number switches off the atomic thermostat.
- Thermostat_Te "number" : to change the temperature of the electronic Berendsen thermostat. The number sets the new atomic temperature in [K]. The negative number switches off the electronic thermostat.
- Thermostat_dt_e "number" : to change the characteristic time of the electronic Berendsen thermostat. The number sets the new characteristic time of the thermostat in [fs]. The negative number switches off the electronic thermostat.
- Verbose : switches the *verbose* option on (see Section IV.6).

- verbose_off : switches the *verbose* option off.

  For example, the following lines in the file shown will reset (1) the duration of simulation to 1000 fs, (2) the printout timestep to 10 fs, (3) the atomic thermostat to the temperature of 300 K, (4) the characteristic time of the atomic thermostat to 150 fs, and (5) switch off the electronic thermostat, see Figure VII.1.

```
time 1000
SAVEdt   10.0
thermostat_Ta    300.0e0
Thermostat_dt_a 150.0
Thermostat_Te    -1
```

**Figure VII.1 Example of the communication file.**

At the end of the simulation, this file is deleted.

### 6) Plotting: OUTPUT_Gnuplot_all.sh

Execute this file to create all the plots of all results of calculations. You can do that even if XTANT-3 is still running, then it will give you transient results. At the end of the simulation run, this command will also be executed automatically.

This is a gnuplot shell script that is created by XTANT-3 to execute all other Gnuplot shell scripts in the folder that are plotting all the essential quantities:

OUTPUT_bands_Gnuplot.sh – plots the bottom of the valence band (VB), top of the VB, bottom of the conduction band (CB), and top of the CB, and chemical potential.

OUTPUT_CB_electron_Gnuplot.sh – plots the density of conduction band electrons.

OUTPUT_coupling_parameter_Gnuplot.sh – plots the electron-phonon coupling parameter.

OUTPUT_deep_shell_holes_Gnuplot.sh – plots the density of deep shell holes in each shell of each atom of the compound.

OUTPUT_electron_Ce.sh – plots the heat capacity of electrons.

OUTPUT_electron_chempotentials.sh – plots band-resolved electron chemical potentials (only if separate thermalization times for VB and CB are used).

OUTPUT_electron_entropy.sh – plots the entropy of electrons (may contain entropies for VB and CV, if separate thermalization times for VB and CB are used).

OUTPUT_electron_distribution_Gnuplot.sh – plots the electron distribution function as an animated gif. Note that the points are plotted in the position of the current energy levels (molecular orbitals), so the shifts of the points reflect the shifts of the orbitals. In case nonequilibrium distributions are used, such as BO or relaxation-time approximation, also the equivalent Fermi distribution will be plotted for comparison. If separate thermalization times for VB and CB are used, there will be also equivalent distributions for the separately thermalized valence and conduction band.

OUTPUT_electron_distribution_on_grid_Gnuplot.sh – plots the electron distribution function on the user-defined grid (*not* the energy levels) as an animated gif. The full distribution is plotted (low- and high-energy electrons), multiplied by the DOS; free-electron DOS is assumed for high-density electrons. Thus, this plot is energy-resolved electron density.

OUTPUT_electrons_and_holes_Gnuplot.sh – plots the high-energy electrons and core holes densities.

OUTPUT_electron_temperatures.sh – plots band-resolved electron kinetic temperatures (only if separate thermalization times for VB and CB are used).

OUTPUT_energies_Gnuplot.sh – plots the total, potential, and atomic energies.

OUTPUT_energy_levels_Gnuplot.sh – plots the electron energy levels (eigenvalues of the TB Hamiltonian). Takes a few minutes to plot.

OUTPUT_mean_displacement_Gnuplot.sh – plots the (mean atomic displacements)^*N* with respect to the initial positions.

OUTPUT_mu_and_Egap.sh – plots the electron chemical potential and band gap.

OUTPUT_Mulliken_charges_Gnuplot.sh – plots charges of different types of atoms.

OUTPUT_optical_coefficients.sh – plots optical R, T, A for specified probe pulse wavelength.

OUTPUT_optical_n_and_k.sh – plots corresponding real and imaginary parts of the refractive index.

OUTPUT_pressure_Gnuplot.sh – plots total pressure in the atomic system in the simulation box.

OUTPUT_stress_tensor_Gnuplot.sh – plots components of the atomic pressure tensor.

OUTPUT_temperatures_Gnuplot.sh – plots the electron and atomic temperatures.

OUTPUT_volume_Gnuplot.sh – plots the volume of the supercell.

In case you set a probe-pulse to be included, additional gnuplot files of the convolved data will be created (see below), that will be named exactly the same way with the word 'CONVOLVED' added at the end, e.g. convolved electron heat capacity would be in a file named

OUTPUT_electron_Ce_CONVOLVED.sh

Note that in the case of the Windows operating system, instead of shell scripts the program will create cmd batch files (with the same name, just a different extension: .cmd instead of .sh). They will need a Windows version of Gnuplot installed[37], and proper paths written in the environment variables[38].

## 7) Output data files

OUTPUT_atomic_coordinates.xyz – contains the atomic positions at each time-step in [A] (saving time-step specified in the input file, not the numerical time-step used in the MD) in the extended XYZ format. The comment line is used to save the supercell vectors, which can be read by many MD visualization programs, such as OVITO[39]. Some optional atomic properties may also be listed in this file to be read by OVITO[40].

[37] http://www.gnuplot.info/
[38] This, and many other useful things, can be done with help of Cygwin: https://www.cygwin.com/
[39] https://www.ovito.org/windows-downloads/
[40] https://www.ovito.org/manual/reference/file_formats/input/xyz.html

OUTPUT_atomic_coordinates.cif – contains the atomic positions at each time-step in [A] (saving time-step specified in the input file, not the numerical time-step used in the MD) in the CIF format, which can be used to construct powder diffraction patterns by standard visualization software such as Mercury[41].

OUTPUT_coordinates_and_velosities.dat – contains the atomic coordinates and velocities for all atoms at each timestep. The first three values and the coordinates in [A], last three are the velocities in each line in [A/fs]. A line describes one atom in the supercell. After all atoms' data for one timestep, there are two empty lines. After that, the next timestep is starting. Use it for a quick look with Gnuplot, e.g.:

sp "OUTPUT_coordinates_and_velosities.dat" i 201 u 1:2:3 pt 6 ps 3

for step number 201, coordinates (columns 1:2:3).

Also, use it for calculations of atomic velocity autocorrelators and phonon spectra (see below).

OUTPUT_Ritchie_CDF_[*material*].cdf – the Ritchie-Howie CDF coefficients used in this material. The file has the format of cdf-file (described above in Section f). The file is only created, if requested by the user with the option 'print_CDF' in the input file. The file contains the following columns:

1) Electronic temperature [K]
2) Electronic heat conductivity [W/(K*m)]
3) Electron-phonon contribution to electronic heat conductivity [W/(K*m)]
4) Electron-electron contribution to electronic heat conductivity [W/(K*m)]
5) Electronic chemical potential [eV]
6) Electronic heat capacity [J/(m^3*K)]

These data are printed out at each timestep; the timestep blocks are separated by two empty lines.

OUTPUT_coupling.dat – electron-ion coupling parameter. Contains the following lines:

1) Time [fs]
2) Total coupling parameter [$W/m^3K$]
3) and further: Partial coupling parameter for each type of pair atoms [*A*]-[*B*] in the compound. For example, for elemental Al targets, there will be one column Al-Al. For compound AlCu, there will be 4 columns: Al-Al, Al-Cu, Cu-Al, and Cu-Cu. Etc.
4) The next columns will be partial coupling for each type of orbital (defined by the basis set) for each kind of atom in the compound. For example, for the $sp^3d^5$ basis set, there will be couplings for the pairs of levels: s-s, s-p, s-d, p-s, p-p, p-d, d-s, d-p, d-d (for each kind of element [*A*]-[*B*] in the compound).

[41] https://www.ccdc.cam.ac.uk/solutions/software/mercury/

OUTPUT_deep_shell_holes.dat – contains timestep [fs], the number of holes in each shell of each kind of atom in the compound in the next columns (normalized to the number of atoms).

OUTPUT_DOS.dat – contains blocks of data separated by two empty lines. Each block contains a few columns:

1) energy in [eV]
2) total DOS in [states/eV]
3) partial DOS (PDOS) for the first atomic shell of the first element in the compound [states/eV]
4) PDOS for the next shell of the first element, etc.

PDOS corresponding to atomic orbital contributions into DOS will be printed out only if the user set the PDOS parameter to 1 (see input file NUMERICAL_PARAMETERS above). The number of columns will depend on the number of elements in the compound and the basis set used. E.g., for one element with a $sp^3d^5$ basis set, there will be three columns with PDOS, corresponding to s, p, and d PDOS. For $N_{elem}$ elements in the compound and $sp^3d^5$ basis set, there will be $3xN_{elem}$ columns; for the $sp^3$ basis set, there will be 2 columns per element: with s and p PDOS, etc.

The file is created only if printing out DOS is set by the user.

OUTPUT_dielectric_function.dat – contains blocks of data separated by two empty lines. Each block contains 16 columns:

1) energy in [eV]
2) the real part of CDF
3) the imaginary part of CDF
4) loss function
5) reflectivity
6) transmission
7) absorption
8) optical n (real part of the refraction coefficient)
9) optical k (imaginary part of the refraction coefficient)
10) dc-conductivity
11) The real part of the (x,x) component of the CDF
12) The imaginary part of the (x,x) component of the CDF
13) The real part of the (y,y) component of the CDF
14) The imaginary part of the (y,y) component of the CDF
15) The real part of the (z,z) component of the CDF
16) The imaginary part of the (z,z) component of the CDF

File created only if printing out the spectrum of optical coefficients is set by the user. From this file, parameters for a chosen photon energy may be extracted in post-processing (see below).

OUTPUT_electron_Ce.dat – electronic heat capacity. Contains the following lines:

1) Time [fs]
2) Total electronic heat capacity [J/m$^3$K]
3) and further: partial electronic heat capacities for each type of orbitals (defined by the basis set) for each kind of atom in the compound.

For example, for the elemental Al target, there will be 3 columns with partial $C_e$: Al_s, Al_p, and Al_d, corresponding to 3s, 3p, and 3d shells.

OUTPUT_electron_chempotentials.dat – this file is created only if separate valence and conduction band thermalizations are used, and contains the following:

1) Time [fs]
2) Global electronic chemical potential [eV]
3) Valence-band electronic chemical potential [eV]
4) Conduction-band electronic chemical potential [eV]

OUTPUT_electron_entropy.dat – electronic heat capacity. Contains the following 3 columns: 1) Time [fs];

2) transient electronic entropy [K/eV];
3) equivalent equilibrium electronic entropy [K/eV] (corresponding to Fermi-Dirac distribution, i.e. maximal possible entropy for the given particle and energy content).

If separate thermalization times for the valence and conduction bands are used, this file will contain additional data for the partial entropies:

4) transient electronic entropy in the valence band [K/eV]
5) equivalent equilibrium electronic entropy in the valence band [K/eV]
6) transient electronic entropy in the conduction band [K/eV]
7) equivalent equilibrium electronic entropy in the conduction band [K/eV]

OUTPUT_electron_distribution.dat – contains electron distribution function on the current energy levels, for each timestep separated by the double empty line. The first column is the energy level in [eV], and the second column is the distribution function. In case nonequilibrium distributions are used, such as BO or relaxation-time approximation, there is the third column, containing equivalent Fermi distribution (Fermi function for equivalent or kinetic electron temperature and chemical potential). If separate thermalization times for VB and CB are used, there will be also columns containing equivalent distributions for the separately thermalized valence and conduction band.

OUTPUT_electron_distribution_on_grid.dat – contains full electron distribution function (energy-resolved density of electrons of both, low- and high-energy electrons, from Boltzmann and MC fractions) on the user-defined grid (*not* current energy levels), for each timestep separated by the double empty line. The first column is the energy grid in [eV], the second column is the electron density in [1/(V*eV)] (where V is the volume of the supercell), the third column is the electron distribution function (the second column divided by the density of states; for MC electrons, the free-electron DOS is assumed with the mass equal to the free-electron mass, whereas for the low-energy electrons, the DOS is counted as the number of energy levels (at the gamma-point) within the given grid interval).

OUTPUT_electron_heat_conductivity.dat – contains blocks of data for each timestep, separated by two empty lines; each block is made of 4 columns:

1) Electronic temperature [K]
2) Electronic heat conductivity [W/k/m]
3) Electronic chemical potential [eV]
4) Electronic heat capacity [J/K/$m^3$]

OUTPUT_electron_hole_numbers.dat – contains the following data:

1) Time [fs];
2) Number of valence band electrons;
3) Number of conduction band electrons;
4) Number of high-energy electrons;
5) Total number of core holes (in all shells summed up);
6) Error in the particle conservation;
7) Number of photons as sampled.

Data are normalized per the number of atoms.

OUTPUT_electron_properties.dat – contains the following data:

17) Time [fs]
18) Number of electrons [%, per atom]
19) Chemical potential [eV]
20) Band gap [eV]
21) Electrons heat capacity [J/$m^3$K]
22) Electron-phonon coupling parameter [W/$m^3$K]
23) Bottom of the VB [eV]
24) Top of the VB [eV]
25) Bottom of the CB [eV]

26) Top of the CB [eV]

27) Mulliken charges for all types of atoms in the modeled compound [electron charge]

OUTPUT_electron_temperatures.dat – this file is created only if separate valence and conduction band thermalizations are used, and contains the following:

1) Time [fs]
2) Global electronic kinetic (equivalent) temperature [K]
3) Valence-band electronic kinetic temperature [K]
4) Conduction-band electronic kinetic temperature [K]

OUTPUT_energies.dat – contains the following data:

1) Time [fs]
2) The energy of electrons [eV/atom]
3) The energy of all core holes [eV/atom]
4) The potential energy of atoms [eV/atom]
5) The kinetic energy of atoms [eV/atom]
6) The total energy of atoms [eV/atom]
7) The total energy of atoms and electrons [eV/atom]
8) The total energy in the system (atoms, electrons, holes) [eV/atom] – should be always conserved, except during an FEL pulse
9) Van der Waals energy (if included in the simulation) [eV/atom]
10) Additional short-range repulsion energy (if included) [eV/atom]

OUTPUT_energy_levels.dat – contains all the eigenstates of the Hamiltonian, at each timestep, separated by two empty lines, in [eV].

OUTPUT_optical_coefficients.dat – Optical coefficients for the given probe photon energy, printed in the same format as 16 columns in the file OUTPUT_dielectric_function.dat, except the first column in time in [fs]. The file is created only if the probe pulse is set by the user.

OUTPUT_pressure_and_stress.dat – Contains:

1) Time [fs]
2) Pressure [GPa]
3) 9 columns with the components of the pressure tensor Pressure(a,b), with a=x,y,z and b=x,y,z, all in [GPa]

OUTPUT_supercell.dat – contains the data for the supercell: time, volume [A$^3$], 9 super-cell vectors [A], and their 9 velocities [A/fs].

OUTPUT_temperatures.dat – contains the following data:

1) Time [fs]
2) Electron temperature [K]

   Note that in case of BO or nonequilibrium simulation, the kinetic (or equivalent; effective; nonequilibrium) electron temperature is printed here: the temperature and chemical potential are defined for the Fermi distribution function that contains the same number of electrons and total energy as in the real simulated system.

3) Kinetic atomic temperature [K]: the first column is the average, then for each element of the compound
4) (Mean atomic displacement)$^N$ [A$^N$]: the first column is the average, then for each element of the compound

At each time instant of the simulation when output files are printed out, XTANT-3 also saves the following backup files: SAVE_atoms.dat, SAVE_supercell.dat, and SAVE_el_distribution.dat. They are consistent with the format that can be used to set initial data, as described above (section Files SAVE_supercell.dat, SAVE_atoms.dat and SAVE_el_distribution.dat). They contain information on the atomic coordinates and velocities, supercell vectors, and electronic distribution function.

### 8) …_CONVOLVED.dat output data files

If in the input file, you set a finite positive duration of the probe pulse, all the data from the output files mentioned above will be additionally convolved with a Gaussian function of a given FWHM. The resulting data will be saved in the new output data files under the same names with the tag 'CONVOLVED' added to them, e.g. temperatures after the convolution will be in the file:

OUTPUT_temperatures_CONVOLVED.dat

These files will be used by gnuplot to prepare convolved figures.

## VIII. Post-processing

XTANT-3 package contains a few programs for post-processing of the output files if required. They are stored in the directory !XTANT_ANALYSIS_SUBROUTINES. To compile them, enter this directory and execute the Make.bat file inside (for Linux-based systems, follow the compilation examples given at the top of each file). It will compile all the post-processing subroutines listed below.

The following post-processing is possible:

### 1) Extracting pair correlation function

If the user did not set to printout the pair correlation function (PCF) in the input files, it can be extracted in the post-processing using the XTANT_atomic_data_analysis.exe, which must be placed into the folder with the output data.

This program requires the following output data to be present:

'OUTPUT_coordinates_and_velosities.dat'

'OUTPUT_supercell.dat'

'OUTPUT_energies.dat'

'OUTPUT_temperatures.dat'

And can be run as follows: XTANT_atomic_data_analysis.exe in the command line. It will construct and print out the pair correlation function at each time instant. It creates the output file OUTPUT_PCF.dat, where the PCF is saved in the same format as OUTPUT_pair_correlation_function.dat described above.

Note, however, that PCF can easily be obtained by standard molecular dynamics visualization software, such as VMD[42], OVITO[43], etc.

## 2) Calculation of velocity autocorrelation and phonon spectra

A program for calculation of atomic velocity autocorrelation functions and phonon spectra XTANT_autocorrelators.exe must be placed into the folder with the output data. To execute, simply run XTANT_autocorrelators.exe in the command line. The program will ask the user to input two parameters: alpha and time step. Alpha is the exponential factor suppressing correlation at long times *exp*(-alpha*t). The time step *tim_step* sets how often to print out autocorrelators and phonon spectra: it will divide the data into *tim_step* steps, and calculate data on them. E.g. if you had a simulation run from 0 to 1000 fs, and set *tim_step*=10, it will print 10 files, at every 200 fs.

The program will create a set of files

OUT_VAF_[*time*].dat

OUT_vibrational_spectrum_[*time*].dat

Where [*time*] is a timestamp at which the velocity autocorrelation function (VAF) and phonon spectrum are calculated. The program uses Fourier transform to get the spectrum from VAF [1].

The files OUT_VAF_[*time*].dat contain two columns:

1) Time [fs]
2) VAF [arb. units]

The files OUT_vibrational_spectrum_[*time*].dat contain two columns:

1) Phonon frequency [THz]
2) Phonon spectrum [arb. units]

## 3) Calculation of electron-ion coupling parameter $g(T_e)$, $C_e(T_e)$, $\mu(T_e)$, $P_e(T_e)$

To calculate the electron-ion coupling parameter as a function of electron temperature $g(T_e)$, electronic heat capacity $C_e(T_e)$, chemical potential $\mu(T_e)$, and electronic pressure $P_e(T_e)$, the following procedure can be used [2]:

It requires several output directories to be present with slightly different simulation parameters.

---

[42] https://www.ks.uiuc.edu/Research/vmd/
[43] https://www.ovito.org/

- They can be automatically set in the INPUT_DATA.txt (or INPUT.txt) input file using the optional command "*Coupling*" (see the end of Section VI.1).

Alternatively, they can be set manually, following the procedure described below:

  - Set a simulation run with the fluence of ~3-5 eV/atom, duration of ~10-30 fs, and low photon energy of ~10 eV, the position of the maximum at 0 fs, and the starting time ~ -300 – -400 fs.
  - Use a small timestep (both, for MD and for printing out) during the pulse. For that, it is convenient to use a file input with the specified time grid (see description of Line 10 in Section VI.2), in which a relatively large time step may be used for the equilibration phase before the pulse arrival (e.g., 1 fs up to -20 fs, and 0.1 fs after -20 fs).
  - Create multiple simulations by either specifying in the INPUT_DATA.txt file the additional option "Coupling" (with a number of simulations recommended 5-10 or more; see for details at the end of Section VI.1), or by manually creating a few additional input files with incrementally increasing numbers to run a few simulations in a sequence (see above how to do that, Section VI.3). In each of these files, vary slightly the parameters of the first pulse, e.g., choose different fluences between 3 and 5 eV/atom, chose different pulse durations between 10-30 fs, initial electronic temperature between 100 and 300 K, and slightly different arrival times of the second pulse between -300 fs to -500 fs. This will randomize the parameters and exclude artificial correlation effects [2]. Recommended number of thusly-created simulation runs: 10 or more.

- Run XTANT-3 to create a few (e.g., 10 or more) output folders with slightly different parameters as said above. Place all output folders into a separate directory, and place the file XTANT_coupling_parameter.exe in the same directory.
- Execute it as follows: XTANT_coupling_parameter.exe alpha max_Te

The two optional parameters alpha and max_Te , and if used, must be set in this fixed order.

alpha : an optional argument meaning the time [fs] from which to start averaging the coupling parameter (used to exclude early times where atoms are not equilibrated yet). If this parameter is absent, the default value is used, which is: $t_0$-2*FWHM (where $t_0$ is the pulse center, and FWHM is its widths, read from the file INPUT_DATA.txt (or INPUT.txt)).

max_Te : an optional parameter, specifying the maximal electronic temperature for grid to span the dependence on the electronic temperature (if not specified, the default value of 25000 K is used).

For example, set alpha=-100 fs and $T_e$=50000 K, i.e. run XTANT_coupling_parameter.exe -100 50000, which will exclude earlier times during which the system was still equilibrating, and create output files with the electronic temperature grid up to 50000 K (note that it is meaningless to set the electronic temperature higher than those achieved in the calculations, defined by the absorbed dose set in the input file).

This program will scan through all the output folders (note that no other folders or files in this directory should start with the word 'OUTPUT', as the program will use them and crush), and use files OUTPUT_electron_properties.dat, OUTPUT_temperatures.dat, OUTPUT_coupling.dat (if exists) and OUTPUT_pressure_and_stress.dat to extract g, $C_e$, μ and Gruneisen parameter as functions of time, then sort them according to the electronic temperatures ($T_e$) at those time, interpolate on a grid of $T_e$, the average overall 10 (or more) simulation runs, and printout averaged values into the following files:

OUT_average_coupling.dat

Which contains 3 columns:

1) Electron temperature [K]
2) Total coupling parameter [W/(m$^3$K)]
3) The standard deviation of the coupling parameter (error bars)

OUT_average_parameters.dat

Which contains 4 columns:

1) Electron temperature [K]
2) Chemical potential [eV]
3) Electron heat capacity [J/(m$^3$K)]
4) The standard deviation of the electron heat capacity (error bars)

OUT_average_partial_couplings.dat

Which contains several columns corresponding to the number of partial couplings for each pair of elements and shells (according to the file OUTPUT_coupling.dat in the same format).

OUT_average_partial_Ce.dat

Which contains several columns corresponding to the number of partial electronic heat capacities for each shell of each element (according to the file OUTPUT_electron_Ce.dat in the same format).

OUT_average_pressure.dat

Which contains 4 columns:

1) Electronic temperature Te in [K]
2) Electron energy in [eV/atom]
3) Pressure in [GPa]
4) Electronic Gruneisen parameter, defined as dP/dE in [Pa/(J/atom)]

Note that an alternative definition of the electronic Gruneisen parameter is P/E (instead of derivative) [139], which can be calculated from columns 2 and 3 in this file. In this case, it is important to subtract the room temperature value from the pressure (since it is rarely exactly zero in the simulation). This definition is useful since the definition based on the derivative often produces too noisy results.

For convenience, it also creates the file collecting the partial electronic heat capacity and electron-ion coupling named

OUT_XTANT3_Cd_partial_Ce_G.dat

Which contains a few columns (depending on the TB basis set, and number of orbitals):

1) Electronic temperature Te in [K]
2) Total electron heat capacity in [J/m$^3$K]

The number of next columns depends on the number of orbitals in TB: for each s, p, and d orbital used, there will be a column with partial electron heat capacity.

Then comes the column with the total electron-ion coupling parameter, rescaled back to normal according to the parameter set in the file NUMERICAL_PARAMETERS.txt. After that, partial couplings for each pair of orbitals are present, where paired orbitals are added together (and rescaled by the same factor as the total one). For example, for the $sp^3d^5$ basis, there will be 6 columns, corresponding to the partial couplings of: s-s, s-p, s-d, p-p, p-d, d-d orbitals. For the calculation of s-p orbital coupling, the two columns from the file OUT_average_partial_couplings.dat (above) are summed: s-p and p-s.

Additionally, two gnuplot scripts are created that use data from the last file:

OUT_gnuplot_Ce.cmd and OUT_gnuplot_G.cmd, which are executed to create the figures

OUT_XTANT3_Cd_partial_Ce.gif and OUT_XTANT3_Cd_partial_G.gif with the calculated partial electron heat capacity and electron-ion coupling.

## 4) Extracting optical parameters for given wavelength from the spectrum

If you run XTANT-3 simulation with printing out the optical spectrum, you can extract optical parameters as a function of time for any photon energy (wavelength) within the spectrum interval, for *s* and *p* polarizations, and given angle of probe incidence and substrate. It uses the RPA approximation for the calculations of the CDF, see Section VI.13.

To run, place the program XTANT_dielectric_function_analysis.exe into the output folder with the results which must contain the file OUTPUT_dielectric_function.dat, and simply call it as: XTANT_dielectric_function_analysis.exe *hw*, where *hw* is the probe photon energy in [eV].

It also requires the following additional input file to be present: OPTICAL_PARAMETERS.dat (place it manually into the same directory with the output data).

```
50.0d0       ! thickness [nm]
90.0d0       ! with respect to surface [deg]
1.0d0    0.0d0          ! n and k of the material above the target
3.6941d0    0.0065435d0 ! n and k of the material below the target
```

**Figure VIII.1 Example of the input file for optical parameters extraction.**

The lines in this file must be as follows:

1) The thickness of the target in [nm]
2) Probe incidence angle with respect to the surface [degree]
3) Optical *n* and *k* of the material above the target (typically, air)
4) Optical *n* and *k* of the substrate (substrate material below the target, may also be air)

The program will create 2 output files:

1) OUTPUT_dielectric_function_[*hw*].dat with the same format as the file OUTPUT_optical_coefficients.dat described above (just by interpolating the data from the file with the dielectric function printed out). The columns are marked in the first line of the file, as follows: time hw Re_eps Im_eps LF R T A n k. The optical coefficients are calculated for s-polarization only. *Note that it might not work well, and the file may contain just zeros!*

2) OUTPUT_dielectric_function_[*hw*]_RTA.dat with recalculated optical parameters taking into account properties of materials above and below the target, following the formalism described in [112]. The columns are marked in the first line of the file, as follows:

1. time [fs]
2. R_s – the first ray reflection for s-polarization
3. T_s – the first ray transmission for s-polarization
4. A_s – the first ray absorption for s-polarization
5. R_p – the first ray reflection for p-polarization
6. T_p – the first ray transmission for p-polarization
7. A_p – the first ray absorption for p-polarization
8. R_s_all – coherently summed all rays' reflection for s-polarization
9. T_s_all – coherently summed all rays transmission for s-polarization
10. A_s_all – coherently summed all rays absorption for s-polarization
11. R_p_all – coherently summed all rays' reflection for p-polarization
12. T_p_all – coherently summed all rays transmission for p-polarization
13. A_p_all – coherently summed all rays absorption for p-polarization

## 5) Calculation of mass spectrum of ablation fragments

If you run a simulation with open boundaries (thin layer, nanoparticle, etc.), you can construct a mass spectrum of ablated fragments. For that, place the program XTANT_fragmentation.exe and the database INPUT_atomic_data.dat (present in the directory !XTANT_ANALYSIS_SUBROUTINES) into the output folder with the results which must contain the file OUTPUT_atomic_coordinates.xyz, and call it as: XTANT_fragmentation.exe *r dt*.

Here *r* set the cut-off radius in [Å] (above which atoms are considered to be separated and belong to a different fragment), and the printout time step *dt* in [fs] (which sets how often the mass spectrum should be printed out; it does not have to coincide with the timestep used in the simulation, but cannot be smaller than that).

Execution of this utility creates the output files:

OUT_fragments_spectrum.dat containing the following columns (first line set the time instants for each column, starting from #2):

Column 1: mass of the fragments in [a.m.u.]

Column 2 and others: number of fragments with the mass, set in column 1, at the time instant, set in the first line.

And OUT_m_over_z_spectrum.dat (only if the file OUTPUT_atomic_coordinates.xyz contains atomic Mulliken charges) containing the following columns (first line set the time instants for each column, starting from #2):

Column 1: mass over charge ratio of the fragments in [a.m.u./electron charge]

Column 2 and others: weighted number of fragments with the m/z, set in column 1, at the time instant, set in the first line (see more information on weights in Section VI.12).

It also creates Gnuplot scripts OUT_fragments.cmd (or OUT_fragments.sh) and OUT_m_over_z.cmd (or OUT_m_over_z.cmd), which plot the color-map plot of the mass spectrum vs. time in the file OUT_fragments.png using the data from the created file OUT_fragments_spectrum.dat, and mass spectrum in m/z vs. time in the file OUT_m_over_z.png using the data from the created file OUT_m_over_z_spectrum.dat.

### 6) Calculation of electron entropy

Although versions of XTANT-3 after 10.02.2023 produce output files with electronic entropy during simulation runs, legacy results may be analyzed with this post-processing utility. If the electronic distribution is printed out (file OUTPUT_electron_distribution.dat), electronic entropy can be calculated. For that, place the compiled post-processing file XTANT_entropy.exe in the folder with the output data, and execute this program. It does not require any additional input. It will create an output file with the electronic entropy OUT_entropy.dat containing three columns:

Column 1: time [fs]

Column 2: electron entropy [K/eV] for the transient distribution function read from the same file

Column 3: equilibrium (equivalent) electron entropy [K/eV] for the Fermi-Dirac distribution function (with the same particle and energy content as the transient one) read from the same file.

Since equilibrium distribution maximizes entropy, a comparison between these two functions demonstrates how far from equilibrium the transient electron distribution is.

The utility also creates and executes gnuplot script OUT_entropy.cmd, which creates a plot OUT_entropy.png with the two functions.

### 7) Analysis of electron distribution

If the distribution function on electronic energy levels (OUTPUT_electron_distribution.dat) or distribution on the grid was printed out (containing both, low-energy and high-energy electrons; file OUTPUT_electron_distribution_on_grid.dat), further analysis of those distributions is possible:

- Both distributions can be averaged over time with a given temporal profile (mimicking, e.g., photo- or Auger-emission from the chosen deep-shell holes)
- The distribution on the grid can be additionally smoothened via convolution with a Gaussian function of a given width in energy space (mimicking the spectral function of a measuring detector)

If one of these files is not present, the analysis will automatically run only for the present file.

The file XTANT_el_distribution_analysis.f90 should be compiled into the post-processing subroutine XTANT_el_distribution_analysis.exe. This subroutine should be placed in the folder with the output data, and executed. To execute, it requires the input file called EL_DISTR.txt to be placed in the same directory. This file must contain three lines with the following parameters:

Line 1: Energy width in [eV]. Set a negative number to exclude convolution in the energy space, if not needed.

Line 1: File name, providing the data on the widths in time to average the distribution with; and the index of the column to use in this file. Set a non-existing file name to exclude averaging in time, if not needed.

Line 1: Cut off for the distribution, below which the distribution is not printed (to exclude too small value, making it easier to print in log-scale, and making the output files smaller)

The results are printed in the 3 files for the distribution on the energy levels (and 3 corresponding Gnuplot scripts, which create 2 png-figure and one animated gif), and 2 output files for distribution on the grid (and 4 Gnuplot scripts to create plots from these files):

a) OUT_fe_average.dat – containing the averaged distribution on the energy levels. The first column is the averaged energy levels in [eV], second is the distribution function. Its corresponding Gnuplot script is OUT_fe_average.cmd (or .sh in Linux) and its plot is in OUT_fe_average.png.

b) OUT_fe_gridded.dat – containing the distribution put on the grid with the step defined by the user here. The first column is the energy grid in [eV], and the second column is the distribution function on this grid. Its corresponding Gnuplot script is OUT_fe_gridded.cmd (or .sh in Linux), and the plot is OUT_fe_gridded.gif.

c) OUT_fe_gridded_average.dat – containing the averaged distribution on the defined here energy grid. The first column is the grid in [eV], second is the averaged distribution function. Its corresponding Gnuplot script is OUT_fe_gridded_average.cmd (or .sh in Linux) and its plot is in OUT_fe_gridded_average.png.

d) OUT_el_distr_vs_time.dat – containing the processed electronic spectra (second column) and distribution function (third column), both convolved in energy space, if requested. In the same format as in the original file OUTPUT_electron_distribution_on_grid.dat. If no convolutions are required (line 1 in EL_DISTR.txt contains a negative number), this file is not created.

Their corresponding gnuplot scripts are: OUT_el_distr_vs_time.cmd (or .sh in Linux), executed automatically to create an animated gif plot of the evolution of the convolved electronic spectrum from the file. The second file is OUT_el_distr_vs_time_norm.cmd executed automatically to create an animated gif plot of the evolution of the convolved distribution function from the file

e) OUT_el_distr_average.dat – containing the electronic spectrum (second column) and the distribution function (third column) averaged over time with the provided weights. If no weights were provided in a file, all weights are assumed to be =1, and the function is just averaged over time.

Their corresponding gnuplot scripts are: OUT_el_distr_average.cmd (or .sh in Linux), which creates a png-plot of the time-averaged spectrum from the file; and OUT_el_distr_average_norm.cmd (or .sh in Linux), which creates a png-plot of the time-averaged distribution from the file.

*Note #1*: files with “fe” (constructed from the distribution on the electron levels) are best for the analysis of low-energy electrons, while high-energy electrons are missing in them. Files with “distr” (constructed from the distribution on the grid) are best for analysis of high-energy electrons, whereas low-energy electrons may need more detailed analysis than the one used in XTANT-3 calculations – the parameters of the grid may need to be changed, etc.

For example, if we want to construct the spectrum of photons emitted by the decay of L-shell holes in simulated aluminum, and measured with the detector with the spectral width of 1.5 eV, we set the following in the file:

```
1.5  0.5       ! [eV] Energy convolution parameter; [eV] energy grid for fe
OUTPUT_deep_shell_holes.dat    5
1.0e-8         ! distribution cutoff
```

We chose this file with the core holes, and set column #5 in it, containing the data on the number of L3-shell holes. This way, the resulting output is the electronic distribution averaged with the weights, corresponding to the transient number of L-shell holes, mimicking the spectrum of photons emitted in radiative decays of these holes. The resulting distribution is convolved with a Gaussian function with a width of 1.5 eV, mimicking the detector resolution. An analogous procedure was used, e.g., in Ref.[29].

Similarly, a spectrum of emitted electrons by a probe pulse can be constructed, by setting up a file with the profile of the probe pulse (e.g., gaussian, centered at a given delay time).

## IX. References

[1] N. Medvedev, V. Tkachenko, V. Lipp, Z. Li, B. Ziaja, Various damage mechanisms in carbon and silicon materials under femtosecond x-ray irradiation, 4open. 1 (2018) 3. https://doi.org/10.1051/fopen/2018003.

[2] N. Medvedev, I. Milov, Electron-phonon coupling in metals at high electronic temperatures, Phys. Rev. B. 102 (2020) 064302. https://doi.org/10.1103/PhysRevB.102.064302.

[3] V. Lipp, V. Tkachenko, M. Stransky, B. Aradi, T. Frauenheim, B. Ziaja, Density functional tight binding approach utilized to study X-ray-induced transitions in solid materials, Sci. Rep. 12 (2022) 1–10. https://doi.org/10.1038/s41598-022-04775-1.

[4] N. Medvedev, H.O. Jeschke, B. Ziaja, Nonthermal phase transitions in semiconductors induced by a femtosecond extreme ultraviolet laser pulse, New J. Phys. 15 (2013) 15016. https://doi.org/10.1088/1367-2630/15/1/015016.

[5] T. Apostolova, E. Artacho, F. Cleri, M. Cotelo, M.L. Crespillo, F. Da Pieve, V. Dimitriou, F. Djurabekova, D.M. Duffy, G. García, M. García-Lechuga, B. Gu, T. Jarrin, E. Kaselouris, J. Kohanoff, G. Koundourakis, N. Koval, V. Lipp, L. Martin-Samos, N. Medvedev, A. Molina-Sánchez, D. Muñoz-Santiburcio, S.T. Murphy, K. Nordlund, E. Oliva, J. Olivares, N.A. Papadogiannis, A. Redondo-Cubero, A. Rivera de Mena, A.E. Sand, D. Sangalli, J. Siegel, A. V. Solov'yov, I.A. Solov'yov, J. Teunissen, E. Vázquez, A. V. Verkhovtsev, S. Viñals, M.D. Ynsa, Tools for investigating electronic excitation: experiment and multi-scale modelling, Universidad Politécnica de Madrid. Instituto de Fusión Nuclear Guillermo Velarde, Madrid, 2021. https://doi.org/10.20868/UPM.book.69109.

[6] B. Rethfeld, D.S. Ivanov, M.E. Garcia, S.I. Anisimov, Modelling ultrafast laser ablation, J. Phys. D. Appl. Phys. 50 (2017) 193001. https://doi.org/10.1088/1361-6463/50/19/193001.

[7] M. V Shugaev, C. Wu, O. Armbruster, A. Naghilou, N. Brouwer, D.S. Ivanov, T.J.Y. Derrien, N.M. Bulgakova, W. Kautek, B. Rethfeld, L. V Zhigilei, Fundamentals of ultrafast laser-material interaction, MRS Bull. 41 (2016) 960–968. https://doi.org/10.1557/mrs.2016.274.

[8] A.P. Caricato, A. Luches, M. Martino, Laser fabrication of nanoparticles, in: Handb. Nanoparticles, Springer International Publishing, 2015: pp. 407–428. https://doi.org/10.1007/978-3-319-15338-4_21.

[9] M. Braun, P. Gilch, W. Zinth, eds., Ultrashort laser pulses in biology and medicine, Springer-Verlag, Berlin Heidelberg, 2008. https://www.springer.com/gp/book/9783540735656.

[10] K. Sugioka, Y. Cheng, Ultrafast lasers—reliable tools for advanced materials processing, Light Sci. Appl. 3 (2014) e149–e149. https://doi.org/10.1038/lsa.2014.30.

[11] A. Ng, Outstanding questions in electron-ion energy relaxation, lattice stability, and dielectric function of warm dense matter, Int. J. Quantum Chem. 112 (2012) 150–160. https://doi.org/10.1002/qua.23197.

[12] J. Rossbach, J.R. Schneider, W. Wurth, 10 years of pioneering X-ray science at the Free-Electron Laser FLASH at DESY, Phys. Rep. 808 (2019) 1. https://doi.org/10.1016/J.PHYSREP.2019.02.002.

[13] J. Bonse, S. Baudach, J. Krüger, W. Kautek, M. Lenzner, Femtosecond laser ablation of silicon-modification thresholds and morphology, Appl. Phys. A. 74 (2002) 19–25. https://doi.org/10.1007/s003390100893.

[14] C.-W. Jiang, X. Zhou, Z. Lin, R.-H. Xie, F.-L. Li, R.E. Allen, Electronic and Structural Response of Nanomaterials to Ultrafast and Ultraintense Laser Pulses, J. Nanosci. Nanotechnol. 14 (2014) 1549–1562. https://doi.org/10.1166/jnn.2014.8756.

[15] N. Medvedev, A.E. Volkov, R. Rymzhanov, F. Akhmetov, S. Gorbunov, R. Voronkov, P.

Babaev, Frontiers, challenges, and solutions in modeling of swift heavy ion effects in materials, J. Appl. Phys. 133 (2023) 100701. https://doi.org/10.1063/5.0128774.

[16] S.P. Hau-Riege, High-intensity X-rays - interaction with matter: processes in plasmas, clusters, molecules and solids, Willey-VCH Verlag, Weinheim, Germany, 2011. http://eu.wiley.com/WileyCDA/WileyTitle/productCd-3527409475.html (accessed October 26, 2015).

[17] R.R. Fäustlin, T. Bornath, T. Döppner, S. Düsterer, E. Förster, C. Fortmann, S.H. Glenzer, S. Göde, G. Gregori, R. Irsig, T. Laarmann, H.J. Lee, B. Li, K.-H. Meiwes-Broer, J. Mithen, B. Nagler, A. Przystawik, H. Redlin, R. Redmer, H. Reinholz, G. Röpke, F. Tavella, R. Thiele, J. Tiggesbäumker, S. Toleikis, I. Uschmann, S.M. Vinko, T. Whitcher, U. Zastrau, B. Ziaja, T. Tschentscher, Observation of Ultrafast Nonequilibrium Collective Dynamics in Warm Dense Hydrogen, Phys. Rev. Lett. 104 (2010) 125002. https://doi.org/10.1103/PhysRevLett.104.125002.

[18] B. Rethfeld, A. Kaiser, M. Vicanek, G. Simon, Ultrafast dynamics of nonequilibrium electrons in metals under femtosecond laser irradiation, Phys. Rev. B. 65 (2002) 214303. https://doi.org/10.1103/PhysRevB.65.214303.

[19] P.L. Silvestrelli, A. Alavi, M. Parrinello, Electrical-conductivity calculation in ab initio simulations of metals:Application to liquid sodium, Phys. Rev. B. 55 (1997) 15515–15522. https://doi.org/10.1103/PhysRevB.55.15515.

[20] C.W. Siders, A. Cavalleri, K. Sokolowski-Tinten, C. Tóth, T. Guo, M. Kammler, M.H. von Hoegen, K.R. Wilson, D. von der Linde, C.P.J. Barty, Detection of nonthermal melting by ultrafast X-ray diffraction, Science. 286 (1999) 1340–1342. https://doi.org/10.1126/science.286.5443.1340.

[21] A. Rousse, C. Rischel, S. Fourmaux, I. Uschmann, S. Sebban, G. Grillon, P. Balcou, E. Förster, J.P. Geindre, P. Audebert, J.C. Gauthier, D. Hulin, Non-thermal melting in semiconductors measured at femtosecond resolution., Nature. 410 (2001) 65–8. https://doi.org/10.1038/35065045.

[22] R.A. Voronkov, N. Medvedev, A.E. Volkov, Superionic State in Alumina Produced by Nonthermal Melting, Phys. Status Solidi - Rapid Res. Lett. 14 (2020) 1900641. https://doi.org/10.1002/pssr.201900641.

[23] R.A. Voronkov, N. Medvedev, A.E. Volkov, Superionic states formation in group III oxides irradiated with ultrafast lasers, Sci. Rep. 12 (2022) 5659. https://doi.org/10.1038/s41598-022-09681-0.

[24] N. Medvedev, A.E. Volkov, Nonthermal acceleration of atoms as a mechanism of fast lattice heating in ion tracks, J. Appl. Phys. 131 (2022) 225903. https://doi.org/10.1063/5.0095724.

[25] P. Stampfli, K. Bennemann, Dynamical theory of the laser-induced lattice instability of silicon, Phys. Rev. B. 46 (1992) 10686–10692. https://doi.org/10.1103/PhysRevB.46.10686.

[26] P. Silvestrelli, A. Alavi, M. Parrinello, D. Frenkel, Ab initio Molecular Dynamics Simulation of Laser Melting of Silicon, Phys. Rev. Lett. 77 (1996) 3149–3152. https://doi.org/10.1103/PhysRevLett.77.3149.

[27] H.O. Jeschke, M.E. Garcia, K.H. Bennemann, Microscopic analysis of the laser-induced femtosecond graphitization of diamond, Phys. Rev. B. 60 (1999) R3701–R3704. https://doi.org/10.1103/PhysRevB.60.R3701.

[28] N. Medvedev, Z. Li, V. Tkachenko, B. Ziaja, Electron-ion coupling in semiconductors beyond Fermi's golden rule, Phys. Rev. B. 95 (2017) 014309. https://doi.org/10.1103/PhysRevB.95.014309.

[29] N. Medvedev, U. Zastrau, E. Förster, D.O. Gericke, B. Rethfeld, Short-time electron dynamics in aluminum excited by femtosecond extreme ultraviolet radiation, Phys. Rev. Lett. 107 (2011) 165003. https://doi.org/10.1103/PhysRevLett.107.165003.

[30] N.S. Shcheblanov, T.E. Itina, Femtosecond laser interactions with dielectric materials: insights of a detailed modeling of electronic excitation and relaxation processes, Appl. Phys. A. 110 (2012) 579–583. https://doi.org/10.1007/s00339-012-7130-0.

[31] S.T. Weber, B. Rethfeld, Phonon-induced long-lasting nonequilibrium in the electron system of a laser-excited solid, Phys. Rev. B. 99 (2019) 174314. https://doi.org/10.1103/PhysRevB.99.174314.

[32] M. Uehlein, S.T. Weber, B. Rethfeld, Influence of Electronic Non-Equilibrium on Energy Distribution and Dissipation in Aluminum Studied with an Extended Two-Temperature Model, Nanomaterials. 12 (2022) 1655. https://doi.org/10.3390/nano12101655.

[33] N. Medvedev, Electronic nonequilibrium effect in ultrafast-laser-irradiated solids, Https://Arxiv.Org/Abs/2302.09098v1. (2023). https://doi.org/10.48550/arxiv.2302.09098.

[34] S.A. Gorbunov, N.A. Medvedev, R.A. Rymzhanov, P.N. Terekhin, A.E. Volkov, Excitation and relaxation of olivine after swift heavy ion impact, Nucl. Instruments Methods Phys. Res. Sect. B Beam Interact. with Mater. Atoms. 326 (2014) 163–168. https://doi.org/10.1016/j.nimb.2013.09.028.

[35] I. Milov, V. Zhakhovsky, D. Ilnitsky, K. Migdal, V. Khokhlov, Y. Petrov, N. Inogamov, V. Lipp, N. Medvedev, B. Ziaja, V. Medvedev, I.A. Makhotkin, E. Louis, F. Bijkerk, Two-level ablation and damage morphology of Ru films under femtosecond extreme UV irradiation, Appl. Surf. Sci. 528 (2020) 146952. https://doi.org/10.1016/j.apsusc.2020.146952.

[36] R. Car, M. Parrinello, Unified Approach for Molecular Dynamics and Density-Functional Theory, Phys. Rev. Lett. 55 (1985) 2471–2474. https://doi.org/10.1103/PhysRevLett.55.2471.

[37] D. Ivanov, L. Zhigilei, Combined atomistic-continuum modeling of short-pulse laser melting and disintegration of metal films, Phys. Rev. B. 68 (2003) 064114. https://doi.org/10.1103/PhysRevB.68.064114.

[38] F. Salvat, M. Fern, PENELOPE-2014 – A code system for Monte Carlo simulation of electron and photon transport, 2015th ed., NUCLEAR ENERGY AGENCY, Organisation for Economic Co-operation and Development, Barcelona, Spain, 2015. https://www.oecd-nea.org/jcms/pl_19590/penelope-2014-a-code-system-for-monte-carlo-simulation-of-electron-and-photon-transport.

[39] N.A. Medvedev, Materials under XUV irradiation: effects of structure, size, and temperature, in: L. Juha, S. Bajt, S. Guizard (Eds.), Opt. Damage Mater. Process. by EUV/X-Ray Radiat., SPIE, 2023: p. 10. https://doi.org/10.1117/12.2664746.

[40] D.P. Kroese, T. Taimre, Z.I. Botev, Handbook of Monte Carlo Methods, Wiley, 2011. https://www.wiley.com/en-au/Handbook+of+Monte+Carlo+Methods-p-9780470177938.

[41] T.M. Jenkins, W.R. Nelson, A. Rindi, Monte Carlo Transport of Electrons and Photons, Springer US, Boston, MA, 1988. https://doi.org/10.1007/978-1-4613-1059-4.

[42] C. Jacoboni, L. Reggiani, The Monte Carlo method for the solution of charge transport in semiconductors with applications to covalent materials, Rev. Mod. Phys. 55 (1983) 645–705. https://doi.org/10.1103/RevModPhys.55.645.

[43] F. James, RANLUX: A Fortran implementation of the high-quality pseudorandom number generator of Lüscher, Comput. Phys. Commun. 79 (1994) 111–114. https://doi.org/10.1016/0010-4655(94)90233-X.

[44] A. Akkerman, T. Boutboul, A. Breskin, R. Chechik, A. Gibrekhterman, Y. Lifshitz, Inelastic Electron Interactions in the Energy Range 50 eV to 10 keV in Insulators: Alkali Halides and Metal Oxides, Phys. Status Solidi Basic Res. 198 (1996) 769–784. https://doi.org/10.1002/pssb.2221980222.

[45] N. Medvedev, A.E. Volkov, Analytically solvable model of scattering of relativistic charged particles in solids, J. Phys. D. Appl. Phys. 53 (2020) 235302. https://doi.org/10.1088/1361-6463/AB7C09.

[46] R. Rymzhanov, N.A. Medvedev, A.E. Volkov, Damage threshold and structure of swift heavy ion tracks in Al2O3, J. Phys. D. Appl. Phys. 50 (2017) 475301. https://doi.org/10.1088/1361-6463/aa8ff5.

[47] N.A. Medvedev, R.A. Rymzhanov, A.E. Volkov, Time-resolved electron kinetics in swift heavy ion irradiated solids, J. Phys. D. Appl. Phys. 48 (2015) 355303. https://doi.org/10.1088/0022-3727/48/35/355303.

[48] U. Fano, Penetration of Protons, Alpha Particles, and Mesons, Annu. Rev. Nucl. Sci. 13 (1963) 1–66. https://doi.org/10.1146/annurev.ns.13.120163.000245.

[49] N. Medvedev, V. Tkachenko, B. Ziaja, Modeling of Nonthermal Solid-to-Solid Phase Transition in Diamond Irradiated with Femtosecond x-ray FEL Pulse, Contrib. to Plasma Phys. 55 (2015) 12–34. https://doi.org/10.1002/ctpp.201400026.

[50] R.H. Ritchie, A. Howie, Electron excitation and the optical potential in electron microscopy, Philos. Mag. 36 (1977) 463–481. https://doi.org/10.1080/14786437708244948.

[51] N. Medvedev, Modeling ultrafast electronic processes in solids excited by femtosecond VUV-XUV laser Pulse, AIP Conf. Proc. 582 (2012) 582–592. https://doi.org/10.1063/1.4739911.

[52] N. Medvedev, F. Akhmetov, R.A. Rymzhanov, R. Voronkov, A.E. Volkov, Modeling time-resolved kinetics in solids induced by extreme electronic excitation, Adv. Theory Simulations. 5 (2022) 2200091. https://doi.org/10.1002/ADTS.202200091.

[53] Y.-K. Kim, M. Rudd, Binary-encounter-dipole model for electron-impact ionization, Phys. Rev. A. 50 (1994) 3954–3967. https://doi.org/10.1103/PhysRevA.50.3954.

[54] D.E. Cullen, EPICS2017: Electron Photon Interaction Cross Sections: w-nds.iaea.org/epics/, Vienna, 2018. https://www-nds.iaea.org/publications/iaea-nds/iaea-nds-224%7B%5C_%7DRev1%7B%5C_%7D2018.pdf.

[55] M.J. Boschini, C. Consolandi, M. Gervasi, S. Giani, D. Grandi, V. Ivanchenko, P. Nieminem, S. Pensotti, P.G. Rancoita, M. Tacconi, An expression for the Mott cross section of electrons and positrons on nuclei with Z up to 118, Radiat. Phys. Chem. 90 (2013) 39–66. https://doi.org/10.1016/J.RADPHYSCHEM.2013.04.020.

[56] F. Salvat, A. Jablonski, C.J. Powell, elsepa—Dirac partial-wave calculation of elastic scattering of electrons and positrons by atoms, positive ions and molecules, Comput. Phys. Commun. 165 (2005) 157–190. https://doi.org/10.1016/j.cpc.2004.09.006.

[57] M.J. Boschini, C. Consolandi, M. Gervasi, S. Giani, D. Grandi, V. Ivanchenko, S. Pensotti, P.G. Rancoita, M. Tacconi, Nuclear and non-ionizing energy-loss for Coulomb scattered particles from low energy up to relativistic regime in space radiation environment, Cosm. Rays Part. Astropart. Phys. - Proc. 12th ICATPP Conf. (2011) 9–23. https://doi.org/10.1142/9789814329033_0002.

[58] S. Agostinelli, J. Allison, K. Amako, J. Apostolakis, H. Araujo, P. Arce, M. Asai, D. Axen, S. Banerjee, G. Barrand, F. Behner, L. Bellagamba, J. Boudreau, L. Broglia, A. Brunengo, H. Burkhardt, S. Chauvie, J. Chuma, R. Chytracek, G. Cooperman, G. Cosmo, P. Degtyarenko, A. Dell'Acqua, G. Depaola, D. Dietrich, R. Enami, A. Feliciello, C. Ferguson, H. Fesefeldt, G. Folger, F. Foppiano, A. Forti, S. Garelli, S. Giani, R. Giannitrapani, D. Gibin, J.J. Gómez Cadenas, I. González, G. Gracia Abril, G. Greeniaus, W. Greiner, V. Grichine, A. Grossheim, S. Guatelli, P. Gumplinger, R. Hamatsu, K. Hashimoto, H. Hasui, A. Heikkinen, A. Howard, V. Ivanchenko, A. Johnson, F.W. Jones, J. Kallenbach, N. Kanaya, M. Kawabata, Y. Kawabata, M. Kawaguti, S. Kelner, P. Kent, A. Kimura, T. Kodama, R. Kokoulin, M. Kossov, H. Kurashige, E.

Lamanna, T. Lampén, V. Lara, V. Lefebure, F. Lei, M. Liendl, W. Lockman, F. Longo, S. Magni, M. Maire, E. Medernach, K. Minamimoto, P. Mora de Freitas, Y. Morita, K. Murakami, M. Nagamatu, R. Nartallo, P. Nieminen, T. Nishimura, K. Ohtsubo, M. Okamura, S. O'Neale, Y. Oohata, K. Paech, J. Perl, A. Pfeiffer, M.G. Pia, F. Ranjard, A. Rybin, S. Sadilov, E. Di Salvo, G. Santin, T. Sasaki, N. Savvas, Y. Sawada, S. Scherer, S. Sei, V. Sirotenko, D. Smith, N. Starkov, H. Stoecker, J. Sulkimo, M. Takahata, S. Tanaka, E. Tcherniaev, E. Safai Tehrani, M. Tropeano, P. Truscott, H. Uno, L. Urban, P. Urban, M. Verderi, A. Walkden, W. Wander, H. Weber, J.P. Wellisch, T. Wenaus, D.C. Williams, D. Wright, T. Yamada, H. Yoshida, D. Zschiesche, Geant4—a simulation toolkit, Nucl. Instruments Methods Phys. Res. Sect. A Accel. Spectrometers, Detect. Assoc. Equip. 506 (2003) 250–303. https://doi.org/https://doi.org/10.1016/S0168-9002(03)01368-8.

[59] A. Ferrari, P.R. Sala, A. Fassò, J. Ranft, Fluka: a multi-particle transport code: http://www.fluka.org/content/manuals/FM.pdf, Geneva, 2005. http://www.fluka.org/content/manuals/FM.pdf.

[60] X-5 Monte Carlo Team, MCNP — A General Monte Carlo N-Particle Transport Code, Version 5 Volume I: Overview and Theory, Revised 10, Los Alamos National Laboratory, University of California, 2003. http://www.nucleonica.net/wiki/images/8/89/MCNPvolI.pdf.

[61] M.L. Knotek, P.J. Feibelman, Stability of ionically bonded surfaces in ionizing environments, Surf. Sci. 90 (1979) 78–90. http://www.sciencedirect.com/science/article/pii/0039602879900116.

[62] Harald O. Jeschke, Theory for optically created nonequilibrium in covalent solids, Technical University of Berlin, 2000. http://www.physics.rutgers.edu/~jeschke/phd.html.

[63] J.G. Powles, G. Rickayzen, D.M. Heyes, Temperatures: old, new and middle aged, Mol. Phys. 103 (2005) 1361–1373. https://doi.org/10.1080/00268970500054664.

[64] R. Darkins, P.-W. Ma, S.T. Murphy, D.M. Duffy, Simulating electronically driven structural changes in silicon with two-temperature molecular dynamics, Phys. Rev. B. 98 (2018) 24304. https://doi.org/10.1103/PhysRevB.98.024304.

[65] S.T. Murphy, S.L. Daraszewicz, Y. Giret, M. Watkins, A.L. Shluger, K. Tanimura, D.M. Duffy, Dynamical simulations of an electronically induced solid-solid phase transformation in tungsten, Phys. Rev. B. 92 (2015) 134110. https://doi.org/10.1103/PhysRevB.92.134110.

[66] A. Jain, A. Sindhu, Pedagogical Overview of the Fewest Switches Surface Hopping Method, ACS Omega. 7 (2022) 45810–45824. https://doi.org/10.1021/acsomega.2c04843.

[67] N. Medvedev, Electron-phonon coupling in semiconductors at high electronic temperatures, Https://Arxiv.Org/Abs/2307.13554. (2023). https://arxiv.org/abs/2307.13554v1 (accessed August 8, 2023).

[68] J.C. Slater, G.F. Koster, Simplified LCAO Method for the Periodic Potential Problem, Phys. Rev. 94 (1954) 1498–1524. https://doi.org/10.1103/PhysRev.94.1498.

[69] P. Koskinen, V. Mäkinen, Density-functional tight-binding for beginners, Comput. Mater. Sci. 47 (2009) 237–253. https://doi.org/10.1016/J.COMMATSCI.2009.07.013.

[70] D. Szczepanik, J. Mrozek, On several alternatives for Löwdin orthogonalization, Comput. Theor. Chem. 1008 (2013) 15–19. https://doi.org/10.1016/J.COMPTC.2012.12.013.

[71] M.J. Mehl, D.A. Papaconstantopoulos, NRL transferable Tight-Binding parameters periodic table: http://esd.cos.gmu.edu/tb/tbp.html, (n.d.). http://esd.cos.gmu.edu/tb/tbp.html (accessed November 21, 2019).

[72] D.A. Papaconstantopoulos, M.J. Mehl, The Slater Koster tight-binding method: a computationally efficient and accurate approach, J. Phys. Condens. Matter. 15 (2003) R413–R440. https://doi.org/10.1088/0953-8984/15/10/201.

[73] C.H. Xu, C.Z. Wang, C.T. Chan, K.M. Ho, A transferable tight-binding potential for carbon, J. Phys. Condens. Matter. 4 (1992) 6047–6054. https://doi.org/10.1088/0953-8984/4/28/006.

[74] I. Kwon, R. Biswas, C. Wang, K. Ho, C. Soukoulis, Transferable tight-binding models for silicon, Phys. Rev. B. 49 (1994) 7242–7250. https://doi.org/10.1103/PhysRevB.49.7242.

[75] Https://dftb.org/, DFTB, (n.d.). https://dftb.org/ (accessed May 20, 2023).

[76] M. Elstner, SCC-DFTB: What Is the Proper Degree of Self-Consistency?†, J. Phys. Chem. A. 111 (2007) 5614–5621. https://doi.org/10.1021/JP071338J.

[77] R.S. Mulliken, Electronic Population Analysis on LCAO–MO Molecular Wave Functions. I, J. Chem. Phys. 23 (1955) 1833–1840. https://doi.org/10.1063/1.1740588.

[78] K.P. Travis, C. Braga, Configurational temperature and pressure molecular dynamics: review of current methodology and applications to the shear flow of a simple fluid, Mol. Phys. 104 (2006) 3735–3749. https://doi.org/10.1080/00268970601014880.

[79] D.C. Rapaport, The Art of Molecular Dynamics Simulation, Cambridge University Press, 2004.

[80] L. Verlet, Computer "Experiments" on Classical Fluids. I. Thermodynamical Properties of Lennard-Jones Molecules, Phys. Rev. 159 (1967) 98–103. https://doi.org/10.1103/PhysRev.159.98.

[81] H. Yoshida, Construction of higher order symplectic integrators, Phys. Lett. A. 150 (1990) 262–268. https://doi.org/10.1016/0375-9601(90)90092-3.

[82] G.J. Martyna, M.E. Tuckerman, Symplectic reversible integrators: Predictor–corrector methods, J. Chem. Phys. 102 (1995) 8071. https://doi.org/10.1063/1.469006.

[83] A. Gonis, M. Däne, Extension of the Kohn-Sham formulation of density functional theory to finite temperature, J. Phys. Chem. Solids. 116 (2018) 86–99. https://doi.org/10.1016/J.JPCS.2017.12.021.

[84] J.F. Littmark, J.P. Ziegler, U. Biersack, The Stopping and Range of Ions in Solids, Pergamon Press, New York, 1985.

[85] M. Toufarová, V. Hájková, J. Chalupský, T. Burian, J. Vacík, V. Vorlíček, L. Vyšín, J. Gaudin, N. Medvedev, B. Ziaja, M. Nagasono, M. Yabashi, R. Sobierajski, J. Krzywinski, H. Sinn, M. Störmer, K. Koláček, K. Tiedtke, S. Toleikis, L. Juha, Contrasting behavior of covalent and molecular carbon allotropes exposed to extreme ultraviolet and soft x-ray free-electron laser radiation, Phys. Rev. B. 96 (2017). https://doi.org/10.1103/PhysRevB.96.214101.

[86] H. Kim, J.-M. Choi, W.A. Goddard, Universal Correction of Density Functional Theory to Include London Dispersion (up to Lr, Element 103), J. Phys. Chem. Lett. 3 (2012) 360–363. https://doi.org/10.1021/jz2016395.

[87] P. Pracht, E. Caldeweyher, S. Ehlert, S. Grimme, A Robust Non-Self-Consistent Tight-Binding Quantum Chemistry Method for large Molecules, (2019). https://doi.org/10.26434/CHEMRXIV.8326202.V1.

[88] R.M. de Oliveira, L.G.M. de Macedo, T.F. da Cunha, F. Pirani, R. Gargano, A Spectroscopic Validation of the Improved Lennard–Jones Model, Mol. 2021, Vol. 26, Page 3906. 26 (2021) 3906. https://doi.org/10.3390/MOLECULES26133906.

[89] H.J.C. Berendsen, J.P.M. Postma, W.F. van Gunsteren, A. DiNola, J.R. Haak, Molecular dynamics with coupling to an external bath, J. Chem. Phys. 81 (1984) 3684–3690. https://doi.org/10.1063/1.448118.

[90] M. Parrinello, A. Rahman, Crystal Structure and Pair Potentials: A Molecular-Dynamics Study, Phys. Rev. Lett. 45 (1980) 1196–1199. https://doi.org/10.1103/PhysRevLett.45.1196.

[91] N. Medvedev, I. Milov, B. Ziaja, Structural stability and electron-phonon coupling in

two-dimensional carbon allotropes at high electronic and atomic temperatures, Carbon Trends. 5 (2021) 100121. https://doi.org/10.1016/J.CARTRE.2021.100121.

[92] N. Medvedev, I. Milov, Electron-Phonon Coupling and Nonthermal Effects in Gold Nano-Objects at High Electronic Temperatures, Materials (Basel). 15 (2022) 4883. https://doi.org/10.3390/MA15144883.

[93] S.A. Gorbunov, N.A. Medvedev, P.N. Terekhin, A.E. Volkov, Electron–lattice coupling after high-energy deposition in aluminum, Nucl. Instruments Methods Phys. Res. Sect. B Beam Interact. with Mater. Atoms. 354 (2015) 220–225. https://doi.org/10.1016/j.nimb.2014.11.053.

[94] Z. Lin, L. Zhigilei, V. Celli, Electron-phonon coupling and electron heat capacity of metals under conditions of strong electron-phonon nonequilibrium, Phys. Rev. B. 77 (2008) 075133. https://doi.org/10.1103/PhysRevB.77.075133.

[95] W.L. McMillan, Transition temperature of strong-coupled superconductors, Phys. Rev. 167 (1968) 331–344. https://doi.org/10.1103/PhysRev.167.331.

[96] P.B. Allen, M.L. Cohen, Pseudopotential calculation of the mass enhancement and superconducting transition temperature of simple metals, Phys. Rev. 187 (1969) 525–538. https://doi.org/10.1103/PhysRev.187.525.

[97] A.M. Brown, R. Sundararaman, P. Narang, W.A. Goddard, H.A. Atwater, Ab initio phonon coupling and optical response of hot electrons in plasmonic metals, Phys. Rev. B. 94 (2016) 075120. https://doi.org/10.1103/PhysRevB.94.075120.

[98] Y. V Petrov, N.A. Inogamov, K.P. Migdal, Thermal conductivity and the electron-ion heat transfer coefficient in condensed media with a strongly excited electron subsystem, JETP Lett. 97 (2013) 20–27. https://doi.org/10.1134/S0021364013010098.

[99] B.Y. Mueller, B. Rethfeld, Relaxation dynamics in laser-excited metals under nonequilibrium conditions, Phys. Rev. B. 87 (2013) 35139. https://doi.org/10.1103/PhysRevB.87.035139.

[100] L. Waldecker, R. Bertoni, R. Ernstorfer, J. Vorberger, Electron-Phonon Coupling and Energy Flow in a Simple Metal beyond the Two-Temperature Approximation, Phys. Rev. X. 6 (2016) 021003. https://doi.org/10.1103/PhysRevX.6.021003.

[101] B. Hüttner, G. Rohr, On the theory of ps and sub-ps laser pulse interaction with metals I. Surface temperature, Appl. Surf. Sci. 103 (1996) 269–274. https://doi.org/10.1016/0169-4332(96)00523-5.

[102] J.L. Hostetler, A.N. Smith, D.M. Czajkowsky, P.M. Norris, Measurement of the Electron-Phonon Coupling Factor Dependence on Film Thickness and Grain Size in Au, Cr, and Al, Appl. Opt. 38 (1999) 3614. https://doi.org/10.1364/AO.38.003614.

[103] Z. Li-Dan, S. Fang-Yuan, Z. Jie, T. Da-Wei, Study on ultra fast nonequilibrium heat transfers in nano metal films by femtosecond laser pump and probe method, Acta Phys. Sin. 61 (2012) 134402. https://doi.org/10.7498/aps.61.134402.

[104] A. Stukowski, Visualization and analysis of atomistic simulation data with OVITO–the Open Visualization Tool, Model. Simul. Mater. Sci. Eng. 18 (2010) 15012. https://doi.org/10.1088/0965-0393/18/1/015012.

[105] W. Humphrey, A. Dalke, K. Schulten, VMD: Visual molecular dynamics, J. Mol. Graph. 14 (1996) 33–38. https://doi.org/10.1016/0263-7855(96)00018-5.

[106] C.F. Macrae, I.J. Bruno, J.A. Chisholm, P.R. Edgington, P. McCabe, E. Pidcock, L. Rodriguez-Monge, R. Taylor, J. van de Streek, P.A. Wood, IUCr, {\textless}i{\textgreater}Mercury CSD 2.0{\textless}/i{\textgreater} – new features for the visualization and investigation of crystal structures, J. Appl. Crystallogr. 41 (2008) 466–470. https://doi.org/10.1107/S0021889807067908.

[107] B. Boulard, J. Kieffer, C.C. Phifer, C.A. Angell, Vibrational spectra in fluoride crystals and

glasses at normal and high pressures by computer simulation, J. Non. Cryst. Solids. 140 (1992) 350–358. https://doi.org/10.1016/S0022-3093(05)80795-1.

[108] S. V Ivliev, The Kubo-Greenwood formula as a result of the random phase approximation for the electrons of the metal, J. Phys. Conf. Ser. 941 (2017) 012047. https://doi.org/10.1088/1742-6596/941/1/012047.

[109] V. Tkachenko, N. Medvedev, Z. Li, P. Piekarz, B. Ziaja, Transient optical properties of semiconductors under femtosecond x-ray irradiation, Phys. Rev. B. 93 (2016) 144101. https://doi.org/10.1103/PhysRevB.93.144101.

[110] F. Trani, G. Cantele, D. Ninno, G. Iadonisi, Tight-binding calculation of the optical absorption cross section of spherical and ellipsoidal silicon nanocrystals, Phys. Rev. B. 72 (2005) 075423. https://doi.org/10.1103/PhysRevB.72.075423.

[111] C.-C. Lee, Y.-T. Lee, M. Fukuda, T. Ozaki, Tight-binding calculations of optical matrix elements for conductivity using nonorthogonal atomic orbitals: Anomalous Hall conductivity in bcc Fe, Phys. Rev. B. 98 (2018) 115115. https://doi.org/10.1103/PhysRevB.98.115115.

[112] P. Yeh, Optical Waves in Layered Media, Volume 61, Wiley, the University of California, Santa Barbara, 2005. http://eu.wiley.com/WileyCDA/WileyTitle/productCd-0471731927.html (accessed December 15, 2015).

[113] Y. Petrov, K. Migdal, N. Inogamov, V. Khokhlov, D. Ilnitsky, I. Milov, N. Medvedev, V. Lipp, V. Zhakhovsky, Ruthenium under ultrafast laser excitation: Model and dataset for equation of state, conductivity, and electron-ion coupling, Data Br. 28 (2020) 104980. https://doi.org/10.1016/j.dib.2019.104980.

[114] V. Recoules, J.P. Crocombette, Ab initio determination of electrical and thermal conductivity of liquid aluminum, Phys. Rev. B - Condens. Matter Mater. Phys. 72 (2005) 104202. https://doi.org/10.1103/PHYSREVB.72.104202/FIGURES/3/MEDIUM.

[115] K.F. Garrity, K. Choudhary, Fast and Accurate Prediction of Material Properties with Three-Body Tight-Binding Model for the Periodic Table, Https://Arxiv.Org/Abs/2112.11585. (2021). https://arxiv.org/abs/2112.11585 (accessed June 3, 2022).

[116] J. Jenke, A.N. Ladines, T. Hammerschmidt, D.G. Pettifor, R. Drautz, Tight-binding bond parameters for dimers across the periodic table from density-functional theory, Phys. Rev. Mater. 5 (2021) 23801. https://doi.org/10.1103/PhysRevMaterials.5.023801.

[117] D. Porezag, T. Frauenheim, T. Köhler, G. Seifert, R. Kaschner, Construction of tight-binding-like potentials on the basis of density-functional theory: Application to carbon, Phys. Rev. B. 51 (1995) 12947–12957. https://doi.org/10.1103/PhysRevB.51.12947.

[118] N. Medvedev, Femtosecond X-ray induced electron kinetics in dielectrics: application for FEL-pulse-duration monitor, Appl. Phys. B. 118 (2015) 417–429. https://doi.org/10.1007/s00340-015-6005-4.

[119] S. Hammes-Schiffer, J.C. Tully, Proton transfer in solution: Molecular dynamics with quantum transitions, J. Chem. Phys. 101 (1994) 4657. https://doi.org/10.1063/1.467455.

[120] J.C. Tully, Molecular dynamics with electronic transitions, J. Chem. Phys. 93 (1990) 1061. https://doi.org/10.1063/1.459170.

[121] M. Harmand, R. Coffee, M. Bionta, M. Chollet, D. French, D.M. Zhu, D.T. Fritz, H. Lemke, N. Medvedev, B. Ziaja, S. Toleikis, M. Cammarata, Achieving few-femtosecond time-sorting at hard X-ray free-electron lasers, Nat Phot. 7 (2013) 215–218. https://doi.org/10.1038/nphoton.2013.11.

[122] C.-C. Fu, M. Weissmann, Tight-binding molecular-dynamics study of amorphous carbon deposits over silicon surfaces, Phys. Rev. B. 60 (1999) 2762–2770.

https://doi.org/10.1103/PhysRevB.60.2762.

[123] C. Molteni, L. Colombo, L. Miglio, Tight-binding molecular dynamics in liquid III-V compounds. I. Potential generation, J. Phys. Condens. Matter. 6 (1994) 5243–5254. https://doi.org/10.1088/0953-8984/6/28/003.

[124] N. Medvedev, H.O. Jeschke, B. Ziaja, Nonthermal graphitization of diamond induced by a femtosecond x-ray laser pulse, Phys. Rev. B. 88 (2013) 224304. https://doi.org/10.1103/PhysRevB.88.224304.

[125] H.J. Monkhorst, J.D. Pack, Special points for Brillouin-zone integrations, Phys. Rev. B. 13 (1976) 5188–5192. https://doi.org/10.1103/PhysRevB.13.5188.

[126] N.A. Medvedev, H.O. Jeschke, B. Ziaja, Non-thermal phase transitions in semiconductors under femtosecond XUV irradiation, SPIE Proc. 8777 (2013) 877709-877709–10. https://doi.org/10.1117/12.2019123.

[127] C.J. Fennell, J.D. Gezelter, Is the Ewald summation still necessary? Pairwise alternatives to the accepted standard for long-range electrostatics, J. Chem. Phys. 124 (2006) 234104. https://doi.org/10.1063/1.2206581.

[128] F. Tavella, H. Höppner, V. Tkachenko, N. Medvedev, F. Capotondi, T. Golz, Y. Kai, M. Manfredda, E. Pedersoli, M.J. Prandolini, N. Stojanovic, T. Tanikawa, U. Teubner, S. Toleikis, B. Ziaja, Soft x-ray induced femtosecond solid-to-solid phase transition, High Energy Density Phys. 24 (2017) 22. https://doi.org/10.1016/j.hedp.2017.06.001.

[129] J. Frenzel, A.F. Oliveira, N. Jardillier, T. Heine, G. Seifert, Semi-relativistic, self-consistent charge Slater-Koster tables for density-functional based tight-binding (DFTB) for materials science simulations., Dresden, 2009. http://www.dftb.org/parameters/download/matsci/matsci-0-3-cc/.

[130] M.J. Mehl, D.A. Papaconstantopoulos, Applications of a tight-binding total-energy method for transition and noble metals: Elastic constants, vacancies, and surfaces of monatomic metals, Phys. Rev. B. 54 (1996) 4519–4530. https://doi.org/10.1103/PhysRevB.54.4519.

[131] H. Shi, P. Koskinen, A. Ramasubramaniam, Self-Consistent Charge Density-Functional Tight-Binding Parametrization for Pt–Ru Alloys, J. Phys. Chem. A. 121 (2017) 2497–2502. https://doi.org/10.1021/acs.jpca.7b00701.

[132] M. Van den Bossche, DFTB-Assisted Global Structure Optimization of 13- and 55-Atom Late Transition Metal Clusters, J. Phys. Chem. A. 123 (2019) 3038–3045. https://doi.org/10.1021/acs.jpca.9b00927.

[133] C. Liu, E.R. Batista, N.F. Aguirre, P. Yang, M.J. Cawkwell, E. Jakubikova, SCC-DFTB Parameters for Fe–C Interactions, J. Phys. Chem. A. 124 (2020) 9674–9682. https://doi.org/10.1021/acs.jpca.0c08202.

[134] M. Van den Bossche, C. Noguera, J. Goniakowski, Understanding the structural diversity of freestanding Al2O3 ultrathin films through a DFTB-aided genetic algorithm, Nanoscale. 12 (2020) 6153–6163. https://doi.org/10.1039/C9NR10487A.

[135] J. Lee, S. Ganguli, A.K. Roy, S.C. Badescu, Density functional tight binding study of β-Ga2O3: Electronic structure, surface energy, and native point defects, J. Chem. Phys. 150 (2019) 174706. https://doi.org/10.1063/1.5088941/198541.

[136] S. Anniés, C. Panosetti, M. Voronenko, D. Mauth, C. Rahe, C. Scheurer, Accessing Structural, Electronic, Transport and Mesoscale Properties of Li-GICs via a Complete DFTB Model with Machine-Learned Repulsion Potential, Materials (Basel). 14 (2021) 6633. https://doi.org/10.3390/ma14216633.

[137] T. Dumitrică, R.E. Allen, Nonthermal transition of GaAs in ultra-intense laser radiation field,

Laser Part. Beams. 20 (2002) 237–242. https://doi.org/10.1017/S026303460220213X.

[138] A. Carlson, An Extended Tight-Binding Approach for Modeling Supramolecular Interactions of Carbon Nanotubes, UNIVERSITY OF MINNESOTA, 2006. http://www.me.umn.edu/%7B~%7Ddtraian/tony-thesis.pdf.

[139] K.P. Migdal, D.K. Il'nitsky, Y. V Petrov, N.A. Inogamov, Equations of state, energy transport and two-temperature hydrodynamic simulations for femtosecond laser irradiated copper and gold, J. Phys. Conf. Ser. 653 (2015) 12086. https://doi.org/10.1088/1742-6596/653/1/012086.